\documentclass[paper]{JHEP3}
\usepackage{epsfig,subfigure}
\usepackage{amssymb}
\usepackage{amsmath}
\usepackage{booktabs,multirow,tabularx}
\usepackage{slashed}

\def\d#1{D_{#1}}
\def\db#1{\bar D_{#1}}

\newcommand{\bqa}{\begin{eqnarray}}
\newcommand{\eqa}{\end{eqnarray}}
\newcommand{\nl}{\nonumber \\}

\def\beq{\begin{equation}}
\def\eeq{\end{equation}}
\def\beqn{\begin{eqnarray}}
\def\eeqn{\end{eqnarray}}
\def\abs#1{\left|#1\right|}

\newcommand\sss{\scriptscriptstyle}
\newcommand\mydot{\!\cdot\!}
\newcommand\ep{\epsilon}
\newcommand\half{\frac{1}{2}}

\newcommand\as{\alpha_{\sss S}}
\newcommand\gs{g_{\sss S}}

\newcommand\MadFKS{{\sc\small MadFKS}}
\newcommand\CutTools{{\sc\small CutTools}}
\newcommand\MadLoop{{\sc\small MadLoop}}
\newcommand\MadGraph{{\sc\small MadGraph}}
\newcommand\HELAS{{\sc\small HELAS}}
\newcommand\Lcut{L-cut}
\newcommand\Lcutting{L-cutting}
\newcommand\qloop{\ell}
\newcommand\bqloop{\bar{\ell}}
\newcommand\tqloop{\tilde{\ell}}
\newcommand\proc{r}
\newcommand\ident{{\cal I}}
\newcommand\amp{{\cal A}}
\newcommand\ampnt{\amp^{(n,0)}}
\newcommand\ampnl{\amp^{(n,1)}}

\newcommand\ampnX{\amp^{(n,{\sss\rm X})}}
\newcommand\ampnRt{\amp^{(n,{\sss R_2})}}
\newcommand\ampnUV{\amp^{(n,{\sss\rm UV})}}
\newcommand\mtop{m_{top}}
\newcommand\mQ{m_Q}
\newcommand\muF{\mu_{\sss F}}
\newcommand\muR{\mu_{\sss R}}
\newcommand\bu{\bar{u}}
\newcommand\bd{\bar{d}}
\newcommand\bb{\bar{b}}
\newcommand\bt{\bar{t}}
\newcommand\bQ{\bar{Q}}
\newcommand\qpart{q^\star}
\newcommand\qbpart{\bar{q}^\star}
\newcommand\diag{{\cal C}}
\newcommand\idiag{C}
\newcommand\unr{{\rm\sss U}}
\newcommand\etal{{\em et~al.}}
\newcommand\pt{p_{\sss T}}
\newcommand\kt{k_{\sss T}}
\newcommand\NC{N_{\sss c}}
\newcommand\CA{C_{\sss A}}
\newcommand\CF{C_{\sss F}}
\newcommand\TF{T_{\sss F}}

\newcommand\xicut{\xi_{cut}}
\newcommand\deltaO{\delta_{\sss O}}
\newcommand\deltaI{\delta_{\sss I}}

\preprint{
 CERN-PH-TH/2011-031\\
 CP3-11-07\\
 ZU-TH 01/11
 }
\title{Automation of one-loop QCD corrections%
}
\author{Valentin Hirschi\\
  ITPP, EPFL, CH-1015 Lausanne, Switzerland
}
\author{Rikkert Frederix\\
Institut f\"ur Theoretische Physik, Universit\"at Z\"urich,
Winterthurerstrasse 190,\\ CH-8057 Z\"urich, Switzerland
}
\author{Stefano Frixione%
  \thanks{On leave of absence from INFN, Sezione di Genova, Italy.}\\
  PH Department, TH Unit, CERN, CH-1211 Geneva 23, Switzerland\\
  ITPP, EPFL, CH-1015 Lausanne, Switzerland
}
\author{Maria Vittoria Garzelli\\
  INFN, Sezione di Milano, I-20133 Milano, Italy\\
  Departamento de F\'\i sica Te\'orica y del Cosmos y CAFPE\\
  Universidad de Granada, E-18071 Granada, Spain
}
\author{Fabio Maltoni\\
  Centre for Cosmology, Particle Physics and Phenomenology (CP3)\\
  Universit\'{e} catholique de Louvain\\
  Chemin du Cyclotron 2, B-1348 Louvain-la-Neuve, Belgium
}
\author{Roberto Pittau%
  \thanks{On leave of absence from Departamento de 
          F\'\i sica Te\'orica y del Cosmos y
          CAFPE, Universidad de Granada.}\\
  PH Department, TH Unit, CERN, CH-1211 Geneva 23, Switzerland
}
\abstract{
 We present the complete automation of the computation of
 one-loop QCD corrections, including UV renormalization, to an
 arbitrary scattering process in the Standard Model. This is achieved
 by embedding the OPP integrand reduction technique, as
 implemented in CutTools, into the MadGraph framework.
 By interfacing the tool so constructed, which we dub MadLoop, with MadFKS, 
 the fully automatic computation of any infrared-safe observable at the 
 next-to-leading order in QCD is attained. We demonstrate the 
 flexibility and the reach of our method by calculating the production 
 rates for a variety of processes at the 7~TeV LHC.
}
\keywords{QCD, NLO Computations, Hadronic Colliders}

\begin{document}

\section{Introduction\label{sec:intro}}
The capability of improving systematically the predictions for any
given observable by means of perturbative techniques has been 
of great importance in helping establish the Standard Model
as the correct theory of electroweak and strong interactions for
sub-TeV energies. Although there exist many different choices for
the expansion parameter of the perturbative series, the most 
common one is by far that of the coupling constant, which
we identify here with that of QCD, $\as$. A power series in
$\as$ is expected to converge well (asymptotically) for
observables that are sufficiently inclusive (such as total rates), 
or are associated with small-probability events (such as large-angle, 
large-energy emissions). The computation of contributions of increasingly 
higher orders in $\as$ for a given observable is analogous to, and goes
hand-in-hand with, the accumulation of data in a real experiment. 
While the latter results
in the decrease of the statistical error associated with the measurement,
the former implies the reduction of the theoretical uncertainty affecting
the prediction, as determined by the dependence on the unphysical
mass scales that enter the computation. 
Precise determinations on both the theoretical and experimental sides have
been essential in the success of physics programmes at particle colliders,
and continue playing a very important role in pinning down potential 
discrepancies between predictions and observations. It should be further
stressed that some discovery strategies pursued by LHC experiments 
(which are thus analogous, in this sense, e.g. to the single-top 
analysis of CDF and D0) make extensive use of theoretical predictions, whose
accuracy is crucial to reduce as much as possible any theoretical
bias on evidence of new physics.

The reduction of the uncertainties that affect theoretical 
predictions is only one of the consequences of computing
higher orders in the perturbative series. In general, one expects to
find corrections that are non-trivial, in the sense
that they cannot possibly be obtained by simply rescaling the 
leading-order (LO) results by a constant. Hadroproduction processes have a 
particularly rich structure from this point of view, since typically
at each order in perturbation theory new partonic channels open up, which
bring about an involved dependence on parton density functions (PDFs).
However, it is easy to realize that the vast majority of these ``dynamical'' 
effects can be approximated in an excellent manner by just keeping the
contributions to the perturbative expansion due to tree-level processes,
subject to suitable cuts (that prevent them from diverging when 
integrated over the phase space). The fact that such an approximation
is actually fairly good is the main reason why these tree-level
computations (which are equivalent, process-by-process, to LO results)
have enjoyed an enormous success in recent years: from the technical
viewpoint, they are incredibly simpler than proper higher-order computations.

It is therefore clear that the decision on whether to perform a fully-fledged
higher order computation for a given observable must be taken after careful
consideration of the benefits versus costs. It is obvious that an increased
precision in the theoretical predictions is always beneficial. Unfortunately,
most of the physics we are interested in studying at the LHC involves
high-multiplicity processes for which even the first non-trivial order
in perturbation theory, i.e.~the next-to-leading order (NLO) one, is 
horrendously complicated to compute. Furthermore, owing to the presence
of many different mass scales, the results will display a non-negligible 
theoretical uncertainty, although in general much reduced w.r.t.~that 
of the LO predictions.

An easy way to solve the issue above is that of reducing to zero the
costs of performing higher-order computations, by letting a computer
do the job through a complete automation of the necessary procedures.
The ultimate goal of the 
work presented in this paper is that of showing that this is an actual
possibility, as far as NLO results are concerned. As is well known,
NLO calculations can be achieved by performing two tasks of overall
comparable complexity. One of them is the computation of the one-loop 
matrix elements. The second of them is the computation of the tree-level 
matrix elements, followed by their combination with the one-loop ones, 
their recasting into terms which are locally non-divergent, and finally
their integration over the phase space.
The complete automation of the latter task was presented in
ref.~\cite{Frederix:2009yq};
that of the former task is the subject of this paper.

The present work has therefore two immediate aims.
\begin{itemize}
\item A technical aim. Namely, the complete automation of the computation 
 of QCD one-loop corrections for any user-defined process in the Standard 
 Model, independently of external particle identities (in particular, whether 
 they are massless or massive). This is achieved by embedding the 
 Ossola-Papadopoulos-Pittau integrand reduction technique~\cite{Ossola:2006us} 
 (through its incarnation in \CutTools) into the \MadGraph\ framework.
 The resulting tool, which we call \MadLoop, is an independent software
 module that, given the process, returns the finite part of the one-loop
 corrections (UV renormalized), and the residues of the double and single 
 infrared poles.
\item A physics aim. By making use of \MadLoop\ and of \MadFKS\ (the 
 automated software developed in ref.~\cite{Frederix:2009yq}), we 
 present results for the total cross sections for a variety of 
 processes at the 7~TeV LHC.
\end{itemize}
We point out that the second item here demonstrates the achievement
of the ultimate goal we have stated above. Namely, that one can 
compute fully-physical NLO results at zero cost, given that the only
human interventions necessary to obtain the cross sections (and any
other infrared-safe observable) are those of specifying the process, 
of defining the input parameters, and possibly of imposing final-state cuts 
where necessary. 

This paper is organized as follows. In sect.~\ref{sec:res} we present a few 
representative physics results. In sect.~\ref{sec:auto} we illustrate the 
main features of the way in which we perform our computations. The current
version of \MadLoop\ has a few limitations, which we describe in
sect.~\ref{sec:limits}. The steps to be taken in order to remove
these limitations, and other future improvements, are reported
in sect.~\ref{sec:outlook}. We give our conclusions 
in sect.~\ref{sec:conclusions}. Appendix~\ref{sec:comps} shows
all comparisons between the results obtained with \MadLoop\ and
those available in the literature. Appendix~\ref{sec:MLtechn}
reports some technical details.

\section{Selected physics results\label{sec:res}}
In this section, we present a sample of the results that
one can obtain with \MadFKS\ and \MadLoop. Although one typically
discusses technical details before reporting their 
phenomenological implications, the ordering adopted in this paper
underscores our belief that, thanks to the automation achieved
here, NLO QCD corrections to Standard Model processes 
are as trivial to calculate as LO results, and must thus 
be treated on equal footing with the latter. This fact has two main 
implications. Firstly, {\em technicalities are independent of the process}, 
and can be understood once and for all. Secondly, the choice of
which processes to consider is essentially limited only by
CPU-time considerations\footnote{With some minor exceptions,
to be discussed in sect.~\ref{sec:limits}.}.

It is important to keep in mind that \MadFKS\ and \MadLoop\ are
fully independent modules, which communicate with a standardized
interface as prescribed by the Binoth-Les Houches accord~\cite{Binoth:2010xt}.
\MadFKS\ has already employed several programs other than \MadLoop\ for the 
computation of one-loop contributions: BlackHat~\cite{Berger:2009zg}, 
Rocket~\cite{Giele:2008bc}, and Samurai~\cite{Mastrolia:2010nb} 
(see e.g.~ref.~\cite{Frederix:2010ne} for a recent application).
Likewise, \MadLoop\ can be called by any code compatible with the
accord of ref.~\cite{Binoth:2010xt}.
It is also appropriate to remind the reader that in the first 
paper on \MadFKS\ (ref.~\cite{Frederix:2009yq}) 
numerical results were presented only for $e^+e^-$ collisions.
However, all the formulae necessary for dealing with hadronic
collisions were laid out in that paper, and their subsequent
implementation in \MadFKS\ has not posed any problem.
In this work, we present for the first time hadroproduction
results obtained with \MadFKS. 

We remind the reader that \MadFKS\ is based on the FKS universal 
subtraction formalism~\cite{Frixione:1995ms,Frixione:1997np}.
The FKS method is very effective in limiting the total number
of subtractions, which scale at most as $n^2$, with
$n$ the number of strongly-interacting particles that enter
the process. Furthermore, the method renders it particularly easy
to further reduce the subtractions to perform, by efficiently exploiting
the symmetries of the matrix elements in the case of identical
final-state particles. In the FKS formalism one introduces three
arbitrary parameters ($\xicut$, $\deltaI$, and $\deltaO$, associated
with soft, initial-state collinear, and final-state collinear singularities
respectively) that control the subtractions, and of which any
physical observable is strictly independent. That such an independence
does indeed occur is a powerful check of the correct implementation
of the method, and we have extensively verified it. We have also 
found that in hadroproduction the same feature holds as we have documented
in ref.~\cite{Frederix:2009yq} for the case of $e^+e^-$ collisions.
Namely, that the convergence of phase-space integration depends
very mildly on $\xicut$, $\deltaI$, and $\deltaO$. This is the signal
that such an integration is very robust, and implies that it is not
necessary to search for the values of the arbitrary parameters introduced
above that are ``optimal'' for physical predictions, since essentially
any values will do.

Given that our goal here is not that of performing any phenomenological
studies, we have considered processes that feature a fair diversity
(pure QCD, with EW vector bosons, with SM Higgs, with massive/massless
$b$ quarks, and at different jet multiplicities) in order to fully test
and prove the flexibility of our setup. We have limited ourselves 
to presenting results for total rates.
For reasons of space, it is impossible to report here even a
small sample of differential distributions. We remark that we
have checked a few standard ones (such as transverse momenta,
rapidities, and pair invariant masses), and have found that the
statistics used to obtain fairly accurate results
for total rates is also sufficient to get reasonably smooth
distributions. This is consistent with past experience with
FKS subtraction, and is ultimately due to the fact that in
this formalism, for any given integration channel, the total number 
of kinematic configurations associated with subtraction counterterms 
is always equal to one, thereby reducing to a minimum the 
probability of mis-binning (see ref.~\cite{Frederix:2009yq}).

\begin{table}
\begin{center}
\begin{tabular}{ll|ll}\toprule
Parameter & value & Parameter & value
\\\midrule
$m_{\sss Z}$ & 91.188 & $\alpha^{-1}$ & 132.50698 
\\
$m_{\sss W}$ & 80.419 & $G_F$         & $1.16639\!\cdot\!10^{-5}$
\\
$m_b$        & 4.75   & ${\rm CKM}_{ij}$ & $\delta_{ij}$
\\
$\mtop$      & 172.5  & $\Gamma_Z$ & 2.4414 
\\
$m_H$        & 120    & $\Gamma_W$ & 2.0476 
\\
$\as^{({\rm NLO},5)}(m_{\sss Z})$ & 0.120179 & 
$\as^{({\rm NLO},4)}(m_{\sss Z})$ & 0.114904
\\
$\as^{({\rm LO},5)}(m_{\sss Z})$  & 0.139387 & 
$\as^{({\rm LO},4)}(m_{\sss Z})$  & 0.133551
\\\bottomrule
\end{tabular}
\end{center}
\caption{\label{tab:params}
General settings of physical parameters used for the computations of the
cross sections in table~\ref{tab:results}, with dimensionful quantities
given in GeV. The upper indices on $\as$ indicate whether the coupling
has been used to obtain an NLO or an LO result, with five or four light
flavours, and the corresponding values are dictated by the choice of
PDFs. Some processes may adopt specific parameter values, different from 
those reported in this table; in particular, the $b$ quark can be 
treated as massless. See the text for details.
}
\end{table}
Physics results are simply obtained by giving in input to
\MadFKS\ and \MadLoop\ the process type, and the QCD and EW
parameters to be used in the runs. Code-writing is limited
to defining the observable(s) one wants to predict, and to
imposing cuts (including the appropriate jet finders).
We simulate $pp$ collisions at 7~TeV. Masses, couplings, and
widths are chosen as reported in table~\ref{tab:params}, with
some process-specific exceptions, to be described below.
For NLO (LO) results with five light flavours, we use the
MSTW2008nlo (MSTW2008lo) PDF set~\cite{Martin:2009iq}, while in the case of
four light flavours we adopt MSTW2008nlo\_nf4 (MSTW2008lo\_nf4).
Each of these sets is associated with a different value of $\as$, which
we report in table~\ref{tab:params}. Jets are defined using
the $\kt$-clustering algorithm~\cite{Catani:1993hr} 
(as implemented in FastJet~\cite{Cacciari:2005hq}),
with $\pt^{(jet)}>25~\textrm{GeV}$ and pseudo-cone 
size $\Delta R=0.7$. Renormalization and factorization scales
are set equal to a common value,
\beqn
\mu\equiv \muR=\muF\,.
\eeqn
Since we present results for 
total cross sections, it is appropriate to assign a fixed (i.e., that
does not depend on the kinematics) value to $\mu$, which is 
process-dependent as reported in table~\ref{tab:results}.
The results are therefore easily reproducible and can be used 
as a standard reference.
In table~\ref{tab:results}, by $n_{lf}$ we have denoted the number of 
quarks whose mass is equal to zero. Thus $n_{lf}$ is equal to five or to four 
when the $b$ quark is considered to be massless or massive, respectively.
In all cases, all six quark flavours have been included in the loops.

\begin{table}
\begin{center}
\begin{small}
\begin{tabularx}{\textwidth}
{lr@{$\,\to\,$}lccr@{$\,\pm\,$}lr@{$\,\pm\,$}X}
\toprule
\multicolumn{3}{c}{Process~~~~~~~~~~~~~~} & 
  $\mu$ & $n_{lf}$ &
  \multicolumn{4}{c}{Cross section (pb)}
\\
\multicolumn{3}{c}{} &&& \multicolumn{2}{c}{LO} & \multicolumn{2}{c}{NLO}
\\\midrule
a.1 & $pp$ & $t\bar{t}$     & $\mtop$   &5& $123.76$ & $0.05$ & $162.08$ & $0.12$ \\
a.2 & $pp$ & $tj$           & $\mtop$   &5& $34.78$ & $0.03$ & $41.03$ & $0.07$ 
\\
a.3 & $pp$ & $tjj$          & $\mtop$   &5& $11.851$ & $0.006$ & $13.71$ & $0.02$
\\
a.4 & $pp$ & $t\bar{b}j$    & $\mtop/4$ &4& $31.37$ & $0.03$ & $32.86$ & $0.04$
\\
a.5 & $pp$ & $t\bar{b}jj$   & $\mtop/4$ &4& $11.91$ & $0.006$ & $7.299$ & $0.05$
\\\midrule
b.1 & $pp$ & $(W^+\to) e^+\nu_e$          & $m_W$ &5& $5072.5$ & $2.9$ & $6146.2$ & $9.8$ 
\\
b.2 & $pp$ & $(W^+\to) e^+\nu_e\,j$       & $m_W$ &5& $828.4$ & $0.8$ & $1065.3$ & $1.8$  
\\
b.3 & $pp$ & $(W^+\to) e^+\nu_e\,jj$      & $m_W$ &5& $298.8$ & $0.4$ & $289.7$ & $0.3$
\\
b.4 & $pp$ & $(\gamma^*/Z\to) e^+e^-$     & $m_Z$ &5& $1007.0$ & $0.1$ & $1170.0$ & $2.4$ 
\\
b.5 & $pp$ & $(\gamma^*/Z\to) e^+e^-\,j$  & $m_Z$ &5& $156.11$ & $0.03$ & $203.0$ & $0.2$ 
\\
b.6 & $pp$ & $(\gamma^*/Z\to) e^+e^-\,jj$ & $m_Z$ &5& $54.24$ & $0.02$ & $54.1$ & $0.6$ 
\\\midrule
c.1 & $pp$ & $(W^+\to) e^+\nu_e b\bar{b}$      & $m_W+2m_b$ &4& $11.557$ & $0.005$ & $22.95$ & $0.07$ 
\\
c.2 & $pp$ & $(W^+\to) e^+\nu_e t\bar{t}$      & $m_W+2\mtop$ &5& $0.009415$ & $0.000003$ & $0.01159$ & $0.00001$ 
\\
c.3 & $pp$ & $(\gamma^*/Z\to) e^+e^- b\bar{b}$ & $m_Z+2m_b$ &4& $9.459$ & $0.004$ & $15.31$ & $0.03$ 
\\
c.4 & $pp$ & $(\gamma^*/Z\to) e^+e^- t\bar{t}$ & $m_Z+2\mtop$ &5& $0.0035131$ & $0.0000004$ & $0.004876$ & $0.000002$ 
\\
c.5 & $pp$ & $\gamma t\bar{t}$                & $2\mtop$     &5& $0.2906$ & $0.0001$ & $0.4169$ & $0.0003$ 
\\\midrule
d.1 & $pp$ & $W^+W^-$     & $2m_W$ &4& $29.976$ & $0.004$ & $43.92$ & $0.03$ 
\\
d.2 & $pp$ & $W^+W^-\,j$  & $2m_W$ &4& $11.613$ & $0.002$ & $15.174$ & $0.008$ 
\\
d.3 & $pp$ & $W^+W^+\,jj$ & $2m_W$ &4& $0.07048$ & $0.00004$ & $0.08241$ & $0.0004$
\\\midrule
e.1 & $pp$ & $HW^+$        & $m_W+m_H$ &5& $0.3428$ & $0.0003$ & $0.4455$ & $0.0003$ 
\\
e.2 & $pp$ & $HW^+\,j$     & $m_W+m_H$ &5& $0.1223$ & $0.0001$ & $0.1501$ & $0.0002$ 
\\
e.3 & $pp$ & $HZ$        & $m_Z+m_H$ &5& $0.2781$ & $0.0001$ & $0.3659$ & $0.0002$ 
\\
e.4 & $pp$ & $HZ\,j$     & $m_Z+m_H$ &5& $0.0988$ & $0.0001$ & $0.1237$ & $0.0001$ 
\\
e.5 & $pp$ & $Ht\bar{t}$ & $\mtop+m_H$ &5& $0.08896$ & $0.00001$ & $0.09869$ & $0.00003$ 
\\
e.6 & $pp$ & $Hb\bar{b}$ & $m_b+m_H$ &4& $0.16510$ & $0.00009$ & $0.2099$ & $0.0006$ 
\\
e.7 & $pp$ & $Hjj$       & $m_H$     &5& $1.104$ & $0.002$ & $1.333$ & $0.002$
\\\bottomrule
\end{tabularx}
\end{small}
\end{center}
\caption{\label{tab:results}
Results for total rates, possibly within cuts, at the 7~TeV LHC,
obtained with \MadFKS\ and \MadLoop. The errors are due to the statistical
uncertainty of Monte Carlo integration. See the text for details.
}
\end{table}
As discussed in ref.~\cite{Frederix:2009yq}, \MadFKS\ allows one to 
integrate all contributions to the NLO cross section in one single 
computation, regardless of whether they have a real-emission or a Born-type 
kinematics. For the results presented here, however, we have integrated 
the one-loop contributions separately from the other ones (i.e., the Born
and real-emission matrix elements, and the subtraction counterterms). 
This is because for a given phase-space point the evaluation of 
virtual corrections performed by \MadLoop\ takes much longer than 
all the other operations carried out by \MadFKS. On the other hand,
no phase-space subtraction is done on virtual corrections, and therefore 
the numerical computations are inherently more stable than those relevant
to the subtracted real-emission contributions. Hence, it turns out
to be more efficient to integrate the one-loop contributions separately 
from all the others, using a reduced statistics (on average, about 
one-tenth\footnote{A more precise figure is difficult to give, since
the total number of integration points per channel is determined
dynamically, in order for the various channels to contribute to the total 
rate with similar absolute accuracies.} of that employed for real corrections).
Even so, for the processes with the highest multiplicities considered here,
the virtual corrections require more computing time than the rest of the
calculation, in order to attain similar integration uncertainties. 
This is actually good news since, as discussed in sects.~\ref{sec:limits}
and~\ref{sec:outlook}, the amount of optimization included in \MadLoop\
is so far very minimal, and hence we expect to reduce the CPU-time 
load in a significant way in the near future.

We finally mention a few technical points relevant to phase-space
integration. All contributions to the cross sections are integrated 
using multi-channel techniques, following the procedure outlined in 
ref.~\cite{Frederix:2009yq}. The sums over colours and helicities
are performed explicitly (we point out that both \MadFKS\ and \MadLoop\ 
are equipped to carry out helicity sums with Monte Carlo methods, but 
this was simply not necessary for the processes considered here). The 
virtual contributions are integrated in a direct manner, i.e.~no 
reweighting by the Born matrix elements has been performed.
Further process-specific comments are given in what follows.
\begin{itemize}
\item A cut
\beqn
m_{e^+e^-}>30 \textrm{ GeV}
\eeqn
has been applied to processes b.4, b.5, b.6, c.3, and c.4.
\item In the case of process c.5, the photon has been isolated
with the prescription of ref.~\cite{Frixione:1998jh}, with
parameters
\beqn
\delta_0=0.4\,,\;\;\;\;\;\;
n=1\,,\;\;\;\;\;\;
\epsilon_\gamma=1\,,
\eeqn
and parton-parton or parton-photon distances defined in the 
$\langle\eta,\varphi\rangle$ plane. The photon is also required to
be hard and central:
\beqn
\pt^{(\gamma)}\ge 20~{\rm GeV}\,,\;\;\;\;\;\;
\abs{\eta^{(\gamma)}}\le 2.5\,.
\eeqn
\item In the case of processes a.3, a.4, a.5, and e.7, diagrams
with EW vector bosons in the loops have been removed, which is
necessary given that a complex-mass scheme is presently not
implemented (see sect.~\ref{sec:limits}). These contributions 
are colour-suppressed.
\item In the case of processes a.3, a.4, and a.5, diagrams
with s-channel $W$'s have been removed, in order to avoid
$t\bt$-type resonances. In the narrow width approximation this 
is a well-defined procedure, and for consistency we thus set $\Gamma_W=0$
for these processes.
\item In the case of processes a.4, a.5, c.1, c.3, and e.6, we do {\em not} 
apply any cuts on $b$ quarks, which is possible since $m_b\ne 0$ implies
the possibility of integrating down to zero transverse momenta.
This is important, firstly because it allows us to test the robustness
of phase-space integration in a very demanding situation (the $b$ quark
being very light), and secondly in view of the matching of these
results with parton shower Monte Carlos, where it gives the possibility
of studying $b$-flavoured hadrons also at small transverse momenta.
\end{itemize}
All the results reported in table~\ref{tab:results} can be computed
by employing up to two hundreds machines running simultaneously 
for two weeks. This running time does not include that required
for actually generating the codes to be run. This latter operation
is not parallelized in the current versions of \MadLoop\ and \MadFKS, 
and one must use one machine per process\footnote{A minimal amount of 
parallelization is in fact included in \MadLoop, since the contributions of
eqs.~(\ref{qqglu})--(\ref{qqeta}) can be run simultaneously -- see 
sect.~\ref{sec:genfil}.}. For the most complicated among the processes 
considered in table~\ref{tab:results}, the running time of the generation
phase amounts to a few hours.

The uncertainties reported in table~\ref{tab:results} are of statistical
origin. In a fully-numerical approach as the one adopted here, another
source of uncertainty is that associated with potential numerical
instabilities. We shall discuss in sect.~\ref{sec:CTinst} the
procedure adopted by \MadLoop\ to treat such instabilities.
Here, we limit ourselves to reporting that they
are very rare, and that the corresponding uncertainties are 
completely negligible w.r.t.~the statistical errors, being 
{\em at least} two orders of magnitude smaller than the latter.

\section{Automation\label{sec:auto}}
In this section, we describe the techniques we have employed in
order to obtain the one-loop contributions to the results given
in sect.~\ref{sec:res}. As discussed in the introduction,
the automation of the computation of one-loop amplitudes has been
achieved by means of a computer program that we call \MadLoop. 
The core of the procedure followed by \MadLoop\ 
is the Ossola-Papadopoulos-Pittau (OPP) reduction
technique. We devote the next section to summarizing it, and will
then describe \MadLoop\ proper in more details.

Before proceeding, it is worth stressing that the very idea of
automating virtual corrections would have been unthinkable
without the introduction of procedures for the computations
of loop tensor integrals that are alternative to the traditional ones based 
on analytic techniques. Although the latter, by a clever use of 
tensor-reduction methods~\cite{Passarino:1978jh,Denner:2005nn,Binoth:2008gx}, 
have helped obtain remarkable results (see 
e.g.~refs.~\cite{Denner:2010jp,Binoth:2009rv} for some recent ones), 
they do not constitute an effective starting point for automation,
the main drawbacks being the need of heavy symbolic manipulations, and 
that of special treatments of unstable decompositions (in particular, 
the analytic approach obliges one to guess {\em a priori}
where numerical instabilities could occur, before taking actions such as
Taylor-expanding small Gram determinants). With some degree of
arbitrariness, we may classify the modern procedures alluded to before 
into two classes, that we call 
Generalized Unitarity (GU)~\cite{Bern:1994zx,Ellis:2007br,Ellis:2008ir}
and Integrand 
Reduction (IR)~\cite{Ossola:2006us,delAguila:2004nf,Mastrolia:2010nb}. 
Both have obtained very significant results: so far, GU- and IR-based efforts
have focused primarily on studies of large-multiplicity final 
states~\cite{Frederix:2010ne,Berger:2010zx} and of massive final 
states~\cite{Bevilacqua:2009zn,Bevilacqua:2010qb} respectively.
As can be seen from sect.~\ref{sec:OPP}, Integrand Reduction is
a procedure independent of the identities of the particles entering
in the process (i.e.~if they are fermions or bosons, or if they are
massless or massive). It is thus perfectly suited to our goal of
performing computations in the most flexible way, which is the reason 
why it has been adopted in \MadLoop.

\subsection{The Ossola-Papadopoulos-Pittau reduction\label{sec:OPP}}
\begin{figure}[htb!]
  \begin{center}
        \epsfig{file=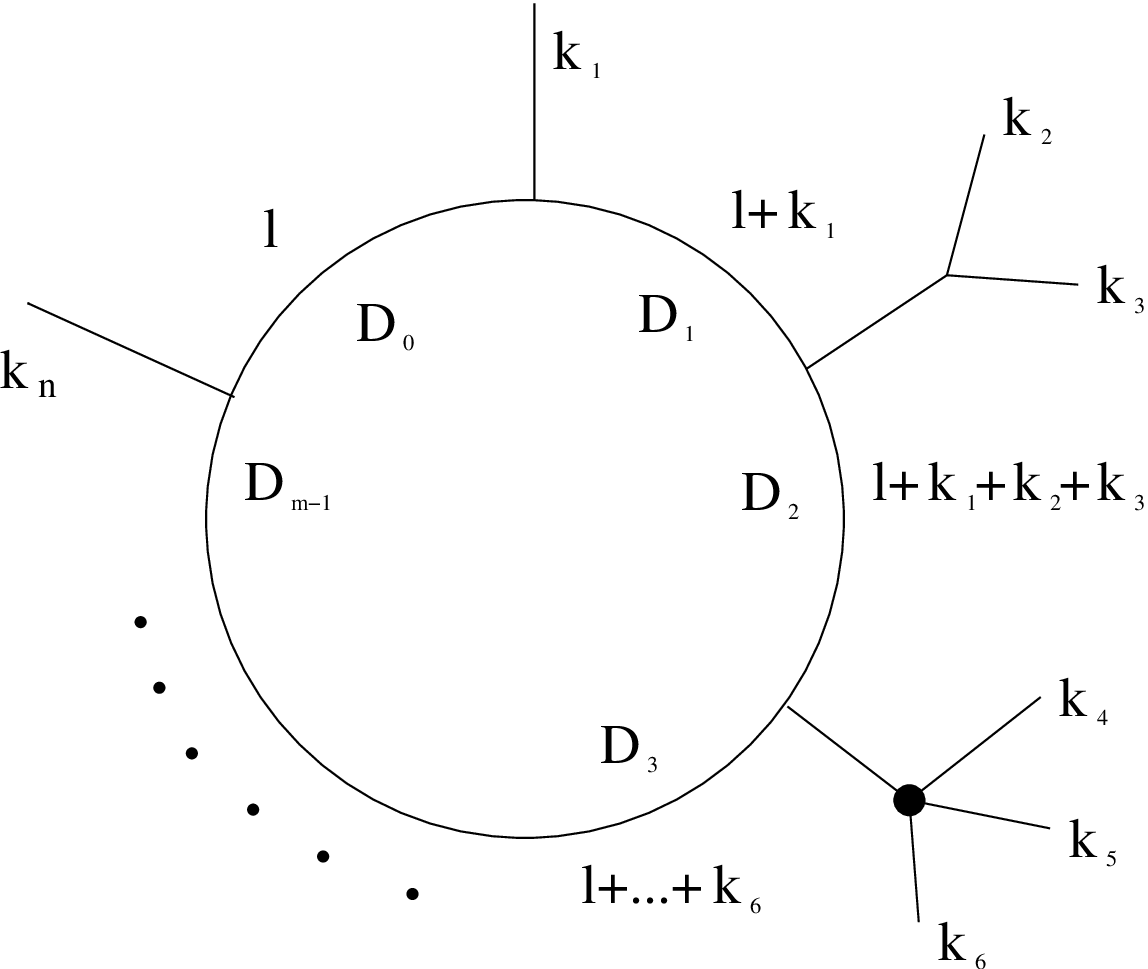, width=0.45\textwidth}
  \end{center}
  \vspace{-20pt}
  \caption{An $n$-point one-loop diagram with $m$ propagators in the loop.
           The dark blob represents a tree structure.}
      \label{closeddiag}
\end{figure}

Let us consider an UV-unrenormalized, $n$-point one-loop amplitude 
$\ampnl_\unr$. We have:
\beqn
\ampnl_\unr=\sum_\alpha \diag_\alpha\,,
\label{Csum}
\eeqn
where the sum runs over all Feynman diagrams relevant to
$\ampnl_\unr$, and $\diag_\alpha$ is the contribution of a given
Feynman diagram after loop integration. The OPP procedure
can be viewed as a linear operator, and therefore in what
follows we shall consider only a given $\diag_\alpha$ -- hence,
the index $\alpha$ will be dropped in order to simplify the notation.
The quantity $\diag$ is in general a tensor in Lorentz 
and colour spaces. The following arguments are however independent 
of the nature of $\diag$, which will therefore be understood;
this is equivalent to fixing the Lorentz and colour indices
in $\diag$, and manipulate the resulting scalar quantity.
It simplifies the present discussion to consider all external momenta
as outgoing:
\beqn
0=k_1+k_2+\cdots k_n\,.
\label{outg}
\eeqn
We consider a diagram with $m$ propagators in the loop; the value
of $m$ need not be specified here, and it suffices to say that
it satisfies the constraint \mbox{$1\le m\le n$}.
It is not restrictive to assume that the external momenta are 
in the same order as in fig.~\ref{closeddiag} (since such a 
configuration can always be obtained through a relabeling).
We denote the loop momentum in $d=4-2\ep$ dimensions by $\bqloop$, and 
decompose it into the sum of a 4-dimensional and of a $(-2\ep)$-dimensional
components, which we denote by $\qloop$ and $\tqloop$ respectively.
Hence:
\beqn
\bqloop=\qloop+\tqloop
\;\;\;\;\;\;{\rm with}\;\;\;\;\;\;
\qloop\mydot\tqloop=0\,.
\label{lbardec}
\eeqn
We introduce the partial sums that enter the propagators that form the 
loop (see fig.~\ref{closeddiag}):
\beqn
p_i=\sum_{j=1}^{M_i} k_j\,,
\;\;\;\;\;\;\;\;
1\le i\le m\,;
\;\;\;\;\;\;\;\;
p_0=p_m\,.
\label{partsums}
\eeqn
The values of the integers $M_i$ depend on the particular diagram
considered (e.g.~in fig.~\ref{closeddiag} we have $M_1=1$, 
$M_2=3$, $M_3=6$), but they must always fulfill the following
conditions:
\beqn
1\le M_i<M_{i+1}\,,
\;\;\;\;\;\;\;\;
M_m=n
\;\;\;\;\;\;\;\;
\Longrightarrow
\;\;\;\;\;\;\;\;
p_0=0\,,
\label{partsums2}
\eeqn
where the last equality of eq.~(\ref{partsums2}) follows from eq.~(\ref{outg}).
The inverses of the loop propagators in $d$ and four dimensions we denote
by $\bar{D}$ and $D$ respectively. Hence:
\beqn
\db{i} = (\bqloop + p_i)^2 - m_i^2  = \d{i}+{\tqloop^2}
\equiv  (\qloop + p_i)^2 - m_i^2+{\tqloop^2}\,,
\;\;\;\;\;\;\;\;
0\le i\le m-1\,,
\label{Dbardef}
\eeqn
which follows from eq.~(\ref{lbardec}), and from the fact that
the $(-2\ep)$-dimensional parts of the external four-vectors are
equal to zero, since the 't~Hooft-Veltman scheme is adopted.
Note that $m_i$ is the mass of the particle flowing in the $i^{th}$
propagator, and therefore in general $p_i^2\ne m_i^2$. As is 
known~\cite{Passarino:1978jh}, the one-loop integral $\diag$ 
can be expressed as a cut-constructible part, i.e.~a linear combination 
of scalar boxes, triangles, bubbles, and tadpoles, plus a 
(non cut-constructible) remainder term $R$, called rational part:
\bqa
\diag &=& 
\sum_{0\le i_0 < i_1 < i_2 < i_3}^{m-1} { d}( i_0 i_1 i_2 i_3) 
\int d^d \bqloop\, \frac{1}{\db{i_0}\db{i_1}\db{i_2}\db{i_3}} \nl
&&+ \sum_{0\le i_0 < i_1 < i_2 }^{m-1}{ c}( i_0 i_1 i_2)
\int d^d \bqloop\, \frac{1}{\db{i_0}\db{i_1}\db{i_2}} \nl
&&+\sum_{0\le i_0 < i_1 }^{m-1} { b}(i_0 i_1) 
\int d^d \bqloop\, \frac{1}{\db{i_0}\db{i_1}} \nl
&&+ \sum_{i_0=0}^{m-1} { a}(i_0) 
\int d^d \bqloop\, \frac{1}{\db{i_0}} \nl
&&+\,\,R\,.
\label{eq:amp}
\eqa
The essence of the OPP method is that of computing $\diag$ by 
determining (in a numerical manner) the set of coefficients
and the rational part
\bqa
d(i_0 i_1 i_2 i_3),\;\;\;\;\; c(i_0 i_1 i_2),\;\;\;\;\;
b(i_0 i_1),\;\;\;\;\;         a(i_0),\;\;\;\;\;
R,
\label{eq:set}
\eqa
and by using the well-known expressions for scalar loop 
integrals~\cite{'tHooft:1978xw,Ellis:2007qk,vanHameren:2010cp}.

The determination of the quantities in eq.~(\ref{eq:set}) is achieved
by working at the integrand level. We start by writing
\beqn
\diag=\int d^d \bqloop\,\bar{\idiag}(\bqloop)\,,
\;\;\;\;\;\;\;\;
\bar{\idiag}(\bqloop)=\frac{\bar{N}(\bqloop)}
{\prod_{i=0}^{m-1}\db{i}}\,,
\label{Cbardef}
\eeqn
which follows from fig.~\ref{closeddiag} and eq.~(\ref{Dbardef}),
and implicitly defines $\bar{N}(\bqloop)$. Next, we decompose
the numerator function $\bar{N}$ into the sum of a four-dimensional 
part and an extra piece (which by definition contains all the 
dependence on $\tqloop$ and on $\ep$; note that there is no
dependence on the $(-2\ep)$-dimensional parts of the external
four-vectors in the 't~Hooft-Veltman scheme): 
\beqn
\bar{N}(\bqloop)=N(\qloop)+\widetilde{N}(\qloop,\tqloop).
\eeqn
We get:
\beqn
\bar{\idiag}(\bqloop)=
\frac{N(\qloop)}{\prod_{i=0}^{m-1}\db{i}}+
\frac{\widetilde{N}(\qloop,\tqloop)}{\prod_{i=0}^{m-1}\db{i}}\,,
\label{integrand}
\eeqn
from which
\beqn
\diag\,\,&=&\diag_{cc+R_1}+R_2\,,
\label{Cdef}
\\
\diag_{cc+R_1}&=&\int d^d \bqloop\,
\frac{N(\qloop)}{\prod_{i=0}^{m-1}\db{i}}\,,
\label{ccpR1def}
\\
R_2&=&\int d^d \bqloop\,
\frac{\widetilde{N}(\qloop,\tqloop)}{\prod_{i=0}^{m-1}\db{i}}\,.
\label{R2def}
\eeqn
Here, $R_2$ contributes only to the rational part $R$, while 
$\diag_{cc+R_1}$ is the sum of a cut-constructible and of a rational
term (called $R_1$); we discuss how to disentangle the latter two in 
what follows. One starts by showing~\cite{Ossola:2006us} that the 
numerator function $N(\qloop)$ can always be cast in the following form:
\beqn
N(\qloop)&=&\!\!\!\!\!
\sum_{0\le i_0 < i_1 < i_2 < i_3}^{m-1}
\Big[
          { d}( i_0 i_1 i_2 i_3 ) +
     { \hat{d}}(\qloop;i_0 i_1 i_2 i_3)
\Big]
\mathop{\prod_{i=0}}_{i\notin\{i_0,i_1,i_2,i_3\}}^{m-1} \!\!\!\!\!\!D_i
\nl
     &+&
\sum_{0\le i_0 < i_1 < i_2 }^{m-1}
\Big[
          { c}( i_0 i_1 i_2) +
     {  \hat{c}}(\qloop;i_0 i_1 i_2)
\Big]
\mathop{\prod_{i=0}}_{i\notin\{i_0,i_1,i_2\}}^{m-1} \!\!\!\!\!D_i
\nl
     &+&
\sum_{0\le i_0 < i_1 }^{m-1}
\Big[
          { b}(i_0 i_1) +
     {  \hat{b}}(\qloop;i_0 i_1)
\Big]
\mathop{\prod_{i=0}}_{i\notin\{i_0,i_1\}}^{m-1} \!\!\!D_i
\nl
     &+&
\sum_{0\le i_0}^{m-1}
\Big[
          { a}(i_0) +
     {  \hat{a}}(\qloop;i_0)
\Big]
\mathop{\prod_{i=0}}_{i\ne i_0}^{m-1} D_i
\,,
\label{eq:oppexp}
\eeqn
where the terms proportional to $\hat{d}$, $\hat{c}$, $\hat{b}$ 
and $\hat{a}$ (called spurious terms) vanish upon integration.
By exploiting the fact that $\d{i} = \db{i}-{\tqloop^2}$ 
(see eq.~(\ref{Dbardef})) in eq.~(\ref{eq:oppexp}), one obtains 
the identity:
\beqn
N(\qloop)=N_{cc}(\qloop,\tqloop^2)+N_{R_1}(\qloop,\tqloop^2)\,,
\label{NNR1}
\eeqn
where we have defined
\beqn
N_{cc}(\qloop,\tqloop^2)=N(\qloop)\Big|_{D_i\to\db{i}}\,.
\label{Ncc}
\eeqn
As the notation suggests, $N_{cc}$ is identical to eq.~(\ref{eq:oppexp}), 
except for the formal replacements of all $D_i$'s with the corresponding
$\db{i}$'s (i.e., the four-dimensional denominators by their
$d$-dimensional counterparts). Equation~(\ref{NNR1}) implicitly
defines $N_{R_1}$. From eq.~(\ref{Dbardef}) we see that:
\beqn
N_{R_1}(\qloop,0)=0\,.
\label{NR1ident}
\eeqn
We can now define the cut-constructible and $R_1$ contributions
separately:
\beqn
\diag_{cc+R_1}&=&\diag_{cc}+R_1
\\
\diag_{cc}&=&\int d^d \bqloop\,
\frac{N_{cc}(\qloop,\tqloop^2)}{\prod_{i=0}^{m-1}\db{i}}\,,
\label{ccdef}
\\
R_1&=&\int d^d \bqloop\,
\frac{N_{R_1}(\qloop,\tqloop^2)}{\prod_{i=0}^{m-1}\db{i}}\,.
\label{R1def}
\eeqn
The rational part $R$ introduced in eq.~(\ref{eq:amp}) is the sum
of $R_1$ and $R_2$, defined in eqs.~(\ref{R1def}) and~(\ref{R2def})
respectively.

The key point is that the coefficients $d$, $c$, $b$, and $a$ which appear 
in eq.~(\ref{eq:oppexp}) are the {\em same} as those which appear in 
eq.~(\ref{eq:amp}), as can be easily seen by inserting eq.~(\ref{eq:oppexp})
into eq.~(\ref{Ncc}), and by using the result so obtained in eq.~(\ref{ccdef}).
This is ultimately the reason why the OPP procedure is carried out
at the integrand level, where the loop momentum \mbox{$(\qloop,\tqloop)$}
is just an external and fixed quantity. Thus, eq.~(\ref{eq:oppexp}) is the 
master equation used in the OPP method for the determination 
of the coefficients $d$, $c$, $b$, and $a$,
which is achieved by solving numerically a system of linear equations.
The idea is that of computing $N(\qloop)$ for suitable values of
$\qloop$, which render the just-mentioned linear system easy to solve.
A pre-condition for this to happen is the fact that the spurious terms
can be determined fully as functions of the external momenta; this has
been proved in ref.~\cite{delAguila:2004nf}.
At this point, the easiest way to proceed is that of computing the
cut-constructible and $R_1$ contributions separately. One starts with
the former, by setting $\tqloop^2=0$ and using eq.~(\ref{NR1ident}).
Then, one determines the two solutions\footnote{There are two (complex)
momenta owing to the quadratic nature of the propagators.} $\qloop^{\pm}$ 
of the equations:
\beqn
\d{i_0}({ \qloop^{\pm}})= \d{i_1}({ \qloop^{\pm}})= 
\d{i_2}({ \qloop^{\pm}})= \d{i_3}({ \qloop^{\pm}}) = 0\,,
\label{quadcut}
\eeqn
for given $i_0$, $i_1$, $i_2$ and $i_3$.
Equation~(\ref{eq:oppexp}) then becomes
\beqn
N(\qloop^\pm) = 
\Big[
{d}( i_0 i_1 i_2 i_3 ) +
     { \hat{d}}({ \qloop^\pm};i_0 i_1 i_2 i_3)
\Big]
\mathop{\prod_{i=0}}_{i\notin\{i_0,i_1,i_2,i_3\}}^{m-1} 
\!\!\!\!\!\!\d{i}(\qloop^\pm)\,,
\eeqn
and one can prove~\cite{Ossola:2006us} that the coefficients of the box 
diagrams are simply given by
\beqn
{d}(i_0i_1i_2i_3) = \frac{1}{2}
\left[
\frac{N({ \qloop^+})}{\prod_{i \ne i_0,i_1,i_2,i_3} \d{i}({\qloop^+})} +
\frac{N({ \qloop^-})}{\prod_{i \ne i_0,i_1,i_2,i_3} \d{i}({\qloop^-})} 
\right].
\label{dsol}
\eeqn
We point out that eq.~(\ref{quadcut}) is nothing but the application
of quadruple unitarity cuts. Once the solutions for the $d$ coefficients
are known thanks to eq.~(\ref{dsol}), the corresponding terms in
eq.~(\ref{eq:oppexp}) are moved to the l.h.s.~there. The procedure is
then iterated by considering triple, double, and single unitarity
cuts in succession, i.e.~values of $\qloop$ such that
three, two, and one denominators vanish respectively.
In exactly the same way one deals with the computation of 
$R_1$~\cite{Ossola:2008xq}. 
The only difference w.r.t.~the case of the cut-constructible
part is that the basis of the scalar loop integrals used in the
two cases is not the same (with that relevant to $R_1$ being
almost trivial).

The procedure described above is implemented in the computer
program \CutTools~\cite{Ossola:2007ax} which, being given in 
input the {\em function} $N(\qloop)$,
the momenta defined in the partial sums of eq.~(\ref{partsums}),
and the masses $m_i$ of the corresponding propagators, returns 
the numerical values of the cut-constructible part and of $R_1$. 
Note that by giving to \CutTools\ the momenta and the masses
entering the loop propagators, rather than the numerical values
of the propagators themselves, one is allowed to bypass the problem
of introducing $d$-dimensional quantities in \MadLoop\ -- these
are completely dealt with internally in \CutTools.
As far as $R_2$ is concerned, this quantity is not returned
by \CutTools. On the other hand, its computation
can be performed by considering tree-level Feynman diagrams,
which get contributions both from the usual rules of the theory
under consideration, and from special $R_2$ functions with up
to four external lines (in any renormalizable theory),
as we discuss in sect.~\ref{sec:R2}. These special
rules can be worked out once and for all from the Lagrangian
of the theory. Both the use of \CutTools\ and the calculation of 
the $R_2$ contribution for a given one-loop amplitude are controlled
by \MadLoop, in a way which we outline in the next section.

\subsection{Organization of the calculation\label{sec:org}}
The input to \MadLoop\ is a $2\to n$ Standard Model partonic 
process\footnote{Here and in what follows, we adopt the notation of 
ref.~\cite{Frederix:2009yq}: $\ident_i$ denotes the identity
of the $i^{th}$ particle that enters the process. Furthermore,
the momenta of the first two particles in eq.~(\ref{proc}) are
incoming, and all the others are outgoing.}
\beqn
\proc=\left(\ident_1,\ident_2,\ident_3,\ldots\,\ident_{n+2}\right)\,,
\label{proc}
\eeqn
which can be either user-defined, or generated by a third-party code 
such as \MadFKS; examples of the two uses are given in 
appendix~\ref{sec:comps} and in sect.~\ref{sec:res} respectively.
The main output of \MadLoop\ is the finite part\footnote{Up to a 
standard pre-factor, see eq.~(\ref{Vexpr}).} of the quantity:
\beqn
V(\proc)=\overline{\mathop{\sum_{\rm colour}}_{\rm spin}}
2\Re\left\{\ampnt(\proc){\ampnl(\proc)}^{\star}\right\},
\label{Vdef}
\eeqn
with $\ampnt$ and $\ampnl$ being the tree-level and one-loop amplitudes
of the process $\proc$, after the latter has been UV-renormalized.
Note that $\ampnl$ is the same quantity as that in eq.~(\ref{Csum}),
except for the fact that in the latter equation the amplitude was 
not UV-renormalized, and that by $n$ we had denoted there the total number 
of particles entering the process (and not only those in the final state
as we do here).
The bar over the sum symbol on the r.h.s of eq.~(\ref{Vdef}) understands
the average factors relevant to the colour and spin degrees of
freedom of initial-state particles.
The finite part of $V$ is convention dependent and, unless otherwise 
specified, we have adopted here the same as in ref.~\cite{Frederix:2009yq},
namely CDR -- see appendix~\ref{sec:comps} for further discussion
on scheme choices.
As a by product, \MadLoop\ also returns the residues of the double and
single infrared poles. The complete information on $\ampnl$ is 
available internally in \MadLoop\ (see app.~\ref{sec:gggZ} for a
case in which we have used such an information), and may be 
given as an additional output if so desired -- this is useful
e.g.~for computing the LO cross section of a loop-induced process,
which is proportional to $\abs{\ampnl}^2$.

Schematically, for a given input process $\proc$, \MadLoop\ goes
through the following steps.
\begin{enumerate}
\item Generates the diagrams relevant to $\ampnl(\proc)$. There are
two classes of them, one for the cut-constructible plus $R_1$ contribution,
and one for the $R_2$ contribution and UV renormalization.
\item Constructs the two integrands associated with the diagrams
determined in item 1. The one relevant to the cut-constructible plus $R_1$ 
contribution is the linear combination of eq.~(\ref{Csum}) at the integrand 
level, whose components are in the form of the first term\footnote{Except
for the fact that the denominators are the $D_i$'s and not the
$\db{i}$'s, since \MadLoop\ works in four dimensions. This is
irrelevant, because the numerical values of the denominators as
evaluated by \MadLoop\ are not used in the computation of the loop 
integrals performed by \CutTools.} on the r.h.s.~of eq.~(\ref{integrand}).
\item For a given $2\to n$ kinematic configuration (user-defined
or generated by \MadFKS), applies the OPP reduction to each of the terms 
of the linear combination determined in item 2. This is achieved by 
passing to \CutTools\ the function $N(\qloop)$ and any other inputs 
it needs (see sect.~\ref{sec:OPP}), and is done separately 
for each helicity configuration. After summing over all
diagrams and helicities, one thus obtains the cut-constructible plus
$R_1$ contribution $\diag_{cc+R_1}$.
\item For the same kinematic configuration as in item 3, computes the 
rational part $R_2$ (which is not returned by \CutTools), 
and performs UV renormalization if necessary. These calculations are
also carried out at fixed helicities, which are summed over as the
final step.
\item Performs sanity checks.
\end{enumerate}
Items 1 and 2 only involve symbolic manipulations, and construct 
the functions whose numerical evaluations will be performed in
items 3--5. In the case of the computation of a physical cross 
section (when \MadLoop\ is called by \MadFKS\ or by an analogous
program), a loop over items 3--5 is performed, with each iteration
of the loop using a different kinematic configuration.

We give a brief description of the least-trivial aspects of this 
procedure in sects.~\ref{sec:genfil} and~\ref{sec:R2}, and summarize
the various techniques specifically developed for \MadLoop\
in sect.~\ref{sec:newstuff}. 
Some further details can be found in appendix~\ref{sec:MLtechn}.
For a discussion on the checks performed by \MadLoop, of which those 
of item 5 above are only a part, see sect.~\ref{sec:checks}.

The current version of \MadLoop\ (and of \MadFKS) uses \MadGraph\ 
v4~\cite{Alwall:2007st}. This implies a few limitations on the use of 
the code. The new version of \MadGraph\ (v5~\cite{MGv5}) is now available, 
and the work to make \MadLoop\ compatible with it has already started. When 
completed, this will allow us to remove most of these limitations -- 
see sects.~\ref{sec:limits} and~\ref{sec:outlook} for a discussion
on this point.

\subsubsection{Generation of one-loop amplitudes
from tree amplitudes\label{sec:genfil}}
Given the fact that \MadLoop\ is based on the \MadGraph\ framework, it
is clear that the most economic way of generating one-loop amplitudes
is that of exploiting as much as possible the capabilities of the latter 
code. These are however limited to constructing tree-level quantities, 
and therefore \MadLoop\ must be able to perform some non-trivial
operations on top on those available from \MadGraph\ in order for us
to achieve our goal. We start by observing that any one-loop diagram
can be turned into a tree-level diagram by simply cutting {\em one} (and
only one) of the propagators entering the loop. It must be clear that 
this cut has nothing to do with the cuts performed when computing 
one-loop integrals with unitarity methods. Thus, in order to avoid 
any confusion, we shall call \Lcut\ the cut we are talking about here.
The tree-level diagram obtained by \Lcutting\ a one-loop diagram will
be called \Lcut\ diagram. In an \Lcut\ diagram, there will be two
particles (that we consider as being in the final state by
definition) which arise from \Lcutting\ the chosen propagator in the loop: 
their identities will be denoted by $\qpart$ and $\qbpart$, and they
will be called \Lcut\ particles. If we consider one-loop corrections
to the process in eq.~(\ref{proc}), the \Lcut\ diagrams we obtain with
the \Lcut\ operation will be {\em a subset} of those relevant to the process:
\beqn
\proc_{\Lcut}=\left(\ident_1,\ident_2,\qpart,\qbpart,
\ident_3,\ldots\,\ident_{n+2}\right)\,.
\label{Lproc}
\eeqn
This discussion suggests to define a procedure which is
the inverse of \Lcutting. Namely, for a given $2\to n$ process
such as that in eq.~(\ref{proc}), we consider all possible 
\Lcut\ processes of the kind of that in eq.~(\ref{Lproc}),
use \MadGraph\ to construct the corresponding amplitudes,
and sew together the two \Lcut\ particles. 
In this way, we achieve the construction of one-loop diagrams
without actually having to start from one-loop topologies,
as done for example by FeynArts~\cite{Hahn:2000kx}. This idea is also at 
the basis of the one-loop computations performed by HELAC-1Loop (see
ref.~\cite{vanHameren:2009dr}).

This construction of one-loop amplitudes by sewing tree-level ones
poses several problems. Firstly, we have to define a 
minimal set of \Lcut\ processes so as not to miss any contributions to 
the one-loop amplitude we are seeking to construct.
Secondly, for a given \Lcut\ process, when sewing together the \Lcut\ 
particles we shall obtain one-loop diagrams with an incorrect
multiplicity: one particular diagram may appear more times than
prescribed by perturbation theory.
We have therefore to discard the one-loop diagrams in excess
after sewing. We call this operation {\em diagram filtering}.
Finally, after filtering the amplitudes for the \Lcut\ diagrams 
we are left with are constructed. However, these amplitudes will not
coincide with the corresponding one-loop amplitudes, because of the
special roles played by the \Lcut\ particles. These are associated with
external wave functions in \Lcut\ diagrams, and with an internal propagator
in one-loop diagrams (therefore, technically the sewing operation corresponds 
to removing the wave functions of the \Lcut\ particles, and to replacing them
with a suitable propagator). \MadGraph\ must therefore be instructed to treat
\Lcut\ particles in a special way -- this includes the fact that
such particles are off-shell and carry complex momenta. 

It is easy to convince oneself that when computing QCD corrections
the \Lcut\ processes one needs to consider correspond to the following
choices of the \Lcut\ particles:
\beqn
(\qpart,\,\qbpart)&=&(g,\,g)
\phantom{;\,(d,\,\bd);\ldots\,(Q,\,\bQ)}
\;\;\;\;\;\;\;\;\,{\rm gluons}\,,
\label{qqglu}
\\
&=&(u,\,\bu);\,(d,\,\bd);\ldots\,(Q,\,\bQ)
\;\;\;\;\;\;\;\;{\rm quarks}\,,
\label{qqqrk}
\\
&=&(\eta,\,\bar{\eta})
\phantom{;\,(d,\,\bd);\ldots\,(Q,\,\bQ)}
\;\;\;\;\;\;\;\;\,{\rm ghosts}\,.
\label{qqeta}
\eeqn
Here, $Q$ denotes the heaviest flavour one wants to circulate in
the loop (in physical applications, $Q$ is typically either a bottom 
or a top quark).

\begin{figure}[htb!]
  \begin{center}
        \epsfig{file=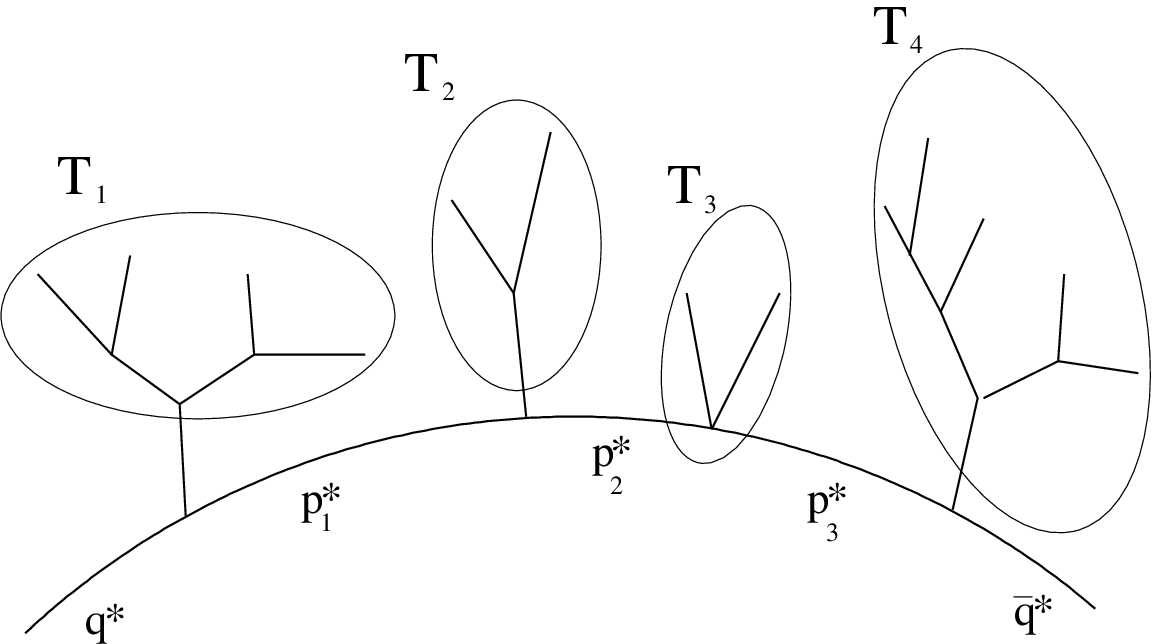, width=0.7\textwidth}
  \end{center}
  \vspace{-20pt}
  \caption{Example of an \Lcut\ diagram that corresponds to a box
           one-loop diagram.}
      \label{Lcutdiag}
\end{figure}
Diagram filtering is performed in the following way. We start from
observing that any \Lcut\ diagram can be depicted as in fig.~\ref{Lcutdiag}.
There, $T_i$ denotes a tree-level structure\footnote{Note that for an
$n$-point one-loop amplitude, $1\le \#\{T_i\}\le n$, with $\#\{T_i\}$ the
number of $T_i$'s.}. The particles whose
propagators enter in the loop have been denoted by $p_j^\star$,
to make the distinction clear from those that contribute to the
tree-level structures $T_i$; the notation is also consistent with
that used for the \Lcut\ particles. \MadLoop\ starts from
obtaining the necessary information on $T_i$ and $p_j^\star$ from \MadGraph\
(such as particle identities and four momenta);
this is done for each \Lcut\ diagram. In this way, each diagram has
an unambiguous representation (its ``identity''), internal to \MadLoop, 
that we can for example identify with the string
\beqn
\qpart\,T_1\,p_1^\star\,T_2\,p_2^\star\,T_3\,p_3^\star\,T_4\,\qpart
\label{IDLcut}
\eeqn
for the case of fig.~\ref{Lcutdiag}. As we shall explain in 
appendix~\ref{sec:filter}, when writing diagram identities one 
need not distinguish between fermions and antifermions in the case
of particles circulating in the loop, and hence no overlines appear 
in eq.~(\ref{IDLcut}) or its analogues (despite the fact that one of 
the two \Lcut\ particles has been correctly denoted by $\qbpart$ on 
the l.h.s.~of eq.~(\ref{qqglu})). At the end of the day, \MadLoop\
will loop over all \Lcut\ diagrams, checking their identities. If a
diagram identity has not been previously found, the diagram
is kept, otherwise it is discarded (i.e., it is filtered out).
It must be stressed that two diagram identities must be considered
equivalent if they are identical up to a cyclic permutation, or to
mirror symmetry, or to a cyclic permutation plus mirror symmetry;
however, mirror symmetry must not be considered in the cases of 
loops that contain only fermions or only ghosts, when one only 
checks equivalence under cyclic permutations. This is because
two \Lcut\ diagrams that differ by a cyclic permutation correspond to
the same one-loop diagram which was \Lcut\ in two different propagators
(see appendix~\ref{sec:filter} for more details).
For example, a cyclic permutation equivalent to eq.~(\ref{IDLcut}) 
could read:
\beqn
p_1^\star\,T_2\,p_2^\star\,T_3\,p_3^\star\,T_4\,q^\star\,T_1\,
p_1^\star\,.
\label{IDcycl}
\eeqn
Likewise, two \Lcut\ diagrams that differ by mirror symmetry correspond to
the same one-loop diagram with the loop momentum flowing in opposite
directions. The identity of the \Lcut\ diagram obtained by mirror
symmetry acting on eq.~(\ref{IDLcut}) reads:
\beqn
\qpart\,T_4\,p_3^\star\,T_3\,p_2^\star\,T_2\,p_1^\star\,T_1\,\qpart\,.
\label{IDmirr}
\eeqn
Equivalence under cyclic permutations and mirror symmetry can be used as a
powerful self-consistency check by \MadLoop\ -- see sect.~\ref{sec:checks}.

\subsubsection{$R_2$ contribution and UV renormalization\label{sec:R2}}
As it was discussed in sect.~\ref{sec:OPP}, \MadLoop\ computes by calling
\CutTools\ the cut-constructible part of, and the $R_1$ contribution
to, the one-loop amplitude $\ampnl(\proc)$ that appears in eq.~(\ref{Vdef})
(item 3 of the list in sect.~\ref{sec:org}). Therefore, in order to obtain 
the full $V(\proc)$, we still need to include the $R_2$ contribution, and 
the UV renormalization (item 4 of the  list in sect.~\ref{sec:org}).
The key observation here is that, thanks to the fact that the $R_2$ part 
can be seen as arising from a set of finite counterterms, its automated 
computation proceeds through the same steps as UV renormalization.

In essence, both the $R_2$ and UV-renormalization contributions to
$V(\proc)$ can be cast in the following form:
\beqn
\overline{\mathop{\sum_{\rm colour}}_{\rm spin}}
2\Re\left\{\ampnt(\proc){\ampnX(\proc)}^{\star}\right\},
\;\;\;\;\;\;\;\;
{\rm X}=R_2\,,\;{\rm UV}\,,
\label{VR2UV}
\eeqn
that is, an interference between the Born amplitude, $\ampnt(\proc)$, 
and another tree-level amplitude, $\ampnRt(\proc)$ or $\ampnUV(\proc)$. 
The latter are the amplitudes of the $2\to n$ process $\proc$, constructed 
with the standard Feynman rules, {\em plus} the rules relevant to either 
the $R_2$ or the UV-renormalization contributions, with the condition that 
the amplitude be exactly one order in $\as$ higher than the Born-level one.
The $R_2$ rules are obtained by explicit loop calculations, and
can be cast in the form of $n$-point functions, with $2\le n\le 4$
(see ref.~\cite{Draggiotis:2009yb}). As far as UV renormalization
is concerned, the additional rules are simply read off the one-loop 
UV counterterms of QCD. The consequence of these facts, together with 
the condition on the power of $\as$ that must appear in the amplitudes,
is that $\ampnRt(\proc)$ and $\ampnUV(\proc)$ are constructed using
the standard vertices and propagators, plus {\em one and only one}
$R_2$ or UV-renormalization $n$-point function.
It is clear, therefore, that in order to compute the quantities
in eq.~(\ref{VR2UV}) we can exploit the ability of \MadGraph\ to
construct tree amplitudes, and the functionalities of \MadLoop\ to  
calculate interference terms, which we already employ when computing 
the cut-constructible and $R_1$ contributions to eq.~(\ref{Vdef}).
Note that the above procedure to compute eq.~(\ref{VR2UV}) should be 
applied to truncated one-loop amplitudes. When generating tree-level
diagrams with \MadGraph, we also get the contributions due to two-point
functions on external legs, which are discarded. As is known,
wave-function renormalization must be carried out through the 
multiplication of the $Z$ factors, which we effectively do according 
to eqs.~(\ref{UVgwf}) and~(\ref{UVextm}). The $R_2$ contributions to
the $Z$ factors are already included in eqs.~(\ref{UVgwf}) 
and~(\ref{UVextm}), and hence their explicit computations are not 
necessary in this case.

The procedure outlined above is fully general. In what follows, we
mention briefly a few technical details related to its current
implementation in \MadLoop, and its being based on \MadGraph\ v4.

In order to compute the $R_2$ contribution, we have implemented new 
\HELAS\ routines, that correspond to the necessary $R_2$ $n$-point 
functions\footnote{This entails writing these routines ``by hand''
once and for all for a given theory. 
It is not difficult to devise a procedure to automate
this writing. See sect.~\ref{sec:outlook} for a discussion on this point.}.
We have restricted ourselves to the functions relevant to the Standard Model
(with a few limitations, see sect.~\ref{sec:limits}), in keeping with
the scope of the present work. The condition on the presence of
one $R_2$ function in $\ampnRt(\proc)$ is conveniently rephrased in
the \MadGraph\ language by associating to each $R_2$ function one power
of a new coupling (whose numerical value is equal to one), which does 
not appear in regular functions, and by requiring the full amplitude to 
have exactly one power of such coupling.

As discussed before, the automation of UV renormalization may 
be achieved in exactly the same way as done for the $R_2$ contribution.
In practice, the cases we consider in this paper allow us 
to make some further simplifications. Before going into that,
let us mention that we automated the UV renormalization of the Standard  
Model, using a scheme which subtracts the massless 
modes according to $\overline{\text{MS}}$, and the
massive ones at zero momentum. The few relevant counterterms
have been typed in once and for all, as done for the $R_2$ functions;
their explicit forms are reported in appendix~\ref{sec:UVcnts}.
As can be seen there, all UV counterterms except the one relevant to
mass renormalization give contributions proportional to the
Born amplitude squared (we have assumed that all contributions
to such an amplitude factorize the same powers of $\as$ and $\alpha$.
See sect.~\ref{sec:limits} for further comments on this point).
We then gather from \MadGraph\ the information on the power of $\as$,
on the number of Yukawa vertices with massless and massive particles,
and on the identities of the external particles; these will serve
for (QCD) charge, Yukawa, and wave-function renormalizations
respectively.

A simple observation allows one to solve the problem of mass
renormalization, which is a procedure that does not factorize
the Born. It turns out that the structure of the UV counterterm
due to mass insertion (eq.~(\ref{UVEnd})) is basically identical
to that of the two-point $R_2$ functions that originate from
bubble diagrams (in the \HELAS\ language, they both
correspond to attaching an extra term -- a fermion propagator
times a pre-factor -- to the propagator present in the \HELAS\
routines that give an off-shell fermion current; this is a
work-around due to the absence of two-point vertices in \MadGraph\ v4). 
Hence, we  simply define an $R_2$-plus-UV function, and therefore we
piggyback UV mass renormalization on the computation of
the $R_2$ contribution.

\subsubsection{Summary of MadLoop features\label{sec:newstuff}}
We list here the features that required substantial writing of computer 
code, either in \MadLoop\ proper, or in one of the modules that
\MadLoop\ uses.
\begin{itemize}
\item The generation of loop diagrams in a way that exploits
 as much as possible the capabilities of \MadGraph\
 (through \Lcut-diagram construction and subsequent filtering).
\item The computation of the resulting one-loop amplitude, in
 a form suited to being given in input to \CutTools. This implies,
 in particular, the removal of the denominators of the propagators
 of loop particles, and the reconstruction of the propagator of
 the sewed \Lcut\ particles.
\item The algorithm that allows \MadGraph\ to deal with two 
 processes simultaneously (the \Lcut\ and Born ones), in order to 
 compute their interference, including the relevant colour algebra
 (in all cases except one-loop computations, \MadGraph\ needs to treat 
 one process at a time, since only an amplitude squared is relevant).
\item The possibility of using complex four-momenta (that circulate
 in the loops).
\item The implementation of ghosts.
\item The implementation of special $R_2$ functions and of UV counterterms, 
 and the computation of the amplitudes relevant to the $R_2$ and 
 UV-renormalization contributions, which use the former functions.
\end{itemize}

In general, we have strived to make \MadLoop\ a black box for \MadFKS\
(or an equivalent calling program) -- the only talk-to is in the
form dictated by the Binoth-Les Houches accord~\cite{Binoth:2010xt}.
Likewise, \CutTools\ is a black box for \MadLoop, so any future
upgrades of the former code (with the condition that its 
inputs be those introduced in sect.~\ref{sec:OPP})
will be compatible with the latter. On the other hand, \MadLoop\
needs information from \MadGraph\ which are not available to the ordinary 
user of the latter program. However, this poses no problem, firstly because 
the amount of code-writing on the \MadGraph\ side is next-to-minimal, 
and secondly (and more importantly) because \MadLoop\ has to be 
considered a module of \MadGraph, to become part of the official release
in due time (hence, new versions of the two codes are being and
will be developed together).

In the same spirit, we point out that the implementation of new
\HELAS\ routines specific to one-loop computations has been done 
in a way which is fully independent of the \MadLoop\ code proper. By
exploiting the capabilities of \MadGraph\ v5, in the future we 
shall generate these routines automatically, but this will not imply 
the modification or the writing anew of {\em any} piece of code in \MadLoop. 

\subsection{Checks\label{sec:checks}}
One of the main advantages of the automation of computations
as performed by \MadLoop\ is that results for individual processes
(irrespective of their complication) are basically guaranteed to be 
correct. The key point is that the computer code which returns the 
relevant one-loop amplitude is written by (in other words, is an output
of) \MadLoop, which uses a fixed number of pre-defined building blocks 
(the \HELAS\ routines), without any involvement from the user.
Therefore, the correctness of any one-loop amplitude follows
from establishing the correctness of \MadLoop. In turn, this
entails two kinds of checks.
\begin{itemize}
\item[{\em a)}] The building blocks used by \MadLoop\ must be 
proven correct.
\item[{\em b)}] The way in which the building blocks are combined
must be proven correct.
\end{itemize}
We point out that this structure of checks is fully analogous
to that one has in the case of \MadGraph, although of course 
the technical aspects are not identical.

It is easy to realize that the checks of item {\em a)} are 
process-independent, while those of item {\em b)} do depend
on the particular one-loop amplitude one wants to compute.
In spite of this, the most straightforward way to carry out
the checks of item {\em a)} is that of comparing the results of
\MadLoop\ with those available in the literature, for a
suitable list of processes. Since the number of building
blocks used by \MadLoop\ is finite, so is such a list.
Clearly, when following this procedure one also performs the 
checks of item {\em b)} for the processes of the list. However,
this does not guarantee that the same checks will be
successful for processes which do not belong to the list
(although it is a powerful hint that this is indeed the
case). Hence, in order to carry out fully the program of item {\em b)},
we have implemented several {\em self-consistency} checks, that
\MadLoop\ will perform for any generated process and for any 
desired $2\to n$ kinematic configurations
among those chosen by the user or by the calling program.
We shall describe these checks later in this section.

The list of processes that we have used in order to compare \MadLoop\ 
results with their analogues available in the literature, and in computer
codes other than \MadLoop, is reported in appendix~\ref{sec:comps}. 
We stress that this 
list is actually redundant as far as the goal of item {\em a)} is 
concerned (i.e.~each \HELAS\ ``one-loop'' function is checked
at least once, but typically more than once). For each process,
we compare the finite part and the residues of the infrared poles
of the quantity $V$ defined in eq.~(\ref{Vdef}) with those of 
other computations. It should be pointed out that, although in some
cases codes other than \MadLoop\ will not return the residues of
the infrared poles, these can in any case be checked against their
analytically-known expressions, which we get from the implementation
in \MadFKS\ of eq.~(B.2) of ref.~\cite{Frederix:2009yq}.
As can be seen from appendix~\ref{sec:comps}, the agreement between
\MadLoop\ and previously-known results is excellent.

We now turn to discussing the self-consistency checks we have
mentioned previously. Some of them are not applicable to specific 
processes; the general strategy we have followed is that of testing
each process generated by \MadLoop\ in the largest possible
number of ways. The checks we have set up are listed in what follows.
\begin{itemize}
\item Alternative diagram filtering. \Lcut\ diagrams whose identities
differ by cyclic permutations or mirror symmetry correspond to
codes which are not written by \MadLoop\ in the same way, in spite
of the fact that they will eventually give exactly the same loop-integrated
results\footnote{This is what \MadGraph\ usually does, and is not specific
to \Lcut\ diagrams. One can see it by comparing the codes written by 
\MadGraph\ that correspond to the same physical process at the LO, but
differ e.g.~in the ordering of final-state particles as they appear in the
input cards.}. We check that this is indeed what happens.
An easy way to do this is that of choosing the \Lcut\ particles
in an order different from that of eqs.~(\ref{qqglu})--(\ref{qqeta}).
This is because \Lcut-diagram identities not filtered out when looping over 
all diagrams will not be the same as those kept with the ``canonical order''
of eqs.~(\ref{qqglu})--(\ref{qqeta}), but rather their equivalents
under cyclic permutations and/or mirror symmetry (for example, this corresponds
to keeping eq.~(\ref{IDLcut}) and discarding eq.~(\ref{IDcycl}), versus
keeping eq.~(\ref{IDcycl}) and discarding eq.~(\ref{IDLcut})).
\item Crossing. We consider the two processes
\beqn
\proc_1=\left(\ident_1,\ident_2,\ldots\ident_i,\ldots\ident_{n+2}\right),
\;\;\;\;\;\;\;\;
\proc_2=\left(\ident_1,\overline\ident_i,\ldots\overline\ident_2,
\ldots\ident_{n+2}\right).
\eeqn
For a given $2\to n$ kinematic configuration
\beqn
k_1+k_2\;\longrightarrow\; k_3+\cdots +k_i+\cdots +k_{n+2}
\eeqn
we check that the corresponding one-loop amplitudes fulfill the
following equation:
\beqn
\ampnl(\proc_1;k_1,k_2,\ldots k_i,\ldots k_{n+2})=
\omega\ampnl(\proc_2;k_1,-k_i,\ldots -k_2,\ldots k_{n+2})\,,
\label{crossing}
\eeqn
with $\omega$ a constant that only depends on the identities
of the particles that are crossed.
This is non trivial, given that \MadLoop\ constructs $\ampnl(\proc_1)$
and $\ampnl(\proc_2)$ in two different ways (which includes the fact that
the \HELAS\ routines used in the constructions of the two amplitudes
may not even be the same). We point out that crossing is a very 
powerful method during the debugging phase, since eq.~(\ref{crossing})
holds diagram by diagram.
\item Dependence on the mass of a heavy quark $Q$. When the one-loop amplitude
for a given process includes diagrams that feature a closed fermion
loop, we can study the dependence of the result on $\mQ$ (by default,
such a dependence is included exactly by \MadLoop). We typically identify
the heavy quark with the top, but this is not mandatory.
In particular, we may check the following two regimes.
\begin{enumerate}
\item Decoupling limit: $\mQ\to\infty$. We compute the amplitude
with the full $\mQ$ dependence for increasingly large values of
$\mQ$, and compare it to the one we obtain by excluding altogether the 
heavy quark from contributing to the loops (in the case $Q=t$, the latter 
is a five-light-flavour amplitude). 
\item Zero-mass limit: $\mQ\to 0$. We compute the amplitude
with the full $\mQ$ dependence for increasingly small values of
$\mQ$, and compare it to the one we obtain by replacing the heavy
quark with an additional massless quark in the loops.
\end{enumerate}
We stress that the comparisons between these two limits of the amplitude
and the $n_{lf}$- and $(n_{lf}+1)$-light-flavours results are not 
straightforward, owing to the possible presence of anomalies 
and to the UV-renormalization scheme adopted here respectively. 
Each process is to be studied as a case on its own,
which is what we do in appendix~\ref{sec:comps}.
\item Gauge invariance. If the process $\proc$ under study involves at 
least a gluon, we check that the corresponding one-loop amplitude satisfies 
the gauge-invariant condition:
\beqn
\ampnl_\mu(\proc;k_1,k_2,\ldots k_i,\ldots k_{n+2})\,k_i^\mu = 0\,,
\label{gaugeinv}
\eeqn
for any gluon momentum $k_i$. The same kind of check is performed
for photons as well.
\item Infrared pole residues. The form of the double and single
infrared poles is analytically known for any process $\proc$
(see e.g.~eq.~(B.2) of ref.~\cite{Frederix:2009yq}).
We compare the residues returned by \MadLoop\ with those
computed with the analytic formulae, implemented in \MadFKS.
We point out that by checking the single pole result we also 
indirectly test the correctness of the UV renormalization procedure.
\end{itemize}

We stress that all these five checks are
local in phase space, i.e.~they are performed for a given
$2\to n$ kinematic configuration. 
In principle, they can therefore be carried out for each 
kinematic configuration when integrating over the phase space;
in practice, this would significantly increase the load on CPU.
Therefore, we have chosen to do these tests for only a few 
(i.e., less than five) kinematic configurations before starting 
the actual integration. It is easy to realize that this is not
restrictive, since given the nature of the checks it is basically impossible
for an one-loop amplitude to pass them for some configurations, and to fail
them for others. An exception to this situation is a failure due
to some numerical instability occuring during the calculation: such
a case is discussed in sect.~\ref{sec:CTinst}.

\subsection{Robustness against numerical instabilities\label{sec:CTinst}}
Once a process has been generated and the checks described
in sect.~\ref{sec:checks} have been performed, the process
is deemed correct by \MadLoop, and phase-space integration
can be carried out. While doing so, kinematic configurations
may be encountered that render it particularly difficult to execute
the necessary numerical manipulations. Numerical instabilities
are typically related either to inaccuracies in the solution of the system
of linear equations performed by \CutTools\ which determines the
quantities in eq.~(\ref{eq:set}), or to the occasional occurrence 
of almost linearly-dependent subsets of external four momenta. 
For each $2\to n$ kinematic and helicity configuration given in input
to \MadLoop, the detection of numerical instabilities relies on 
two self-diagnostic tests performed by \CutTools. When potential
problems are encountered, \MadLoop\ adopts a procedure that
either fixes them, or provides one with an estimate of the
resulting uncertainties. We describe this procedure in what
follows. Before proceeding, however, we would like to stress
that numerical instabilities are a fairly rare occurrence indeed.
For all but two of the processes that we consider in table~\ref{tab:results}, 
their fraction is less than $10^{-5}$ of the total number of
kinematic configurations generated during phase-space integration.
Processes b.3 and b.6 have slightly larger fractions ($3\mydot 10^{-4}$
and $3\mydot 10^{-3}$ respectively), but as was mentioned in 
sect.~\ref{sec:res} the associated uncertainties are completely negligible
w.r.t.~the statistical errors on the final results.

The first test performed by \CutTools\ is based on the fact that the 
set in eq.~(\ref{eq:set}) can also be obtained in an independent way 
by first shifting all masses in the $D_i$'s that appear in
eq.~(\ref{eq:oppexp}) by a common amount (see ref.~\cite{Pittau:2010tk}):
\bqa
\label{eq:shift}
m_i^2 \to m_i^2+M^2\,.
\eqa
The value of $M$ is arbitrary, but one finds it convenient to relate
it to the typical scale of the process; we choose 
\mbox{$M={\cal O}(\sqrt{\hat{s}}/10)$}, with $\sqrt{\hat{s}}$ the
partonic c.m.~energy. This results in a new set
\bqa
d^\prime(i_0 i_1 i_2 i_3),\;\;\;\;\; c^\prime(i_0 i_1 i_2),\;\;\;\;\;
b^\prime(i_0 i_1),\;\;\;\;\;         a^\prime(i_0),\;\;\;\;\;
R^\prime,
\label{eq:setprime}
\eqa
that allows an independent determination of the cut-constructible and $R_1$
contributions, which we denote by $\diag_{cc+R_1}^\prime$. \CutTools\ then
checks whether the following inequality is satisfied:
\bqa
\abs{\diag_{cc+R_1}-\diag_{cc+R_1}^\prime}\ll \abs{\diag_{cc+R_1}}\,. 
\label{eq:esta}
\eqa
When this is not the case, the sets in eqs.~(\ref{eq:set}) 
and~(\ref{eq:setprime}) are determined again by \CutTools, but 
using routines written in multi-precision (except for the function 
$N(\qloop)$, which is still in double precision). The inequality 
in eq.~(\ref{eq:esta}) is then checked with these multi-precision
results. It turns out that the vast majority of kinematic configurations
that give results that do not pass the test of eq.~(\ref{eq:esta}) 
in double precision, do so when multi-precision calculations are
carried out. 

As second test, by using the coefficients $d$, $c$, $b$, and $a$ 
previously determined, \CutTools\ chooses an arbitrary loop momentum 
$\qloop$ (again related to the typical scale of the process, e.g.
\mbox{$\qloop_i={\cal O}(\sqrt{\hat{s}}/5)$} and
\mbox{$\qloop^2={\cal O}(\hat{s}/100)$}), and computes the
two sides of eq.~(\ref{eq:oppexp}), which must agree at 
a user-defined level. This second test 
is typically able to detect instabilities that would not appear
if also the numerator function $N(\qloop)$ were computed 
in multiple precision. 

If a kinematic configuration passes both these tests it is defined
stable by \CutTools, and the computation proceeds to the next step.
Otherwise, i.e.~when at least one of the two tests fails, the kinematic 
configuration is called an {\em exceptional phase-space point} (EPS), 
and the problem then becomes that of estimating its contribution
to the one-loop amplitude.

It should be clear that a given kinematic configuration is an 
EPS only in connection with a given diagram: other diagrams typically
display no problems, and are reliably computed by \CutTools.
One can therefore trust the result for the total one-loop integral
in the case in which the unstable diagram does not contribute
significantly to the total. This can be determined by checking that
the resulting one-loop amplitude is gauge invariant (which is
possible only if at least one gluon is involved in the process), 
i.e.~that eq.~(\ref{gaugeinv}) is fulfilled to an accuracy which 
is the same as that which holds for stable kinematic configurations, 
and which we determine at the beginning of the run. When this
is the case, we say that the EPS is rescued to stability, \MadLoop\ 
accepts the result given by \CutTools, and proceeds to the next kinematic
configuration. 

In the present version of the code, in order to be very conservative
we have chosen not to rescue to stability any EPS's at this stage.
What we do, instead, is to estimate the contribution of the EPS's
to the one-loop amplitude, and to associate an uncertainty with this
procedure. We start by observing that in basically all cases EPS's 
originate from the inability of performing all the relevant computations 
in multiple precision\footnote{As described before, in our setup this 
applies to $N(\qloop)$, since \CutTools\ does use multi-precision 
routines to rescue potential EPS's.} (or, perhaps more importantly,
from the unwillingness to do so, since multi-precision calculations are
very demanding from the CPU viewpoint). This is because EPS's are the
result of the lack of cancellations between pairs of numbers, which would
take place beyond double-precision accuracy. These cancellations are
so delicate that even a small change in the kinematic configuration
can bring them back into the double-accuracy regime. This suggests
that, given an EPS, one can slightly deform it, and use the (now
generally stable) result given by \CutTools\ as an estimate of
the true result associated with the EPS.

In order to carry out this deformation, we consider the EPS in the 
partonic c.m.~frame, and rescale there the $z$ components\footnote{In
practice, we also consider deformations along the other two three-axes.
In order to simplify the notation, we limit ourselves to discussing
here the case of the $z$ axis; this does not entail any loss of
generality.} of the particle three-momenta:
\beqn
k_i^3\;\longrightarrow\;(1+\lambda_\pm) k_i^3\,.
\label{kresc}
\eeqn
The transverse momenta are left invariant, 
and the energy components (and $\sqrt{\hat{s}}$) are
adjusted so as to impose on-shellness conditions and four-momentum 
conservation. In eq.~(\ref{kresc}), $\lambda_\pm$ are two small and 
arbitrary real numbers. Typically, $\lambda_-=-\lambda_+$, but this is not 
strictly necessary. In this way, we obtain two kinematic configurations, 
that we call shifted EPS's, and compute the associated Born and one-loop 
contributions. Let us denote by
\beqn
\ampnt_{\lambda=0}\,,
\;\;\;\;\;\;
\ampnt_{\lambda_\pm}\,,
\;\;\;\;\;\;
V_{\lambda_\pm}^{\rm FIN}\,,
\eeqn
the Born amplitude computed with the original EPS, the two Born amplitudes
computed with the shifted EPS's, and the two finite parts of the quantity
defined in eq.~(\ref{Vdef}) computed with the shifted EPS's, respectively
(needless to say, the computation of the Born can always be fully trusted,
regardless of whether a kinematic configuration is an EPS).
It will also be convenient to define the one-loop contributions normalized
to the corresponding Born amplitudes squared, i.e.
\beqn
v^{\rm FIN}=
\frac{V^{\rm FIN}}{\abs{\ampnt}^2}\,,
\;\;\;\;\;\;\;\;
v_{\lambda_\pm}^{\rm FIN}=
\frac{V_{\lambda_\pm}^{\rm FIN}}{\abs{\ampnt_{\lambda_\pm}}^2}\,.
\label{vsmall}
\eeqn
We write our estimate $V_{\lambda=0}^{\rm FIN}$ for the finite part of $V$
computed with the EPS as follows:
\beqn
V_{\lambda=0}^{\rm FIN}=\abs{\ampnt_{\lambda=0}}^2\left(c\pm \Delta\right)\,.
\label{VinEPS}
\eeqn
We point out that eq.~(\ref{VinEPS}) is used with all 
stable kinematic configurations in the evaluation of
the total rate and differental distributions. In doing so, we compute
three integrals (associated with the central value $c$ and the extrema
$c\pm\Delta$), whose spread will be interpreted as the uncertainty
associated with the presence of EPS's. In this procedure, stable 
configurations can be thought of having the same form as in 
eq.~(\ref{VinEPS}), with $\Delta=0$. The $\Delta$'s associated with
different configurations (stable or EPS) are combined in quadrature.
The values of $c$ and $\Delta$ depend on the nature of the \CutTools\ 
results for $\ampnl_{\lambda_\pm}$, according to the following
three possibilities.
\begin{itemize}
\item Both shifted EPS's are classified by \CutTools\ as
stable points, or can be rescued to stability by the gauge-invariance
check. We then set:
\beqn
c&=&\half\left(v_{\lambda_+}^{\rm FIN}+v_{\lambda_-}^{\rm FIN}\right)\,,
\\
\Delta&=&\abs{v_{\lambda_+}^{\rm FIN}-v_{\lambda_-}^{\rm FIN}}\,.
\eeqn
\item One of the shifted EPS's (say, that with $\lambda_+$ to be
definite) is classified by \CutTools\ as stable, or can be rescued to 
stability by the gauge-invariance check, while the other
shifted EPS is an EPS that cannot be rescued to stability by 
the gauge-invariance check. We then set:
\beqn
c=v_{\lambda_+}^{\rm FIN}\,.
\label{SU}
\eeqn
The value of $\Delta$ is set in the same way as for the case
immediately below, and we shall give its definition later.
\item Both shifted EPS's are classified by \CutTools\ as EPS's,
and cannot be rescued to stability by the gauge-invariance check.
In this case, we set:
\beqn
c={\rm med}\Big(\left\{v^{\rm FIN}\right\}_{\rm stable}\Big)\,,
\label{UU}
\eeqn
where ${\rm med}()$ denotes the median of its argument, which is a set 
of real numbers. Such a set in eq.~(\ref{UU}) is that of the values of
the ratios of the finite part of $V$ over the Born
amplitude squared, computed for all the {\em stable} phase-space
points encountered so far.
\end{itemize}
As far as the uncertainty $\Delta$ is concerned, associated with 
the central values defined in eqs.~(\ref{SU}) and~(\ref{UU}),
we define it as follows:
\beqn
\Delta=
{\rm med}\Bigg(\Bigg\{\abs{v^{\rm FIN}-
{\rm med}\Big(\left\{v^{\rm FIN}\right\}_{\rm stable}\Big)}
\Bigg\}_{\rm stable}\Bigg),
\label{Deltadef}
\eeqn
which is the median absolute deviation of the same set as that used
in eq.~(\ref{UU}). The use of the median and of the median absolute 
deviation in eqs.~(\ref{UU}) and~(\ref{Deltadef}) in place of the more 
common mean and standard deviation is discussed below.

A couple of comments are in order. Firstly, the effect of shifting the
kinematic configuration has been tested on stable points. One can
see that the ratio $V/\abs{\ampnt}^2$ depends very weakly on the 
value of $\lambda_\pm$, while the dependence on $\lambda_\pm$ of either 
of the two quantities that appear in this ratio is larger. This is one 
of the motivations for considering this ratio (rather than $V$) when
computing the r.h.s.~of eq.~(\ref{VinEPS}). Consistently with this,
we point out that the phase-space weight associated with 
$V_{\lambda=0}^{\rm FIN}$ is that of the original (unshifted)
EPS -- by using the shifted-EPS phase-space weights one would
introduce an unnecessary dependence on $\lambda_\pm$.
Secondly, the ratio $V/\abs{\ampnt}^2$ allow us to use
stable-point results for estimating EPS results,
as done in eqs.~(\ref{SU})--(\ref{Deltadef}), because by
dividing by the Born amplitude squared one becomes largely insensitive
to the structure of the Born itself. The actual values of $\lambda_\pm$
must be chosen so as the shifted EPS's have a good chance of being
classified as stable by \CutTools, or of being rescued by means of
the gauge-invariance test, while still being small. Empirically, 
we have determined that \mbox{$\lambda_\pm=\pm {\cal O}(0.01)$} 
are sufficient to this end.

Owing to the logarithms that appear in one-loop amplitudes,
the ratio $V/\abs{\ampnt}^2$ may be very large in certain corners
of the phase space. This is not a problem in the course of a numerical
integration, because these corners have small measures (which is 
equivalent to saying that the logarithm has an integrable singularity),
and hence large ratios will appear quite infrequently, thus giving
an overall small contribution to the total integral. In spite of being
rare, however, arbitrarily large ratios can still drive the mean 
of all ratios to arbitrarily large values. In a statistical
language, this implies that the mean is not a robust measure of central 
tendency, being sensitive to outliers (i.e., to measurements that
lie in the tails of distributions, or to deviations w.r.t.~the
assumed probability distribution). This would not be an issue if one were 
able to estimate the mean of the stable values by using large statistics,
and before encountering the first EPS: unfortunately, neither of
these conditions is true. The quantity that measures the
central tendency with the largest robustness is the median, which
is the reason why we have chosen it in eq.~(\ref{UU}).
The analogous quantity that measures the variability of a sample
is the median absolute deviation (which is the analogue of the
standard deviation), adopted in eq.~(\ref{Deltadef}). It should be
kept in mind that the median and the mean are always less than one 
standard deviation apart, and this implies that they are statistically
equivalent for our purposes. Furthermore, since the median absolute deviation
is larger than the standard deviation, eq.~(\ref{Deltadef}) represents a 
conservative choice for the estimation of the uncertainty associated
with the procedure described here. It should be finally remarked that
by choosing $c$ in eq.~(\ref{VinEPS}) equal to the mean, the total
integral (i.e., the sum over all kinematic configurations) would be
strictly equal to the exact one, provided that the Riemann sums computed
for the numerical evaluation of such an integral were associated with
phase-space partitions of equal-volume cells. This is not the case
when adaptive-integration algorithms are used, as in the module
Vegas~\cite{Lepage:1977sw} that we employ. However, it is easy to realize
that when using Vegas the total integral is equal to the mean of the 
integrand, times the Vegas weight (which plays the role of the jacobian 
associated with a change of variables). Therefore, in \MadLoop\ what we
actually use are the quantities defined in eq.~(\ref{vsmall}), times
the Vegas weight. We have refrained from introducing the latter in the
equations above just in order to simplify the discussion and the notation.

In conclusion, with the procedure described in this section we are able to
detect, cure or approximate on the fly all possible numerical instabilities.
In the case in which an approximation is necessary, we give an estimate
of its impact. We point out again that EPS's almost never appear,
and those which are still EPS after the kinematic shifts of 
eq.~(\ref{kresc}) are {\em extremely} rare. As mentioned at
the beginning of this section, among the processes considered 
in table~\ref{tab:results}, only b.3 and b.6 have a fraction
of EPS's larger than $10^{-4}$, of which 99.9\% are recovered to
stability after shifting. The uncertainties due to EPS's on the
cross sections of processes b.3 and b.6 are equal to $4\mydot 10^{-4}$~pb
and $2\mydot 10^{-4}$~pb respectively.

\section{Limitations of the current implementation\label{sec:limits}}
We have presented in sect.~\ref{sec:res} a few examples that prove
that \MadLoop\ is able to cope with a large variety of process.
There are obviously computations that \MadLoop\ cannot perform,
the reason being due either to the physics of the process,
or to computer-related issues. Before going into the details,
let us remind the reader that \MadLoop\ currently computes QCD
one-loop corrections to tree-level SM processes; in other words, other 
kind of corrections (e.g.~EW ones), and in general corrections to 
processes defined in other theories (e.g.~SUSY) are not yet 
possible\footnote{We stress explicitly that this limitation applies
to any theory with effective vertices. This implies, in particular,
that we cannot compute NLO corrections to processes that contain a $ggH$ 
vertex, such as $gg\to H+X$. Note that in the SM, NLO corrections to 
$gg$-induced Higgs production is a two-loop computation, for which 
no general numerical algorithm is presently available.} -- see 
sect.~\ref{sec:outlook} for an outlook on this point.

We start with physics-related limitations. There are four of 
them\footnote{The only limitation in this list that applies also
to \MadFKS\ is the one in item 3.},
with the first three that prevent one from generating some classes
of processes (in the sense that results cannot be obtained for any 
phase-space point), and a fourth one that does not allow one to
integrate straightforwardly over the phase space.
We list them in what follows.
\begin{enumerate}
\item A process cannot be generated if, at the Born level,
it contains a four-gluon vertex.
\item The presence of EW massive vector bosons in the loops imply that
certain processes cannot be generated, depending on the number of
the bosons and on the identities of the other loop particles.
\item A process cannot be generated if all contributions to the
Born amplitude squared do not factorize the same powers of $\as$ 
and $\alpha$.
\item Finite-width effects, due to intermediate massive unstable (i.e.,
that can decay) particles that can also enter in the loops, are not 
implemented.
\end{enumerate}
We now briefly comment on these four points in turn. 
The condition in item 1 is due to the fact that the four-gluon
vertex is represented internally by \MadGraph\
in a very involved way, owing to the non-trivial mixture between its
Lorentz and colour structures. This implies that the $R_2$ vertices
with an analogous form would require a substantial amount of
debugging before being deemed correct since, although their
implementation does not pose any problems of principle, it 
is technically error-prone\footnote{The cut-constructible 
and $R_1$ contributions can be computed for the processes we are
discussing here, but this is not particularly helpful from 
the physics point of view, since it is only after adding the $R_2$
contribution that one obtains a gauge-invariant result.}.
It appears that the investment in time required for the implementation
of four-gluon-type $R_2$ vertices is not justified in light of the fact 
that \MadGraph\ v5 is now available, and when \MadLoop\ will be made
compatible with it, this task will become trivial. 

The restriction in item 2 is due to the fact that \CutTools\ gives
a sensible answer only if the largest power of the loop momentum
$\qloop$ in $N(\qloop)$ is smaller than or equal to the number of
denominators (propagators are always manipulated so as their
denominators are quadratic in the loop momentum). 
It is therefore clear that the term
\mbox{$k^\mu k^\nu/m^2$} that appears in the propagators of the 
massive EW vector bosons {\em may} lead to violations of the above
condition. Obviously, this happens because \MadGraph\ v4 adopts
the unitarity gauge. This difficulty will be lifted
when \MadGraph\ v5 will be used, since this will allow us to use
the Feynman gauge, in which the term mentioned above is not present.

The case of item 3 applies to processes that feature interference
between QCD and EW contributions at the Born level (e.g., in the 
$q\bar{q}\to t\bt$ process one may want to include the diagrams that
contain $Z$ and $\gamma^*$ exchanges, which then interfere with the
diagram that contains a gluon exchange). This implies
that the counterterms needed for UV renormalization cannot 
be proportional to the Born amplitude squared, as assumed
in appendix~\ref{sec:UVcnts} (except for the case of mass 
insertion). Hence, in this situation the UV renormalization
procedure as set up in the current version of \MadLoop\ fails. 
Although it is not difficult to fix this\footnote{For example, by 
performing the renormalization along the general lines described in 
sect.~\ref{sec:R2}, i.e.~analogously to what is done at present for 
the $R_2$ contribution.}, in practice it is not particularly 
useful, since the case discussed here is that of a double 
perturbative expansion in $\as$ and $\alpha$. Hence, consistency 
would dictate that EW corrections be computed as well, which is something 
that neither \MadLoop\ nor \MadFKS\ are equipped to do at the moment.

Finally, SM cases relevant to item 4 are those of the top quark and
of the massive EW vector bosons. When considering diagrams which 
include one or more propagators of one or more of these particles,
there may be configurations of external momenta which end up 
in putting some of them on their mass shells. This thus results
in a divergence, which can be avoided by giving finite widths 
to the unstable particles. In the case in which the relevant particles 
also enter in the loops, the use of non-zero widths is non trivial, 
and consistency dictates the use of a scheme like the complex-mass
one~\cite{Denner:2006ic}. At present, such a scheme is not implemented 
in \MadGraph\ (on the other hand, \CutTools\ is already
able to treat complex masses). 

Let us now turn to computer-related limitations. A minor one is
the condition that each \Lcut\ process have less than $10^4$ diagrams;
this condition will be removed in future versions of \MadLoop.
Furthermore, we cannot restrict the particle types flowing in the 
internal lines of the diagrams through the input cards, as done in \MadGraph.
More relevant is the fact that, since the present one is the first 
version of \MadLoop, it has been constructed with a very
minimal amout of optimization in order not to complicate the
code structure with features not dictated by physics. This implies
that large-multiplicity processes\footnote{Exactly how large depends
on the nature of the final state. A large number of jets will use
a lot more CPU than the same number of e.g.~vector bosons, mainly
because of gluon-dominated subprocesses.} cannot be computed even 
in a few days on a ${\cal O}(100)$-machine cluster. At present,
the optimization is limited to caching the wave functions of external
particles, and to scanning (diagram-by-diagram) the helicity space for the 
first few phase-space points, in order not to consider in the rest of the 
run those helicity-configuration/diagram combinations that give a 
contribution exactly equal to zero to the one-loop matrix elements.
The possibility of summing over helicities using a Monte Carlo procedure
is also implemented.

As a general and concluding remark on CPU-driven issues, it is
clear that, regardless of the amount of optimization done on the
code, the use made of Feynman diagrams in the current
version introduces a factorially-growing complexity in the
calculation. However, it is easy to realize that the FKS subtraction
method is completely independent of Feynman-diagram techniques, and
that to a large extent this is also the case for the OPP reduction procedure.
For comments on this point and in general on the optimization of future
versions of \MadLoop, see sect.~\ref{sec:outlook}.

\section{Outlook\label{sec:outlook}}
What we have discussed so far proves that the combination of 
OPP reduction and FKS subtraction is sufficient to deal with
any kind of SM process in a fully-automated manner. The limitations 
of the present version of \MadLoop, which we have listed
in sect.~\ref{sec:limits}, are due either to features inherited
from \MadGraph\ v4, or to the lack of optimization of the code.
We shall now sketch our short- and mid-term plans for
improving \MadLoop.

From the physics viewpoint, the most serious limitations at
present are those reported in items 1 and 2 of sect.~\ref{sec:limits}.
These limitations will be fully removed when \MadLoop\ will use
\MadGraph\ v5 rather than v4, which we expect to happen in the next 
few months, given that all the necessary ingredients are already available 
and reasonably well tested. The reason for this is due to 
the capability of \MadGraph\ v5 of {\em constructing} \HELAS\
routines\footnote{This has to be compared to v4 and earlier versions, 
in which those routines had to be written by hand.} starting from a set 
of rules. Therefore, all $R_2$ \HELAS\ routines
(which removes the limitation of item 1) and all regular \HELAS\ 
EW routines defined in any $R_\xi$ gauge (and in particular in the Feynman 
gauge, which removes the limitation of item 2) will become available
in a straightforward way\footnote{The $R_2$ vertices for the EW theory
have been derived in refs.~\cite{Garzelli:2009is,Garzelli:2010qm}.}. 
The chief advantage of this procedure is
that it will render the construction of the building blocks of an
amplitude virtually error-free.

From the computing viewpoint, the most notable lack of optimization
at present is due to the insufficient caching of information.
In particular, for any given $2\to n$ kinematic configuration
it is only the numerical values of the external wave functions 
that are saved in memory. This implies that the internal propagators
and vertices that enter the tree structures $T_i$'s attached
to the loops (see sect.~\ref{sec:genfil}) are recomputed each time
\CutTools\ changes the loop momentum, in spite of the fact that they
do not depend on such momentum. The storage of the values of the
$T_i$'s is technically almost trivial and we shall do that in the
near future, again exploiting the features of \MadGraph\ v5, whose
framework is more naturally suited than that of \MadGraph\ v4
to caching operations. We finally point out that at present the
contributions due to different massless quarks circulating in 
closed-fermion loops are computed independently from each other,
which is quite inefficient in view of the fact that some of them
will give identical results (all of them in the case of pure-QCD
processes). The optimization of this aspect of the computation
does not pose significant problems.

Our plans for the mid-term future are described in what follows. 

First of all, the efficiency of \MadLoop\ will be further increased, 
by using \CutTools\ in a more rational way than at present. One simply
observes that the OPP procedure need not be necessarily applied 
on a diagram-by-diagram basis, but one can sum diagrams whose
amplitudes have the same denominators, and perform the OPP reduction 
on this sum (this amounts to identify all the $\bar{\idiag}$'s of 
eq.~(\ref{Cbardef}) with the same combination of $\db{i}$'s, and to give in 
input to \CutTools\ {\em the sum} of the corresponding $N(\qloop)$ functions).
It is easy to realize that the diagrams that can be summed together
have the same number of $T_i$'s, and that their colour structures 
along the loop are identical (these colour structures are equivalent
to the set of the colour indices of the $T_i$'s and of the loop particles that 
appear in the \Lcut-diagram identities). Hence, the sum of the $N(\qloop)$ 
functions will contain the sum of all $T_i$'s functions with the same
index $i$ (i.e \mbox{$\sum_\alpha T_i^{(\alpha)}$}, with $\alpha$ running
over the relevant Feynman diagrams), whose numerical values are different,
but which have the same external four-momenta, and the same colour indices 
on the leg attached to the loop. At this point, one will cache the
values of \mbox{$\sum_\alpha T_i^{(\alpha)}$} for all $i$'s, and the
caching procedure will be identical to the one discussed above for the
case of a diagram-by-diagram OPP reduction. We expect a significant
reduction in computing time w.r.t.~that of the single-diagram approach, 
especially in the case of ``small'' loops with ``large'' $T_i$'s.

This would be the end of story in the context of a calculation entirely
based on Feynman diagrams. On the other hand, it is easy to realize
that the \mbox{$\sum_\alpha T_i^{(\alpha)}$} combinations can be
interpreted as (tree-level) currents attached to the loops. Therefore,
their computations can be performed with more efficient methods than
Feynman diagrammatics. In particular, in the case of many-gluon $T_i$'s 
we plan to employ recursion relations, whose implementation in
\MadGraph\ v5 is well under way. However, apart from processes with very 
large multiplicities, we expect that the gain in efficiency due to the caching
of tree structures will be larger than that due to the use of recursion
relations. This is because recursion relations cannot be applied to 
the \Lcut\ diagrams as a whole but only to the $T_i$'s, since 
the loop propagators play a special role in \CutTools, and must always
be treated in a Feynman-diagrammatic way. On the other hand, recursion
relations are a very powerful tool to reduce the computing time
of the real-emission contributions to the cross section.
We point out that their use is fully compatible with the implementation
of the FKS subtraction done in \MadFKS, since the latter only depends 
on the knowledge of the identities of the particles entering the process,
and treats the matrix elements as black boxes. 

Turning now to plans more directly related to physics, we point out 
that the use of \MadGraph\ v5 will render the implementation of
a complex-mass scheme an easy task. This will eliminate the limitation
reported as item 4 in sect.~\ref{sec:limits}. Although such a scheme
is not expected to pose any technical problems, it has not yet
been tested in \MadGraph\ (while it is fully operational in \CutTools). 
It is also worth mentioning that the possibility of computing NLO
corrections to processes that involve a $ggH$ effective vertex
only requires the implementation of a few trivial $R_2$ functions --
as shown in ref.~\cite{Frederix:2009yq}, \MadFKS\ is already capable
of handling such a situation.

In a longer-term perspective, the possibility of using complex masses 
and a renormalizable gauge will pave the way to automate the computation
of EW corrections to any SM process. This will therefore allow us
to treat consistently a double perturbative expansion (in $\as$
and $\alpha$), and ultimately to remove the limitation reported 
as item 3 in sect.~\ref{sec:limits}. In order to achieve this goal,
some technical work is still necessary on both \MadLoop\ and \MadFKS\ 
(where it is essentially trivial, since it just amounts to including the
subtraction of QED singularities).

It is also interesting to notice that the automation of one-loop
computations achieved here is based on a few components that
have required analytical work: the scalar integrals, and the
UV and $R_2$ counterterms. While the former are not related to
any particular theory, and their validity is thus universal,
the latter are theory-specific: e.g., QCD and EW counterterms
are different. It would be extremely useful to be able to
obtain the counterterms directly from the Lagrangian of the
theory, in the same way as is done now for the usual Feynman
rules. This will clearly open up the possibility of automating
one-loop corrections in any renormalizable theory.

\section{Conclusions\label{sec:conclusions}}
The work we have presented in this paper is based on the strategic assumption 
that, for the word ``automation'' to have its proper meaning, the only 
operation required from a user is that of typing-in the process to be 
computed, and other analysis-related information (such as final-state cuts).
In particular, the code that achieves the automation may only differentiate
between processes depending on their general characteristics, but 
must never work on a case-by-case basis. Furthermore, such 
a differentiation must be hidden to the user. This approach guarantees
a maximum amount of flexibility, which we believe is clearly shown
by the variety of the physics results presented in sect.~\ref{sec:res}.
It also ensures that, after an initial validation phase (which we
have described in details here -- see in particular sect.~\ref{sec:checks}
and appendix~\ref{sec:comps}), the computations of new processes will be much
more likely to be correct than if they were performed from scratch
using traditional methods.

An essential component of the physics results of sect.~\ref{sec:res}
is the calculation of the subtracted real-emission contribution,
its combination with the one-loop part, and their integration over
the phase space: we have performed these operations with \MadFKS.
We point out that \MadFKS\ and \MadLoop\ are fully independent,
and can be used in connection with other codes.

The drawback of the automation strategy adopted here is that, for any 
given process, the amount of computer time required for evaluating the
corresponding cross section will be typically larger than that one
would have needed by using dedicated codes. Furthermore,
one may not be able to compute classes of processes.
We have discussed the limitations of our code in sect.~\ref{sec:limits}.
It is very encouraging that they are quite few, and the majority
of them will be eliminated in the next version of \MadLoop, which
is foreseen to appear in the near future, as we have discussed
in sect.~\ref{sec:outlook}.

In summary, we have shown that NLO computations in QCD are essentially
on equal footing with LO ones, up to some minor improvements that 
we have illustrated above. This highly non-trivial situation has
been achieved thanks to a detailed understanding of perturbative techniques,
which has allowed one to define process- and multiplicity-independent
subtraction formalisms, and to new methods for the computation of
the virtual contributions to cross sections.

\section*{Acknowledgments}
We thank Tim Stelzer for his help with technical aspects of \MadGraph.
We are grateful to John Campbell for his assistance with MCFM, 
to Adrian Signer for having provided us with the code MENLO PARK,
and to Stefan Dittmaier for discussions concerning $t\bt j$ production.
All the authors who are not based at CERN would like to thank
the CERN TH group for the continuous hospitality during the
course of this work.
This research has been supported by the Swiss National Science Foundation
(NSF) under contract 200020-126691, and in part by the Interuniversity 
Attraction Pole Program -- Belgium Science Policy P6/11-P and the IISN 
convention 4.4511.10. M.V.G., F.M. and R.P. thank the 
financial support of the MEC project FPA2008-02984 (FALCON).
The research of R.P. has been supported by the Spanish Ministry of
education under contract PR2010-0285.

\appendix
\section{Comparisons with existing results\label{sec:comps}}
In this appendix, we present the comparisons between \MadLoop\
results, and those obtained either with public computer codes, or by 
implementing ourselves analytical results published in the literature.
As discussed in sect.~\ref{sec:checks}, these comparisons are
an essential part of the validation of \MadLoop, and allow us
to check all the building blocks used by the code to construct
the one-loop amplitudes relevant to arbitrary processes.

We compute the quantity\footnote{Except in one case, when we
shall consider the square of a non-divergent one-loop amplitude --
see sect.~\ref{sec:gggZ}.}
defined in eq.~(\ref{Vdef}), which can be re-expressed as follows:
\beqn
V(\proc)=\frac{(4\pi)^\ep}{\Gamma(1-\ep)}
\left(\frac{\muF^2}{Q^2}\right)^\ep
F\left(\frac{c_{-2}}{\ep^2}+\frac{c_{-1}}{\ep}+c_0\right)\,.
\label{Vexpr}
\eeqn
We shall also denote by
\beqn
a_0=\overline{\mathop{\sum_{\rm colour}}_{\rm spin}}\abs{\ampnt}^2
\label{a0def}
\eeqn
the Born matrix element squared, summed/averaged over spin and colour 
degrees of freedom. The constant $F$ in eq.~(\ref{Vexpr})
may clearly be absorbed into the coefficients $c_i$;
we have introduced it in order to facilitate the comparison between
\MadLoop\ results and those of other codes. In the following, where
not explicitly indicated otherwise, its value has been set equal 
to one. The expression in eq.~(\ref{Vexpr}) understands that we may treat 
as independent the Ellis-Sexton scale $Q$, the factorization scale $\muF$, 
and the renormalization scale $\muR$ (which is the argument of $\as$,
contained implicitly in $c_i$). As explained in ref.~\cite{Frederix:2009yq} 
(see in particular appendices~B and~C there), this is most conveniently 
done by computing the one-loop contribution by setting all scales equal 
to the Ellis-Sexton scale there, and by introducing in the short-distance
cross section a compensating contribution, proportional to the Born
amplitude squared and which depends linearly on \mbox{$\log\muF/Q$}
and on \mbox{$\log\muF/\muR$}. In our setup, we have included the latter
contribution in \MadFKS, and consistently with this choice only 
the Ellis-Sexton scale $Q$ enters the computations performed by \MadLoop.
It should be remarked that most of the codes we have used in this
appendix as benchmarks for the validation of \MadLoop\ are forced
in any case to set
\beqn
\mu\equiv Q=\muF=\muR\,,
\label{eqscales}
\eeqn
and hence the compensating factor in \MadFKS\ is equal to zero.
In a few cases we were able to relax the condition of eq.~(\ref{eqscales}),
and thus to test the full scale dependence of our computations.

All the comparisons we present here are local, i.e.~performed at
fixed $2\to n$ configurations. We typically show the results for
one such configuration, although as a safety measure we have
checked a few more of them. We did not attempt to choose the same
set of input parameters for different processes, since priority
was given to using the codes we compare our results to with their
defaults, in order to limit as much as is possible the number of
operations we perform on them. This consideration is particularly
important when it comes to choosing the kinematic configuration(s)
we use in our comparisons. When using a public code which embeds
the one-loop amplitudes in a cross-section integrator, we run the
code and save in a file the kinematic configurations and the 
corresponding one-loop results. We then pick up at random a few 
of them, and give them in input to \MadLoop. This procedure guarantees
that the public code is not modified except for a few trivial
output statements.

As discussed in sect.~\ref{sec:checks}, it is not always possible
to compare our results for $c_{-2}$ and $c_{-1}$ with those of
public codes. However, we do always compare them with their known
analytic forms\footnote{This is done when single poles do not
contain UV contributions, namely if UV renormalization is
performed. Although we do this by default, there are cases in which
we compare to computations which do not include UV renormalization -- 
see sects.~\ref{uubbbb}, \ref{uuttbb}, and~\ref{qqwwbb}. In these
situations, the comparisons with analytical results for pole residues 
are not carried out.}, and in doing so the agreement we find is at the
level of the $12^{th}$ digit or better for all processes.
The corresponding comparisons with public codes are often worse
than this, being at the level of single-precision computations
of real-number algebra. The reason for this is indeed a single-precision
to double-precision conversion done by the computer, since some of
the parameters in the input cards of \MadLoop\ are in
single precision (whereas all computations are performed in
double precision). These differences are obviously completely
irrelevant in cross section calculations, and are mentioned 
here for the sake of completeness. 

In the context of the computation of a cross section, the infrared
divergences of $V(\proc)$ cancel those present in the subtracted
real-emission contribution. Since the latter is computed using
Conventional Dimensional Regularization (CDR), this is the scheme
of choice for $V(\proc)$ as well. On the other hand, other schemes
are more suitable to one-loop computations. The 't~Hooft-Veltman scheme
is used by \CutTools: the finite part $c_0$ computed in such a scheme
is identical to that computed in CDR. The residues $c_{-2}$ and $c_{-1}$ 
in the 't~Hooft-Veltman scheme can be obtained from those 
in CDR by setting the number of space-time dimensions equal to four
there (in CDR one has $d=4-2\ep$). Another popular infrared scheme
is Dimensional Reduction (DR). The difference between the finite
parts $c_0$ computed in CDR and DR is proportional to the Born,
and can thus be easily accounted for. The relevant formulae can
be found in eqs.~(B.3) and~(B.4) of ref.~\cite{Frederix:2009yq}.
For more detailed discussions on infrared schemes, 
see e.g.~ref.~\cite{Kunszt:1993sd}. In this appendix, we shall
use either the 't~Hooft-Veltman or the DR scheme for the finite parts,
while pole residues will always be given in the 't~Hooft-Veltman scheme.

When considering closed fermion loops with one EW vector boson 
leg there is possibly an anomalous contribution. By default, 
we shall always include complete quark families in the loops,
thereby avoiding this problem. In this appendix, we have studied
cases in which the anomaly cancellation may not be immediately evident 
(this occurs in connection with the decoupling limit -- see 
sect.~\ref{sec:checks}). We have discussed the relevant processes
in some details.

Unless otherwise indicated, all dimensionful quantities that
appear in this appendix are given in GeV.

\subsection{QCD processes}
All processes considered in this section are pure-QCD ones,
i.e.~EW-boson exchanges are excluded from computations.

\subsubsection{The process $u\bu\to d\bd$\label{qqqpqp}}
The following set of parameters is used:
\begin{center}
\begin{tabular}{cl|cl}\toprule
Parameter & value & Parameter & value
\\\midrule
$\as$ & \texttt{0.13} & $n_{lf}$ & \texttt{2}
\\
$\mu$ & \texttt{91.188}    &  & 
\\\bottomrule
\end{tabular}
\end{center}
The kinematic configuration considered is:
\begin{small}
  \begin{align*}
    p_u=&\textrm{( 102.6289752320661}&&\textrm{, 0}&&\textrm{, 0}&&\textrm{, \phantom{-}102.6289752320661}&&\textrm{)}\\
    p_{\bar{u}}=&\textrm{( 102.6289752320661}&&\textrm{, 0}&&\textrm{, 0}&&\textrm{, -102.6289752320661}&&\textrm{)}\\
    p_d=&\textrm{( 102.6289752320661}&&\textrm{, -85.98802977488269}&&\textrm{, -12.11018104528534}&&\textrm{, \phantom{-}54.70017191625945}&&\textrm{)}\\
    p_{\bar{d}}=&\textrm{( 102.6289752320661}&&\textrm{, \phantom{-}85.98802977488269}&&\textrm{, \phantom{-}12.11018104528534}&&\textrm{, -54.70017191625945}&&\textrm{)}
  \end{align*}
\end{small}
The finite part is given in the 't~Hooft-Veltman scheme. We compare the 
\MadLoop\ results with those of the code of ref.~\cite{Frixione:1997np},
which implements the formulae given in ref.~\cite{Ellis:1985er}.
We obtain what follows:
\begin{center}
\begin{tabular}{cll}\toprule
  $u\bar{u}\to d\bar{d} $& \MadLoop\ & Ref.~\cite{Frixione:1997np}
\\\midrule
$a_0$ & \texttt{\phantom{-}0.76152708418254678E+000}&\texttt{\phantom{-}0.76152695293848227E+000}
\\
$c_{-2}$ & \texttt{-0.08403255449056724E+000}&\texttt{-0.08403254000812221E+000}
\\
$c_{-1}$ & \texttt{-0.10222774402685941E+000}&\texttt{-0.10222772640859645E+000}
\\
$c_0$ & \texttt{-0.44023547006851060E-001}&\texttt{-0.44023539433227843E-001}
\\\bottomrule
\end{tabular}
\end{center}
Since the code of ref.~\cite{Frixione:1997np} allows one to set the
mass scales entering the process independently from each other, we have
also found excellent agreement with \MadLoop\ plus the compensating 
contribution computed by \MadFKS\ for $\muR\neq\muF\neq Q$; we refrain 
from reporting the results of these tests here.

\subsubsection{The process $dg \to dg$}
The same parameters as in sect.~\ref{qqqpqp} are chosen.
The kinematic configuration considered is:
\begin{small}
  \begin{align*}
    p_d=&\textrm{( 220.9501779577791}&&\textrm{, 0}&&\textrm{, 0}&&\textrm{, \phantom{-}220.9501779577791}&&\textrm{)}\\
    p_g=&\textrm{( 220.9501779577791}&&\textrm{, 0}&&\textrm{, 0}&&\textrm{, -220.9501779577791}&&\textrm{)}\\
    p_d=&\textrm{( 220.9501779577791}&&\textrm{, \phantom{-}119.9098300357375}&&\textrm{, \phantom{-}183.0492135511419}&&\textrm{, -30.55485589367430}&&\textrm{)}\\
    p_g=&\textrm{( 220.9501779577791}&&\textrm{, -119.9098300357375}&&\textrm{, -183.0492135511419}&&\textrm{,  \phantom{-}30.55485589367430}&&\textrm{)}
  \end{align*}
\end{small}
The comparison between \MadLoop\ and the code of ref.~\cite{Frixione:1997np}
reads as follows:
\begin{center}
\begin{tabular}{cll}\toprule
  $dg\to dg $& \MadLoop\ & Ref.~\cite{Frixione:1997np}
\\\midrule
$a_0$ & \texttt{\phantom{-}13.032125409659082E+000}&\texttt{\phantom{-}13.032125409659088E+000}
\\
$c_{-2}$ & \texttt{-2.3368499538132292E+000}&\texttt{-2.3368499538132284E+000}
\\
$c_{-1}$ & \texttt{\phantom{-}2.1147910298734116E+000}&\texttt{\phantom{-}2.1147910298729693E+000}
\\
$c_0$ & \texttt{-1.8580245435782883E+000}&\texttt{-1.8580245414019134E+000}
\\\bottomrule
\end{tabular}
\end{center}
For the present process, we also performed the crossing checks as discussed
in sect.~\ref{sec:checks}, by comparing it to the $gg\to d\bar{d}$ and
$d\bar{d} \to gg$ processes. We have found perfect agreement.

\subsubsection{The processes $d\bd\to t\bt$ and $gg\to t\bt$}
We compare the \MadLoop\ results for top-pair production with
those of MCFM~\cite{Campbell:1999ah}. The following input parameters
are used:
\begin{center}
\begin{tabular}{cl|cl}\toprule
Parameter & value & Parameter & value
\\\midrule
$\as$ & \texttt{0.13} & $n_{lf}$ & \texttt{5}
\\
$m_{top}$ & \texttt{172.5} & $\mu$ & \texttt{91.188}
\\\bottomrule
\end{tabular}
\end{center}
Note that the top quark is considered a stable particle, and hence
its width is set to zero. The phase-space point used for this check is:
\begin{small}
  \begin{align*}
    p_1=&\textrm{( 63.71791270829688}&&\textrm{, 0}&&\textrm{, 0}&&\textrm{, \phantom{-}63.717912708296879}&&\textrm{)}\\
    p_2=&\textrm{( 814.2396220727112}&&\textrm{, 0}&&\textrm{, 0}&&\textrm{, -814.2396220727112}&&\textrm{)}\\
    p_t=&\textrm{( 663.0455079348429}&&\textrm{, -54.65940267927511}&&\textrm{, \phantom{-}25.31239299113409}&&\textrm{, -637.3733035297141}&&\textrm{)}\\
    p_{\bar{t}}=&\textrm{( 214.9120268461650}&&\textrm{, \phantom{-}54.65940267927511}&&\textrm{, -25.31239299113409}&&\textrm{, -113.14840583470011}&&\textrm{)}
  \end{align*}
\end{small}
with $p_1$ and $p_2$ the momenta of the initial-state partons coming
from the left ($g$ or $d$) and from the right ($g$ or $\bd$) respectively.
The finite part is given in the DR scheme. We obtain what follows:
\begin{center}
\begin{tabular}{cll}\toprule
  $d\bar{d}\to t \bar{t} $& \MadLoop\ & MCFM
\\\midrule
$a_0$ & \texttt{\phantom{-}1.1446316446116180E+000}&\texttt{\phantom{-}1.1446316446116067E+000}
\\
$c_{-2}$ & \texttt{-6.3153578543239441E-002}&\texttt{\phantom{-}--}
\\
$c_{-1}$ & \texttt{\phantom{-}9.4367389242209110E-002}&\texttt{\phantom{-}--}
\\
$c_0$ & \texttt{-0.3252731490962368E+000}&\texttt{-0.3252731490962548E+000}
\\\midrule
 $g g \to t \bar{t}$ &&
\\\midrule
$a_0$ & \texttt{\phantom{-}1.3065171790431791E+000}&\texttt{\phantom{-}1.3065171790431644E+000}
\\
$c_{-2}$ & \texttt{-0.1621921604777766E+000}&\texttt{\phantom{-}--}
\\
$c_{-1}$ & \texttt{\phantom{-}0.1441104689442122E+000}&\texttt{\phantom{-}--}
\\
$c_0$ & \texttt{-2.1116148093780568E-002}&\texttt{-2.1116148095016787E-002}
\\\bottomrule
\end{tabular}
\end{center}
The residues of the double and single poles have been checked
against those returned by \MadFKS, and perfect agreement has been found.
This is not entirely trivial, because of the role played by UV
renormalization, and in particular by the insertion of the UV mass counterterm
on the virtual top-quark line that appears in the $gg$ channel, which
could not be checked in the case of dijet production.

\subsubsection{The process $ug \to t\bar{t}u$}
The process $pp\to t\bar{t}+1j$ has been first computed by
Dittmaier, Uwer and Weinzierl~\cite{Dittmaier:2007wz,Dittmaier:2008uj}. 
Recently the results of this calculation have been verified by 
HELAC-1Loop~\cite{Bevilacqua:2010ve} and by Melnikov and
Schultze~\cite{Melnikov:2010iu}. Here we consider the $ug \to t\bar{t}u$ 
channel, and compare our results with those of ref.~\cite{Dittmaier:2008uj}.
The parameters we use are: 
\begin{center}
\begin{tabular}{cl|cl}\toprule
Parameter & value & Parameter & value
\\\midrule
$\as$ & \texttt{0.1075205492734706} & $n_{lf}$ & \texttt{5}
\\
$m_{top}$ & \texttt{174.0} & $\mu$ & \texttt{174.0}
\\
$F$ & $\Gamma(1-\epsilon)\,\Gamma(1+\epsilon)\,g_{\sss S}^6\,a_0$ &  & 
\\\bottomrule
\end{tabular}
\end{center}
where the value of $F$ has been chosen in order to follow the conventions
of the appendix of ref.~\cite{Dittmaier:2008uj}.
The kinematic configuration we adopt is:
\begin{small}
  \begin{align*}
    p_u&=\textrm{(500, 0, 0, 500)}\\
    p_g&=\textrm{(500, 0, 0, -500)}\\
    p_t&=\textrm{(458.5331753852783, 207.0255169909440, 0, 370.2932732896167)}\\
    p_{\bar{t}}&=\textrm{(206.6000026080000, -10.65693677252589, 42.52372780926147, -102.3998210421085)}\\
    p_u&=\textrm{(334.8668220067217, -196.3685802184181, -42.52372780926147, -267.8934522475083)}
  \end{align*}
\end{small}
We obtain:
\begin{center}
\begin{tabular}{cll}\toprule
$ug \to t\bar{t}u$ & \MadLoop\ & Ref.~\cite{Dittmaier:2008uj}
\\\midrule
$a_0$ & \texttt{\phantom{-}1.607845322071586E-005}&\texttt{\phantom{-}1.607845322071585E-005}
\\
$c_{-2}$ & \texttt{-9.697041910469525E-002}&\texttt{-9.69704191047088E-002}
\\
$c_{-1}$ & \texttt{-5.643095699401332E-003}&\texttt{-5.6430956994203E-003}
\\
$c_0$ & \texttt{\phantom{-}4.003849386476366E-001}&\texttt{\phantom{-}4.003849386477017E-001}
\\\bottomrule
\end{tabular}
\end{center}

\subsubsection{The process $u\bu\to b\bb b\bb$ with massless $b$\label{uubbbb}}
This six-quark amplitude is one of the results presented in 
ref.~\cite{vanHameren:2009dr} by A.~van Hameren \etal\footnote{The authors
of ref.~\cite{vanHameren:2009dr} have compared their results against those
obtained by T.~Binoth \etal\ in ref.~\protect\cite{Binoth:2009rv}.}.
The results by \MadLoop\ are thus a check of the implementation of the
\CutTools\ software in a framework different from that of 
HELAC-1Loop~\cite{Bevilacqua:2010ve}. The following parameters are used:
\begin{center}
\begin{tabular}{cl|cl}\toprule
Parameter & value & Parameter & value
\\\midrule
$g_{\sss S}$ & \texttt{1} & $n_{lf}$ & \texttt{5}
\\
$\mtop$ & \texttt{174.0} & $m_b$ & \texttt{0}
\\
$\mu$ & \texttt{500.0} & $F$ & $1\big/(64\pi^2)$
\\\bottomrule
\end{tabular}
\end{center}
where the value of $\mtop$ reminds one that top quarks enter the loops.
The kinematic configuration considered is:
\begin{small}
  \begin{align*}
    p_u=&\textrm{( 250}&&\textrm{, 0}&&\textrm{, 0}&&\textrm{, \phantom{-}250}&&\textrm{)}\\
    p_{\bar{u}}=&\textrm{( 250}&&\textrm{, 0}&&\textrm{, 0}&&\textrm{, -250}&&\textrm{)}\\
    p_{b}=&\textrm{( 147.5321146846735}&&\textrm{, \phantom{-}24.97040523056789}&&\textrm{, -18.43157602837212}&&\textrm{, \phantom{-}144.2306511496888}&&\textrm{)}\\
    p_{\bar{b}}=&\textrm{( 108.7035966213640}&&\textrm{, \phantom{-}103.2557390255471}&&\textrm{, -0.5484684659584054}&&\textrm{, \phantom{-}33.97680766420219}&&\textrm{)}\\
    p_b=&\textrm{( 194.0630765341365}&&\textrm{, -79.89596300367462}&&\textrm{, \phantom{-}7.485866671764871}&&\textrm{, -176.6948628845280}&&\textrm{)}\\
   p_{\bar{b}}=&\textrm{( 49.70121215982584}&&\textrm{, -48.33018125244035}&&\textrm{, \phantom{-}11.49417782256567}&&\textrm{, -1.512595929362970}&&\textrm{)}
  \end{align*}
\end{small}
The finite part is given in the 't~Hooft-Veltman scheme. 
For consistency with ref.~\cite{vanHameren:2009dr}, no UV counterterms 
are included\footnote{We thus set $\ep_{\sss IR}=\ep_{\sss UV}$,
and shall do the same in sects.~\ref{uuttbb} and~\ref{qqwwbb}.}, and the 
residue of the single pole. We obtain:
\begin{center}
\begin{tabular}{cll}\toprule
  $u\bar{u}\to b\bar{b} b\bar{b} $& \MadLoop\ & Ref.~\cite{vanHameren:2009dr}
  \\\midrule
$a_0$ & \texttt{\phantom{-}5.75329342809431E-009}&\texttt{\phantom{-}5.753293428094391E-009}
\\
$c_{-2}$ & \texttt{-9.205269484950836E-008}&\texttt{-9.205269484951069E-008}
\\
$c_{-1}$ & \texttt{-2.404679886707934E-007}&\texttt{-2.404679886692200E-007}
\\
$c_0$ & \texttt{-2.553568662825831E-007}&\texttt{-2.553568662778129E-007}
\\\bottomrule
\end{tabular}
\end{center}
We have also considered different crossings of this process, moving
either a $b$ or a $\bb$ quark to the initial state,
and found full consistency among \MadLoop\ results.

\subsubsection{The process $u\bu \to t\bt b\bb$ with 
massless $b$\label{uuttbb}}
This process has been first computed by Bredenstein \etal\ in
ref.~\cite{Bredenstein:2008zb}. Here, we compare \MadLoop\ with 
the result of A.~van Hameren \etal~\cite{vanHameren:2009dr},
similarly to what was done in sect.~\ref{uubbbb}.
The following set of parameters is used:
\begin{center}
\begin{tabular}{cl|cl}\toprule
Parameter & value & Parameter & value
\\\midrule
$g_{\sss S}$ & \texttt{1} & $n_{lf}$ & \texttt{5}
\\
$\mtop$ & \texttt{174.0} & $m_b$ & \texttt{0} 
\\
$F$ & $1\big/(16\pi^2)$ & $\mu$ & \texttt{500.0} 
\\\bottomrule
\end{tabular}
\end{center}
The top width and the mass of the $b$ quark are set equal to zero. 
The kinematic configuration considered is:
\begin{small}
  \begin{align*}
    p_u=&\textrm{( 250}&&\textrm{, 0}&&\textrm{, 0}&&\textrm{, \phantom{-}250}&&\textrm{)}\\
    p_{\bar{u}}=&\textrm{( 250}&&\textrm{, 0}&&\textrm{, 0}&&\textrm{, -250}&&\textrm{)}\\
    p_{t}=&\textrm{( 190.1845561691092}&&\textrm{, \phantom{-}12.99421901255723}&&\textrm{, -9.591511769543683}&&\textrm{, \phantom{-}75.05543670827210}&&\textrm{)}\\
    p_{\bar{t}}=&\textrm{( 182.9642163285034}&&\textrm{, \phantom{-}53.73271578143694}&&\textrm{, -0.2854146459513714}&&\textrm{, \phantom{-}17.68101382654795}&&\textrm{)}\\
    p_b=&\textrm{( 100.9874727883170}&&\textrm{, -41.57664370692741}&&\textrm{, \phantom{-}3.895531135098977}&&\textrm{, -91.94931862397770}&&\textrm{)}\\
   p_{\bar{b}}=&\textrm{( 25.86375471407044}&&\textrm{, -25.15029108706678}&&\textrm{, \phantom{-}5.981395280396083}&&\textrm{, -0.7871319108423604}&&\textrm{)}
  \end{align*}
\end{small}
The finite part is given in the 't~Hooft-Veltman scheme. 
For consistency with ref.~\cite{vanHameren:2009dr}, no UV counterterms 
are included except the one relevant to top-mass renormalization. We obtain:
\begin{center}
\begin{tabular}{cll}\toprule
  $u\bar{u}\to t\bar{t} b\bar{b} $& \MadLoop\ & Ref.~\cite{vanHameren:2009dr}
  \\\midrule
$a_0$ & \texttt{\phantom{-}2.201164677187738E-008}&\texttt{\phantom{-}2.201164677187727E-008}
\\
$c_{-2}$ & \texttt{-2.347908989000171E-007}&\texttt{-2.347908989000179E-007}
\\
$c_{-1}$ & \texttt{-2.082520105664531E-007}&\texttt{-2.082520105681483E-007}
\\
$c_0$ & \texttt{\phantom{-}3.909384299566400E-007}&\texttt{\phantom{-}3.909384299635230E-007}
\\\bottomrule
\end{tabular}
\end{center}

\subsection{Processes with a single vector boson}

\subsubsection{The process $u\bd\to e^+\nu_e$}

It is a matter of trivial algebra to show that for this process one has:
\beqn
c_0=\frac{\as}{2\pi}C_F\big(-\log^2(\mu^2/s)-
3\log(\mu^2/s)+\pi^2-8\big)\,a_0\,,
\eeqn
where $s$ is the parton center-of-mass energy squared.
For this simple case, it suffices thus to compute the ratio $c_0/a_0$,
which only depends on $s$, $\mu$ and $\as$. These parameters
are chosen as follows:
\begin{center}
\begin{tabular}{cl|cl}\toprule
Parameter & value & Parameter & value
\\\midrule
$\as$ & \texttt{0.118} & $\mu$ & \texttt{91.1876}
\\
$\sqrt{s}$ & \texttt{200} & & 
\\\bottomrule
\end{tabular}
\end{center}
We find:
\begin{center}
\begin{tabular}{cll}\toprule
  $u\bd\to e^+\nu_e$& \MadLoop\ & Analytic result
\\\midrule
$c_0/a_0$ & \texttt{0.10303097397333823}&\texttt{0.10303099058268723}
\\\bottomrule
\end{tabular}
\end{center}

\subsubsection{The processes $u\bd\to\nu_ee^+ g$ and $ug\to\nu_ee^+ d$}
These two processes contribute to the NLO corrections to $W+1$ jet.
We compare the \MadLoop\ results with the implementation of
MCFM~\cite{Campbell:1999ah}. The following input parameters
are used:
\begin{center}
\begin{tabular}{cl|cl}\toprule
Parameter & value & Parameter & value
\\\midrule
$\as$ & \texttt{0.118} & $n_{lf}$ & \texttt{5}
\\
$m_Z$ & \texttt{91.1876} & $\mu$ & \texttt{91.1876}
\\
$m_W$ & \texttt{80.44} & $\alpha^{-1}$ & \texttt{132.6844139}
\\
$\sin^2\theta_W$ & $1-m_W^2/m_Z^2$ &  & 
\\\bottomrule
\end{tabular}
\end{center}
The kinematic configuration is:
\begin{small}
  \begin{align*}
    p_u&=\textrm{(214.56992446426548, 0, 0, 214.56992446426548 )}\\
    p_{\bar{d},g}&=\textrm{(76.595570417607490, 0, 0, -76.595570417607490)}\\
    p_{\nu_e}&=\textrm{(186.05703769425895, -25.245095379680929, 11.566386894022147, 183.97316415458937)}\\
    p_{e^+}&=\textrm{(34.360975783073229, 23.891509313117499, 15.166967889135465, 19.489369526901015)}\\
    p_{g,d}&=\textrm{(70.747481404540792, 1.3535860665634296, -26.733354783157612, -65.488179634832392)}
  \end{align*}
\end{small}
The finite part is given in the DR scheme. We obtain what follows:
\begin{center}
\begin{tabular}{cll}\toprule
  $u\bar{d}\to \nu_ee^+ g$& \MadLoop\ & MCFM
\\\midrule
$a_0$ & \texttt{\phantom{-}0.93604465169278606}&\texttt{\phantom{-}0.93604465172998652}
\\
$c_{-2}$ & \texttt{-9.96153744476348524E-002}&\texttt{\phantom{-}--}
\\
$c_{-1}$ & \texttt{-0.17006206699872445}&\texttt{\phantom{-}--}
\\
$c_0$ & \texttt{\phantom{-}0.18788412330548998}&\texttt{\phantom{-}0.18788412331301654}
\\\midrule
  $u g \to \nu_ee^+ d$&&
\\\midrule
$a_0$ & \texttt{\phantom{-}0.25449996399907276}&\texttt{\phantom{-}0.25449996400918706}
\\
$c_{-2}$ & \texttt{-2.70842947126827029E-002}&\texttt{\phantom{-}--}
\\
$c_{-1}$ & \texttt{-4.43691342449630960E-002}&\texttt{\phantom{-}--}
\\
$c_0$ & \texttt{\phantom{-}4.58106840475423702E-002}&\texttt{\phantom{-}4.58106840493787554E-002}
\\\bottomrule
\end{tabular}
\end{center}
The residues of the double and single poles have been checked
against those returned by \MadFKS, and perfect agreement has been found.

\subsubsection{The processes $d\bd (\to \gamma^*/Z)\to e^-e^+ g$ and 
$dg (\to \gamma^*/Z)\to e^-e^+ d$\label{Z1j}}
These two processes contribute to the NLO corrections to $Z+1$ jet.
We compare the \MadLoop\ results with the implementation of
MCFM~\cite{Campbell:1999ah}. 
The amplitudes in MCFM do not include closed fermion loops attached to
the vector boson, which simplifies the calculation slightly. This means
that in \MadLoop\ we have to remove these contributions
(with the \texttt{UserFilter} function in the \texttt{Diagram.cpp}
file). It is straightforward to remove the triangle loop diagrams
attached to the EW vector boson. There is, however, one subtlety: 
\texttt{UserFilter} does not affect the $R_2$ contributions. Even though
there is no $R_2$ contribution coming from closed fermion loops when
summing over a complete family (as one should do to have a proper
anomaly cancellation), when including only five massless quarks as done
in MCFM, the $R_2$ contribution is non zero. We therefore
have to set the one $R_2$ diagram with the ggV interactions to zero by
hand.

\begin{figure}[htb!]
  \begin{center}
        \epsfig{file=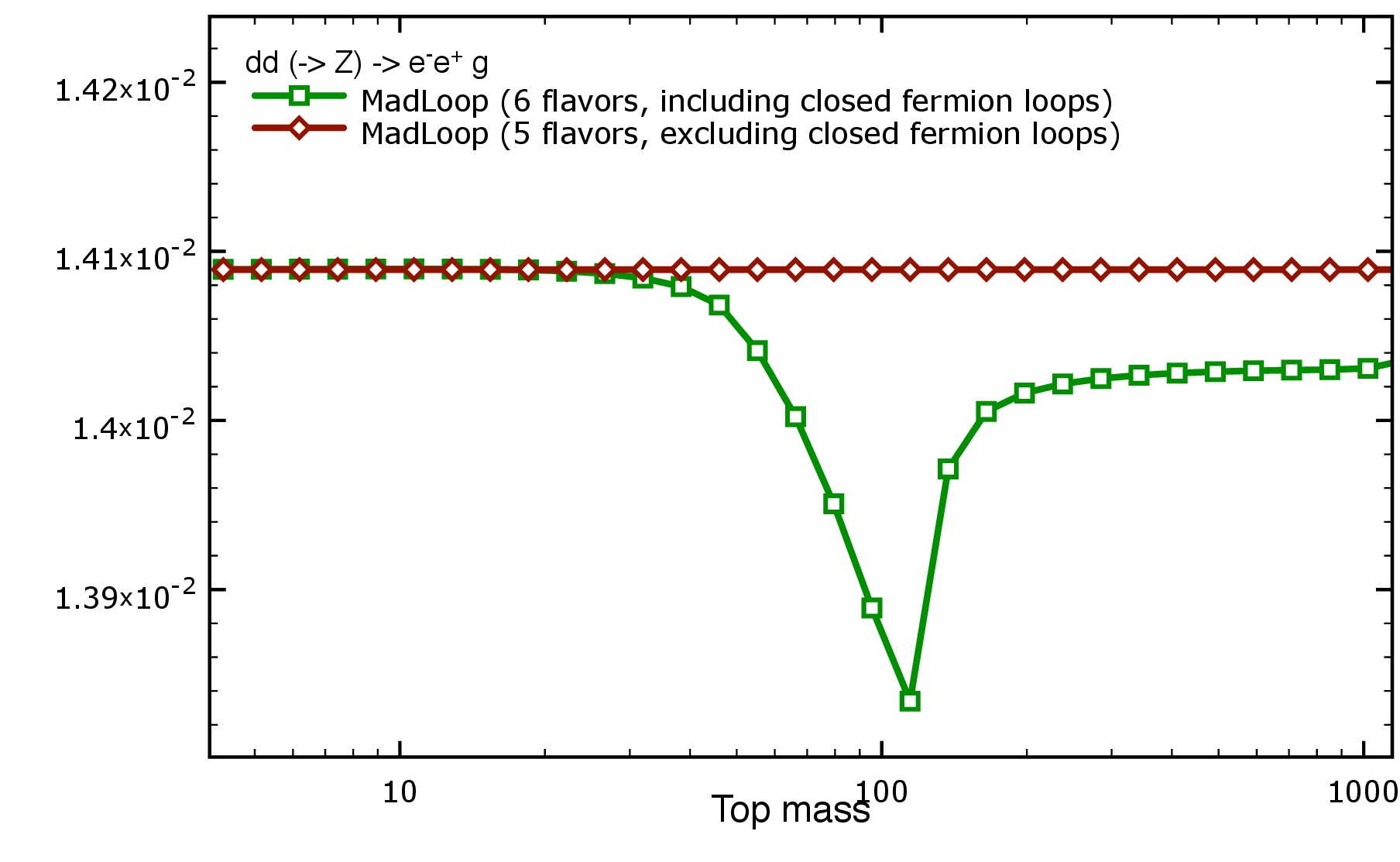, width=0.72\textwidth}
  \end{center}
  \vspace{-20pt}
  \caption{The finite part $c_0$ for $d\bd(\to\gamma^*/Z)\to e^-e^+ g$
    and the kinematic configuration reported in the text, as a function
    of $\mtop$, with five and six flavours circulating in the loop.
    In the five-flavour case all contributions due to closed fermion
    loops have been excluded. See the text for details.}
      \label{dd_Zg}
\end{figure}

The following input parameters are used:
\begin{center}
\begin{tabular}{cl|cl}\toprule
Parameter & value & Parameter & value
\\\midrule
$\as$ & \texttt{0.118} & $n_{lf}$ & \texttt{5}
\\
$m_Z$ & \texttt{91.1876} & $\mu$ & \texttt{91.1876}
\\
$m_W$ & \texttt{80.44} & $\alpha^{-1}$ & \texttt{132.6844139}
\\
$\sin^2\theta_W$ & $1-m_W^2/m_Z^2$ & $\Gamma_Z$ & \texttt{2.4952}
\\\bottomrule
\end{tabular}
\end{center}
The kinematic configuration is:
\begin{small}
  \begin{align*}
    p_d&=\textrm{(219.81636757818666, 0, 0, 219.81636757818666)}\\
    p_{\bar{d},g}&=\textrm{(78.514049708950481, 0, 0, -78.514049708950481)}\\
    p_{e^-}&=\textrm{(190.91987238779512, -28.468337054964493, 10.154026810698143, 188.51219376322723)}\\
    p_{e^+}&=\textrm{(36.663063494801236, 27.114750988401063, 16.579327972459467, 18.278303740841352)}\\
    p_{g,d}&=\textrm{(70.747481404540792, 1.3535860665634296, -26.733354783157612, -65.488179634832392)}
  \end{align*}
\end{small}
The finite part is given in the DR scheme. We obtain what follows:
\begin{center}
\begin{tabular}{cll}\toprule
  $d\bar{d}(\to \gamma^*/Z)\to e^-e^+ g$& \MadLoop\ & MCFM
\\\midrule
$a_0$ & \texttt{\phantom{-}6.76069763764682863E-002}&\texttt{\phantom{-}6.76069763035082871E-002}
\\
$c_{-2}$ & \texttt{-7.19484295416346949E-003}&\texttt{\phantom{-}--}
\\
$c_{-1}$ & \texttt{-1.21173922992901528E-002}&\texttt{\phantom{-}--}
\\
$c_0$ & \texttt{\phantom{-}1.40892172309674130E-002}&\texttt{\phantom{-}1.40892172165083020E-002}
\\\midrule
  $dg(\to \gamma^*/Z)\to e^-e^+ d$&&
\\\midrule
$a_0$ & \texttt{\phantom{-}1.71059868986021858E-002}&\texttt{\phantom{-}1.71059868803008740E-002}
\\
$c_{-2}$ & \texttt{-1.82044658566124528E-003}&\texttt{\phantom{-}--}
\\
$c_{-1}$ & \texttt{-2.91385295414951957E-003}&\texttt{\phantom{-}--}
\\
$c_0$ & \texttt{\phantom{-}3.17029216184223396E-003}&\texttt{\phantom{-}3.17029215866125201E-003}
\\\bottomrule
\end{tabular}
\end{center}
In order to further test the internal consistency of \MadLoop, we
have redone the calculation by including the contributions of the
closed fermion loops, that have been neglected in the comparison
with MCFM. In doing so, we are also able to perform the 
$\mtop$-dependence studies introduced in sect.~\ref{sec:checks}.
To be definite, we consider only the process 
\mbox{$d\bd(\to\gamma^*/Z)\to e^-e^+ g$}.
In fig.~\ref{dd_Zg}, we present the results for the finite part
$c_0$ as a function of $\mtop$, for the kinematic configuration
given above. We compare the five-flavour calculation
(which is independent of $\mtop$, and from which we remove the contribution
of closed fermion loops) with the six-flavour calculation that retains the 
full $\mtop$ dependence (and where closed fermion loops are included).
We see that in the decoupling limit ($\mtop\to\infty$) the six-flavour 
result does not agree with the five-flavour one. 
This difference is due to the non-anomalous part of the $b$-quark triangle 
diagrams, which contribute only to the six-flavour result (having been
excluded by hand from the five-flavour one), since in the decoupling
limit the third fermion family effectively includes only the $b$ quark
(the anomalous part, being mass-independent, is zero also in the
decoupling limit).

The limit for $\mtop\to 0$ is equal to the five-flavour 
result even though we are using a renormalization scheme in which we
subtract the top quark loop at zero momentum. This means that 
the heavy-flavour contributions to the 
UV counterterms relevant to strong coupling and gluon wave function 
renormalization diverge, but for this particular process there are an equal
number of powers of $\as$ as of external gluons, so that these divergences
cancel (see eqs.~(\ref{UVas}) and~(\ref{UVgwf})). 
Furthermore, due to the fact that the mass difference between
the top quark and the bottom quark goes to zero as well, the contribution 
of triangle diagrams is exactly zero at $\mtop=0$, and hence the six-flavour
result coincides with the five-flavour one.

\subsubsection{The processes $dc(\to W^-)\to e^-\bar{\nu}_e uc$ and 
$dg(\to W^-)\to e^-\bar{\nu}_e ug$\label{dcenuuc}}
The virtual corrections to these processes can be inferred from 
those relevant to $e^+e^-\to 4$ partons that have been first  
calculated by Bern, Dixon and Kosower (BDK)~\cite{Bern:1997sc}. 
Here we compare the \MadLoop\ results against the implementation of
the crossings of the BDK amplitudes in MCFM~\cite{Campbell:2002tg}.
The BDK amplitudes and their implementations in MCFM assume five
massless quark flavours, plus the top quark which is taken to be massive. 
However, only terms up to $1/\mtop^2$ have been kept, with higher
inverse powers of $\mtop$ being neglected. 
The \MadLoop\ implementation also features five massless quarks
plus a massive top quark, but the dependence on $\mtop$ is 
retained in full. Therefore, we only expect agreement with MCFM for large
top-quark masses. 
The following input parameters are used:
\begin{center}
\begin{tabular}{cl|cl}\toprule
Parameter & value & Parameter & value
\\\midrule
$\as$ & \texttt{0.118} & $n_{lf}$ & \texttt{5}
\\
$m_Z$ & \texttt{91.1876} & $\mu$ & \texttt{91.1876}
\\
$m_W$ & \texttt{80.44} & $\alpha^{-1}$ & \texttt{132.6844139}
\\
$\sin^2\theta_W$ & $1-m_W^2/m_Z^2$ & $\Gamma_W$ & \texttt{2.1054}
\\\bottomrule
\end{tabular}
\end{center}
and a diagonal CKM matrix. The kinematic configuration is:
\begin{small}
  \begin{align*}
    p_{d}&=\textrm{(77.882588584131682, 0, 0, 77.882588584131682)}\\
    p_{c,g}&=\textrm{(324.00529231091792, 0, 0, -324.00529231091792)}\\
    p_{e^-}&=\textrm{(41.228205880918381, -10.045117616293838, 17.003544184693592, -36.190330993144983)}\\
    p_{\bar{\nu}_e}&=\textrm{(169.31980614449293, -82.680596722191993, -81.515421497409065, -123.24103105934591)}\\
    p_u&=\textrm{(142.62173792637984, 51.789721902948564, 71.086870231086095, -112.27395831226823)}\\
    p_{c,g}&=\textrm{(48.718130943258451, 40.935992435537273, -6.5749929183706302, 25.582616637972905)}
  \end{align*}
\end{small}
\begin{figure}[htb!]
  \begin{center}
    \subfigure[]{
      \epsfig{file=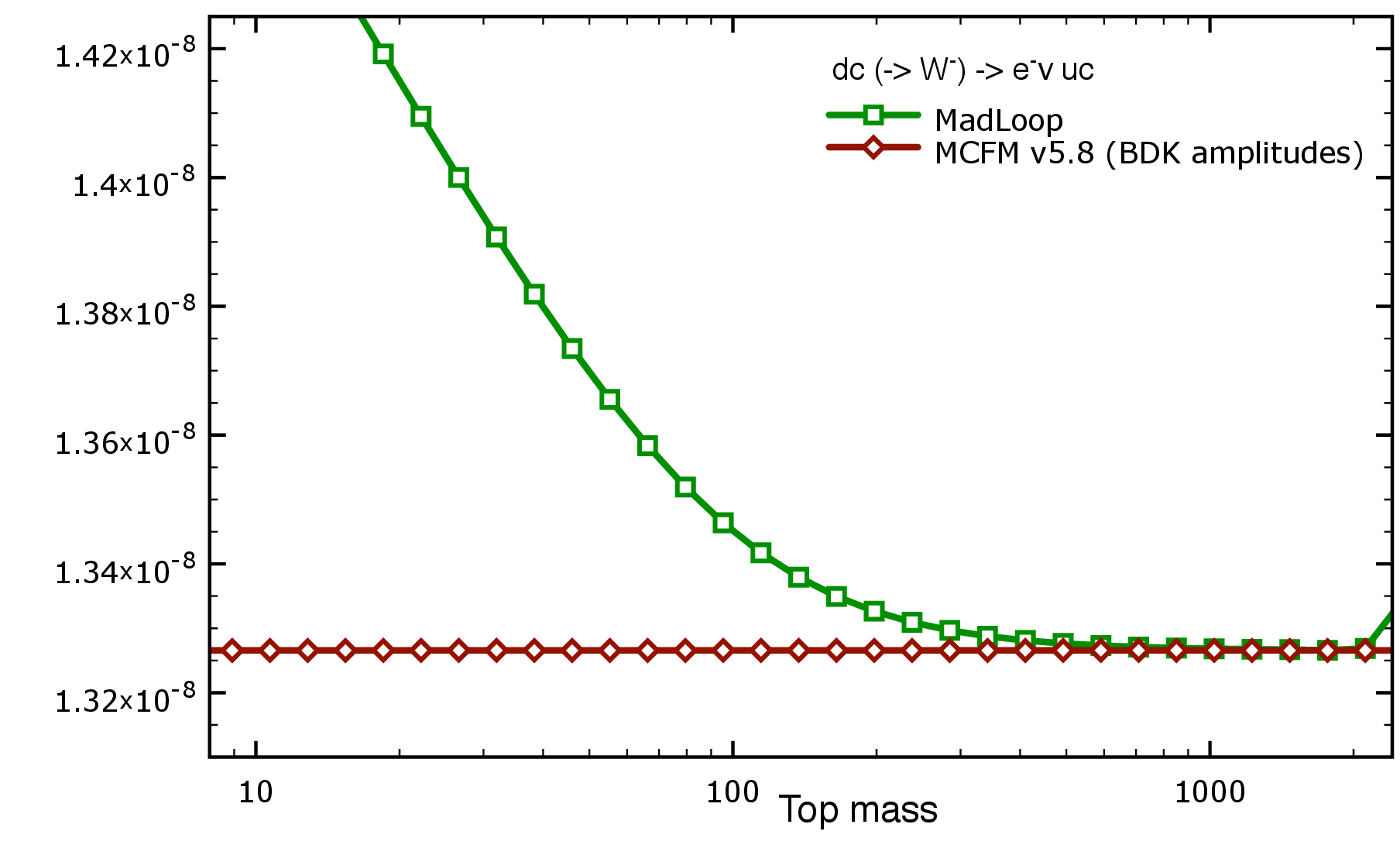, width=0.72\textwidth}
    }
    \subfigure[]{
      \epsfig{file=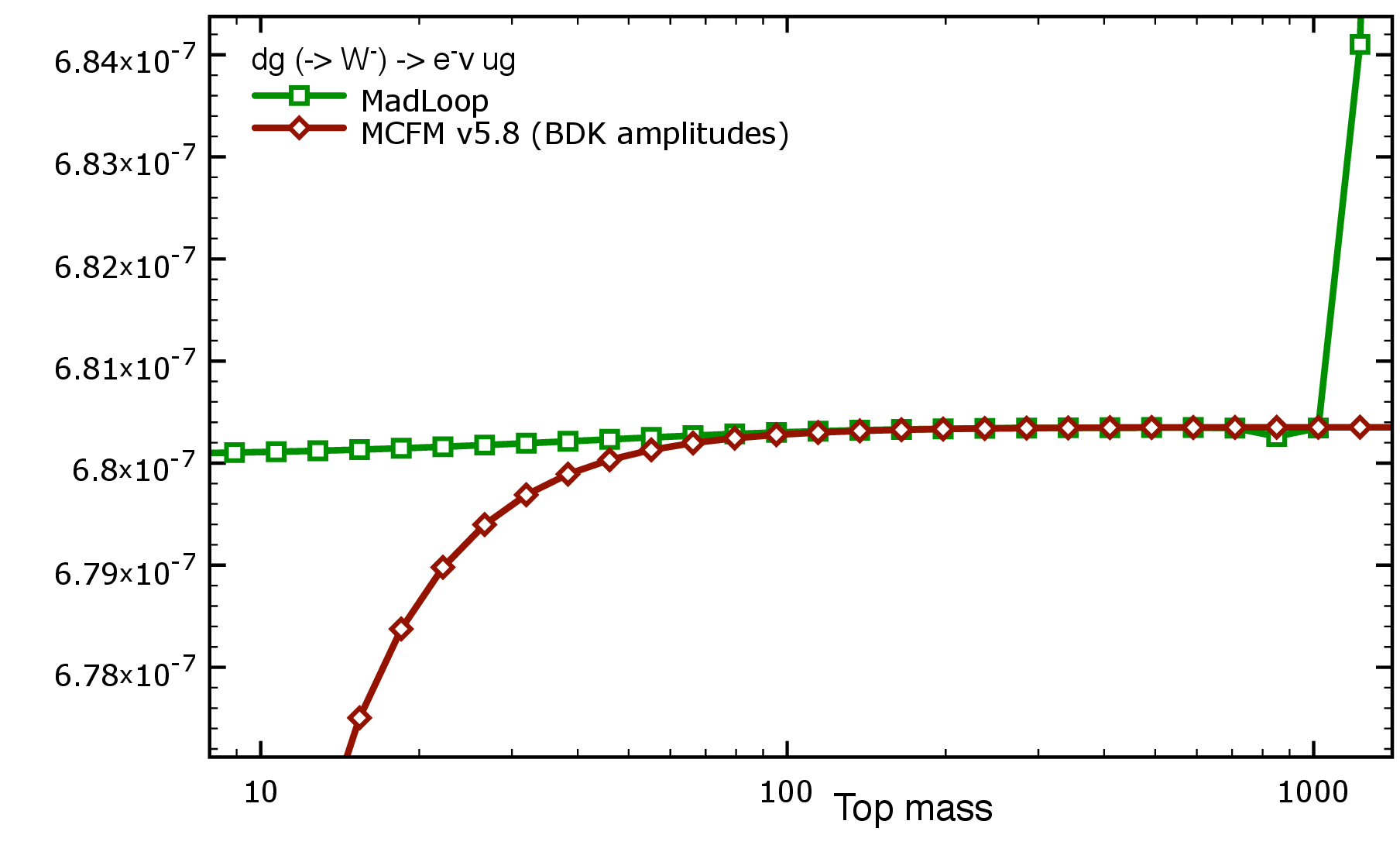, width=0.72\textwidth}
    }
  \end{center}
  \vspace{-20pt}
  \caption{The finite part $c_0$ for (a) $dc(\to W^-)\to e^-\bar{\nu}_e uc$ 
  and (b) $dg(\to W^-)\to e^-\bar{\nu}_e ug$, as a function of $\mtop$.
  \MadLoop\ (boxes) and MCFM (diamonds) results are shown.}
      \label{w2j}
\end{figure}
We present the results in the form of two plots where the finite part $c_0$
(in the DR scheme) is shown as a function of $\mtop$.

For the process $uc(\to W^-)\to e^-\bar{\nu}_e uc$, fig.~\ref{w2j}(a),
we find a relatively large dependence on the top quark mass in \MadLoop, 
while the MCFM result is a constant (as explained in ref.~\cite{Bern:1997sc}). 
However, from the plot it is clear that the two results converge
to the same number in the limit $\mtop\to\infty$. For 
\mbox{$\mtop\approx 1.7$~TeV}, the relative difference between \MadLoop\
and MCFM is smaller than $10^{-5}$. For yet larger top masses numerical 
instabilities in \MadLoop\ render it impossible to reach such level
of precision. These instabilities are due to large cancellations between 
the $R_2$ and the cut-constructible-plus-$R_1$ contributions. For a
sensible computation of their sum one would need to obtain these two
contributions with an accuracy beyond double precision.

For the process $dg(\to W^-)\to e^-\bar{\nu}_e ug$, fig.~\ref{w2j}(b),
the situation is similar. At $\mtop=600$~GeV, the relative difference 
between \MadLoop\ and MCFM is of the order of $10^{-7}$. For larger masses, 
\MadLoop\ displays the same numerical instabilities as those mentioned
above. 

We conclude this section by pointing out that the results presented
here do not allow one to assess the impact of terms of order $1/\mtop^4$ 
and higher on observable cross sections. In fact, although for the
physical value of the top mass these terms seem generally to have a 
small effect (${\cal O}(1\%)$), they do depend on the partonic channel
and the kinematic configuration chosen. A firm conclusion can thus be
reached only through a comparison at the level of integrated cross sections.

\subsubsection{The process $ug(\to Z/\gamma^*)\to e^-e^+ ug$\label{ugeeug}}
The virtual corrections to this process can be inferred from 
those relevant to $e^+e^-\to 4$ partons that have been first 
calculated by Bern, Dixon and Kosower (BDK)~\cite{Bern:1997sc}. 
Here we compare the \MadLoop\ results against the implementation of
the crossings of the BDK amplitudes in MCFM~\cite{Campbell:2002tg},
where they contribute to the $pp\to Z/\gamma^\star+2j$ cross section.
We start by restricting ourselves to testing the pure vector
coupling, and hence we can switch off completely the $Z$-boson 
exchange contributions in both \MadLoop\ and MCFM.
We point out that the same remark on the $\mtop$-dependence
of the BDK amplitudes applies here as in sect.~\ref{dcenuuc}.

For the present comparison we use the same input parameters
as in sects.~\ref{Z1j} and~\ref{dcenuuc}, and choose the following 
kinematic configuration:
\begin{small}
  \begin{align*}
    p_u&=\textrm{(79.343740010234328, 0, 0, 79.343740010234328)}\\
    p_g&=\textrm{(330.04970916921303, 0, 0, -330.04970916921303)}\\
    p_{e^-}&=\textrm{(44.502332440251230, -9.0128674723142481, 20.893959782906148, -38.244846151337029)}\\
    p_{e^+}&=\textrm{(173.55124786955787, -83.712846866171589, -85.405837095621621, -125.76978133334636)}\\
    p_u&=\textrm{(142.62173792637984, 51.789721902948564, 71.086870231086095, -112.27395831226823)}\\
    p_g&=\textrm{(48.718130943258451, 40.935992435537273, -6.5749929183706302, 25.582616637972905)}
  \end{align*}
\end{small}
\begin{figure}[htb!]
  \begin{center}
    \subfigure[]{
      \epsfig{file=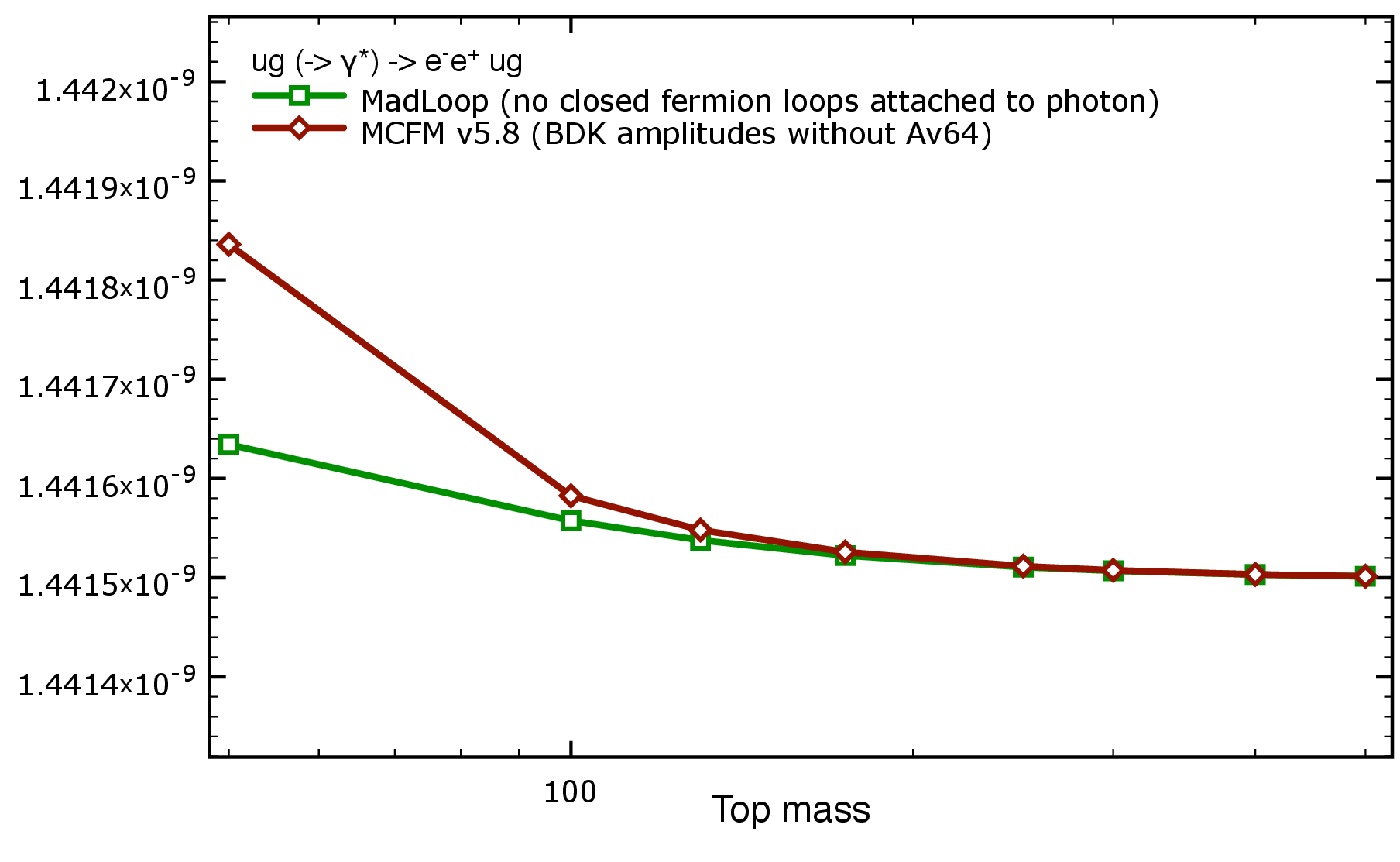, width=0.72\textwidth}
    }
    \subfigure[]{
      \epsfig{file=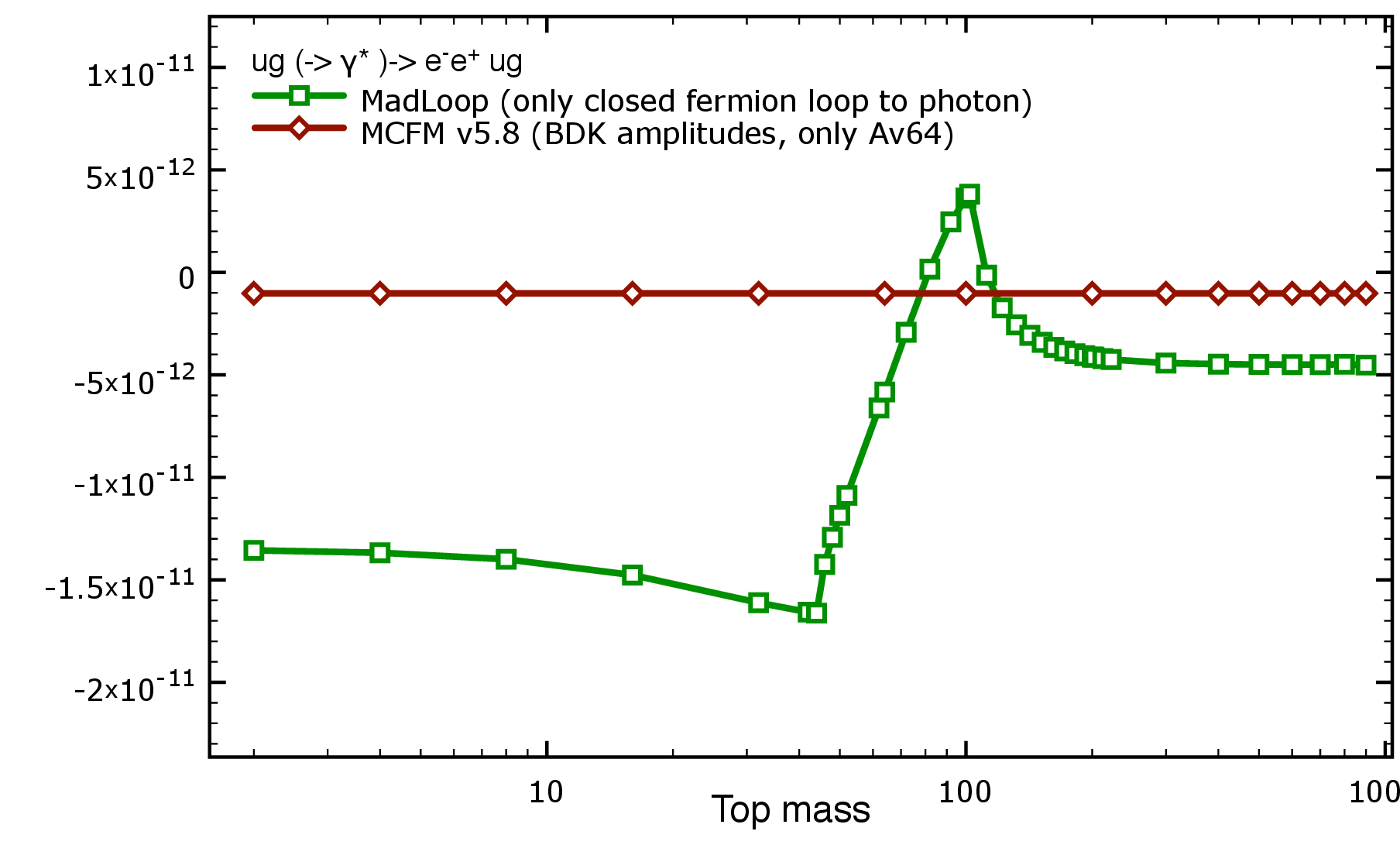, width=0.72\textwidth}
    }
  \end{center}
  \vspace{-20pt}
  \caption{Contributions to the finite part $c_0$ for 
   $ug(\to\gamma^*)\to e^-e^+ ug$,
   as a function of $\mtop$; \MadLoop\ (boxes) and MCFM (diamonds) 
   results are shown. Panel (a) is the contribution without closed 
   fermion loops attached to the photon, and panel (b) shows only the 
   contribution from those loops. The sum of the two is equal to $c_0$.}
      \label{ug_gammaug}
\end{figure}
The results are presented in fig.~\ref{ug_gammaug} as a function
of $\mtop$. Figure~\ref{ug_gammaug}(b) shows the contribution 
to the finite part $c_0$ due {\em only} to the closed fermion 
loops attached to the photon (up to the interference with the Born,
this contribution is called $A^v_{6;4}$ in ref.~\cite{Bern:1997sc}).
Figure~\ref{ug_gammaug}(a) shows all the other contributions to $c_0$.
The \MadLoop\ and MCFM results in fig.~\ref{ug_gammaug}(a) are seen
to be in excellent agreement for large top quark masses (at 
$\mtop=500$~GeV there is a relative difference of order $10^{-7}$).

The situation of fig.~\ref{ug_gammaug}(b) is different, since the
\MadLoop\ and MCFM results are in clear disagreement. The contribution
to $c_0$ considered here is due to six box diagrams, whose sum is
gauge invariant. Moreover, it is a finite contribution, so it does not 
require UV counterterms (nor terms to switch from the 't~Hooft-Veltman 
scheme to the DR scheme). We have checked that for $\mtop\to\infty$ 
the ratio between \MadLoop\ and MCFM is not an overall constant, but 
depends on the kinematic configuration chosen. We have also checked
that in such a decoupling limit the \MadLoop\ result agrees with the
one that we obtain by running \MadLoop\ with five massless flavours
only. Furthermore, in the limit $\mtop\to 0$ the \MadLoop\ result
coincides with that obtained by running \MadLoop\ with six massless
flavours. These tests clearly suggest that the box contribution is
either wrongly implemented in MCFM, or has been wrongly extracted
from MCFM by us. 

We have also carried out the analogue of the 
comparison shown in fig.~\ref{ug_gammaug}(b), by considering
the axial (for a $Z$ exchange) rather than the vector coupling.
We have found the same pattern as for the vector coupling, namely
a disagreement with MCFM\footnote{Triggered by the present results, 
John Campbell has reconsidered the implementation of BDK amplitudes in 
MCFM, and found a mistake in the vector part. To the best of our 
knowledge, the disagreement on the axial part persists to this day.}.
In order to confirm that there is no problem with
the computation of these box diagrams in \MadLoop, we have checked
them against a different implementation of the BDK amplitudes -- see
sect.~\ref{menlo_parc_gg}.

\subsubsection{Closed fermion loops contributing to 
$e^+e^-\to d\bd gg$\label{menlo_parc_gg}}
As anticipated in sect.~\ref{ugeeug}, we compare here 
the \MadLoop\ results for the BDK box and triangle 
diagrams~\cite{Bern:1997sc} (with the latter contributing only 
to the axial part),
with that of their implementation in the program
MENLO PARC~\cite{Signer:1996bf,Dixon:1997th}. We do so by
considering the process $e^+e^-\to d\bd gg$. We remind the reader
that the BDK amplitudes retain the top-mass dependence in the
loops only up to terms of order $1/\mtop^2$.

For our comparisons with use $\mu=m_Z=91.187$ and $\as=0.118$.
Furthermore, we set the coupling constants of the photon to the
fermions (both quarks and leptons) equal to one. For the $Z$ boson
couplings to fermions, we adopt a pure axial coupling, with magnitude $1$
for leptons and up-type quarks, and $-1$ for down-type quarks.
We choose the following kinematic configuration:
\begin{small}
  \begin{align*}
    p_{e^+}&=\textrm{(45.5935, 0, 0, 45.5935)}\\
    p_{e^-}&=\textrm{(45.5935, 0, 0, -45.5935)}\\
    p_{\bar{d}}&=\textrm{(30.844322198071779, -19.016873847975504, 2.6503380947327226, 24.139312933309792)}\\
    p_d&=\textrm{(11.409429861138499, -1.5750381678441148, -9.8113210223060605, 5.6064538099701196)}\\
    p_g&=\textrm{(21.321201379655506, 12.486676226087143, -12.407109827066350, -12.030800922457644)}\\
    p_g&=\textrm{(27.612046561134214, 8.1052357897324754, 19.568092754639686, -17.714965820822268)}
  \end{align*}
\end{small}
and present the result in the 't~Hooft-Veltman scheme. 

We start by considering the intermediate (off-shell) photon case. In such a
way we test only the vectorial couplings and we need only include
the six boxes in \MadLoop, which correspond to the $A^v_{6;4}$ term 
of ref.~\cite{Bern:1997sc}. In the program MENLO PARC, $A^v_{6;4}$ 
can be easily computed by setting the \texttt{color part} parameter 
equal to \texttt{vect}. As can be seen from fig.~\ref{ug_gammaug}(b),
this term is independent of the top mass. Thus, in order to compare it
to the prediction of \MadLoop, we can either consider the decoupling
limit, or exclude the top-quark contribution to the loops; we 
adopt the latter option here\footnote{We have also computed the decoupling
limit, and found a perfectly regular numerical behaviour.}. We find 
an excellent agreement:
\begin{center}
\begin{tabular}{cll}\toprule
$e^+e^-\to\gamma^*\to \bar{q}qgg$& \multirow{2}*{\MadLoop}& \multirow{2}*{Refs.~\cite{Signer:1996bf,Dixon:1997th}}
\\
(closed fermion loops) &&
\\\midrule
$c_0$ & \texttt{-4.50041164255807986E-007}&\texttt{-4.5004116164533885E-007}
\\\bottomrule
\end{tabular}
\end{center}

\begin{figure}[htb!]
  \begin{center}
        \epsfig{file=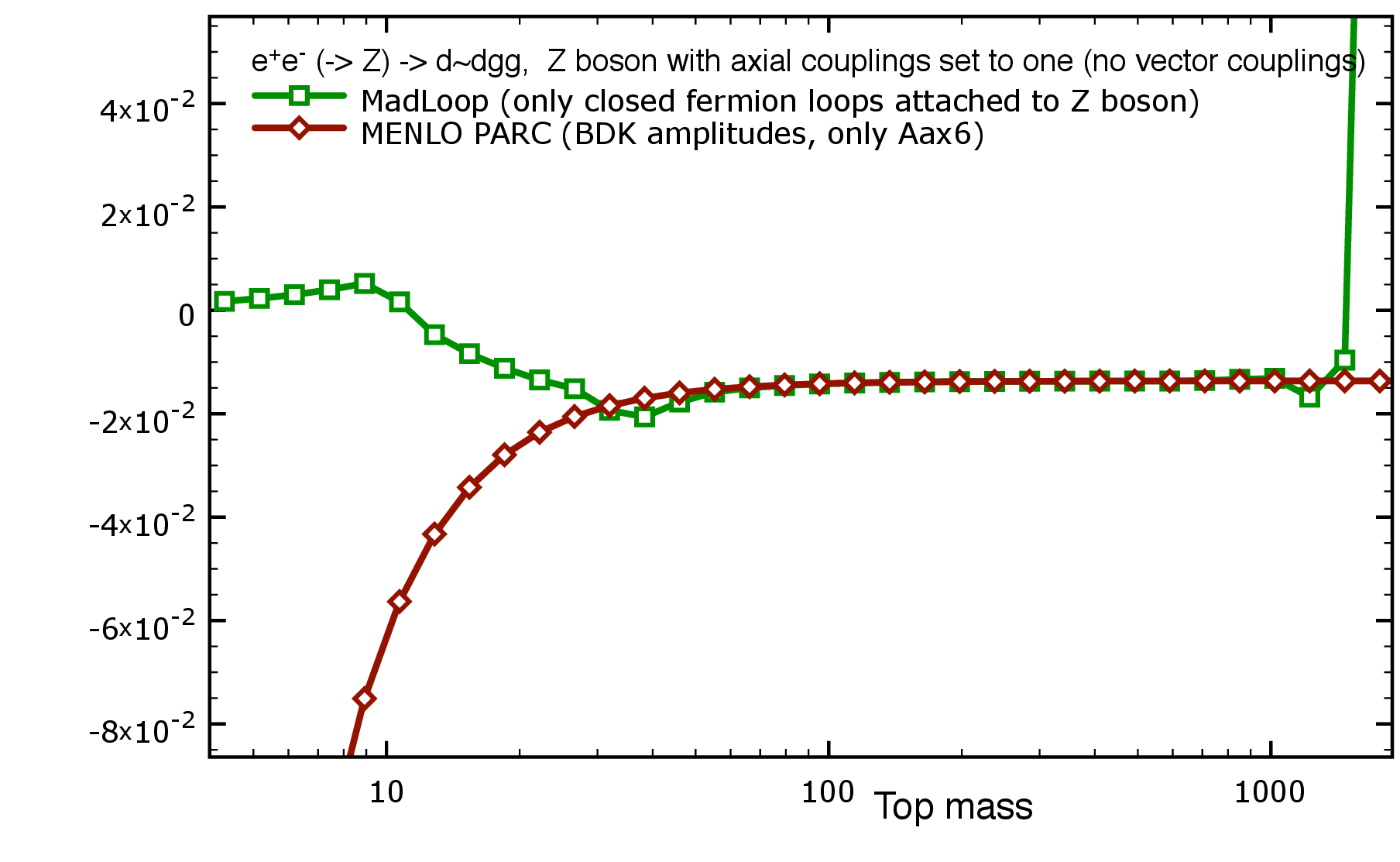, width=0.72\textwidth}
  \end{center}
  \vspace{-20pt}
  \caption{Pure-axial contributions to the finite part $c_0$ for 
   $e^+e^-\to d\bd gg$, as a function of $\mtop$; \MadLoop\ (boxes) 
   and MENLO PARK (diamonds) results are shown.}
      \label{menlo_parc_qqgg}
\end{figure}
We next consider the case of a purely-axial intermediate $Z$ boson.
This corresponds to the $A^{ax}_{6;4}$ and $A^{ax}_{6;5}$ terms 
of ref.~\cite{Bern:1997sc}, which can be obtained from MENLO PARC
by setting the parameter \texttt{color part} equal to \texttt{axal}.
The results are presented in fig.~\ref{menlo_parc_qqgg} as a function
of the top mass, for the same kinematic configuration as was used above.
As in the case of the purely vector couplings, we find excellent
agreement between the two results. In particular, for large $\mtop$
the relative difference between \MadLoop\ and MENLO PARK is 
of the order of $10^{-4}$. When \mbox{$\mtop={\cal O}(1~{\rm TeV})$}
numerical instabilities in \MadLoop\ spoil the accuracy of the
comparison; as in previous cases, these instabilities are due to
large cancellations between the cut-constructible-plus-$R_1$ 
and $R_2$ contributions.
For $\mtop\to0$, the BDK amplitudes diverge (due to the fact that terms
are kept only up to $1/\mtop^2$), while \MadLoop\ reproduces the correct 
result. This is equal to zero, since in this limit
the bottom and top quark loop contributions exactly cancel each other.

We have explicitly checked that \MadLoop\ is self-consistent, and
consistent with the results given in sect.~\ref{ugeeug}, by 
crossing the antiquark and one of the gluons to the initial state, and
the $e^+e^-$ pair to the final state.

\subsubsection{The $gg \to Zg$ one-loop amplitude squared\label{sec:gggZ}}
As a final check on the computations of closed fermion loop
diagrams with one external EW boson leg, we consider the
process $gg \to Zg$, whose amplitude was first computed by Van 
der Bij and Glover in ref.~\cite{vanderBij:1988ac}\footnote{The results 
also agree with an independent calculation by F.~Tramontano.}.
At variance with the cases discussed in 
sect.~\ref{ugeeug} and~\ref{menlo_parc_gg}, this process 
does not have a Born-level contribution. So we shall not
compute here the quantity $V$ defined in eq.~(\ref{Vdef}),
but rather the one-loop amplitude squared ($\abs{\ampnl}^2$),
summed/averaged over spins and colours\footnote{We also point out
that in the present case the $Z$ is on-shell, while the intermediate
vector bosons were off-shell in the cases 
discussed in sect.~\ref{ugeeug} and~\ref{menlo_parc_gg}.}.
This implies that for the present case we had to hack the \MadLoop\
code, in order for it to compute the square of an amplitude rather
than the interference of two amplitudes.
As far as the computation of ref.~\cite{vanderBij:1988ac} is concerned,
the helicity amplitudes presented in appendix B of that paper are
typed in {\tt Mathematica}, which is then used to perform all
subsequent analytical and numerical manipulations, and to obtain
the results denoted by ``Ref.~\cite{vanderBij:1988ac}'' in what follows.
We stress that the results of ref.~\cite{vanderBij:1988ac} are given
for one quark flavour circulating in the loop. 

\begin{figure}[htb!]
  \begin{center}
       \epsfig{file=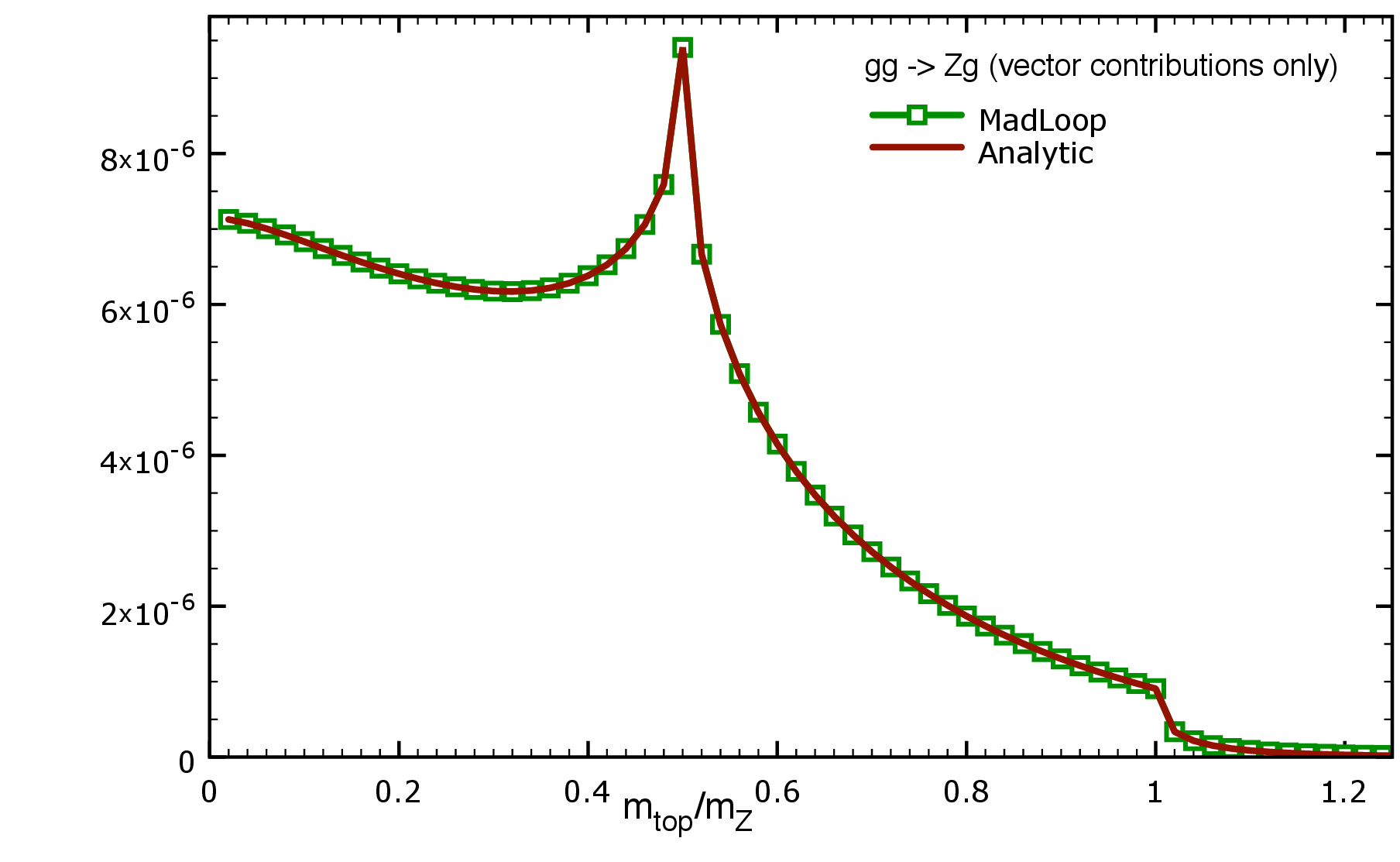, width=0.72\textwidth}
  \end{center}
  \vspace{-20pt}
  \caption{Square of the one-loop amplitude for $gg\to Zg$ (vectorial
  coupling only), as a function of $\mtop/m_Z$, with only the top quark
  circulating in the loop.}
      \label{gg_Zg_vector}
\end{figure}

We use the following input parameters:
\begin{center}
\begin{tabular}{cl|cl}\toprule
Parameter & value & Parameter & value
\\\midrule
$\as$ & \texttt{0.1176} & $\alpha^{-1}$ & \texttt{128.0}
\\
$M_Z$ & \texttt{91.188} & $\mtop$ & \texttt{80.0}
\\
$\sin^2 \theta_W $ & \texttt{0.23122} & $\mu$ & \texttt{91.188}
\\\bottomrule
\end{tabular}
\end{center}
We choose a kinematic configuration such that the Mandelstam variables
are $t=u=-\frac{3}{2}M^2_Z$, as defined in ref.~\cite{vanderBij:1988ac}. 
This can be obtained using e.g.~the following four-momenta:
\begin{small}
  \begin{align*}
    p_{1}=&\textrm{( 91.188}&&\textrm{, 0}&&\textrm{, 0}&&\textrm{, \phantom{-}91.188}&&\textrm{)}\\
    p_{2}=&\textrm{( 91.188}&&\textrm{, 0}&&\textrm{, 0}&&\textrm{, -91.188}&&\textrm{)}\\
    p_Z=&\textrm{( 113.985}&&\textrm{, -48.35973987212917}&&\textrm{, -48.35973987212917}&&\textrm{, 0}&&\textrm{)}\\
    p_{g_f}=&\textrm{( 68.391}&&\textrm{, \phantom{-}48.35973987212917}&&\textrm{, \phantom{-}48.35973987212917}&&\textrm{, 0}&&\textrm{)}
  \end{align*}
\end{small}
Note that $\sqrt{s}=2M_Z$, and hence we have adopted a value of $\mtop$
that is not effectively close to the decoupling limit -- this is useful
lest we have to deal with very small amplitude values. In any case,
the dependence on $\mtop$ has been studied as well, as can be 
seen in fig.~\ref{gg_Zg_vector}.
The axial and the vector coupling lead to amplitudes separately gauge
independent, which do not interfere (simply because of the symmetry and
antisymmetry of their colour factors respectively); we thus check them
independently. We start here with the vector coupling, for which we can 
limit ourselves to considering only the top quark running in the loop,
since the corresponing amplitude is not anomalous.
The resulting Feynman diagrams are then six (finite) massive fermion-loop 
boxes. We obtain:
\begin{center}
\begin{tabular}{cll}\toprule
$gg \to Zg$ vector & \MadLoop\ & Ref.~\cite{vanderBij:1988ac}
\\\midrule
$c_0$ & \texttt{\phantom{-}1.41346852305044352E-006}&\texttt{\phantom{-}1.4134685231123695E-006}
\\\bottomrule
\end{tabular}
\end{center}
We have also computed these results by varying $\mtop$.
The comparison between \MadLoop\ and ref.~\cite{vanderBij:1988ac} is
presented in fig.~\ref{gg_Zg_vector}. The thresholds at $\mtop/m_Z=0.5$ and
$\mtop/m_Z=1$ can be easily understood in terms of the optical theorem. 

\begin{figure}[htb!]
  \begin{center}
       \epsfig{file=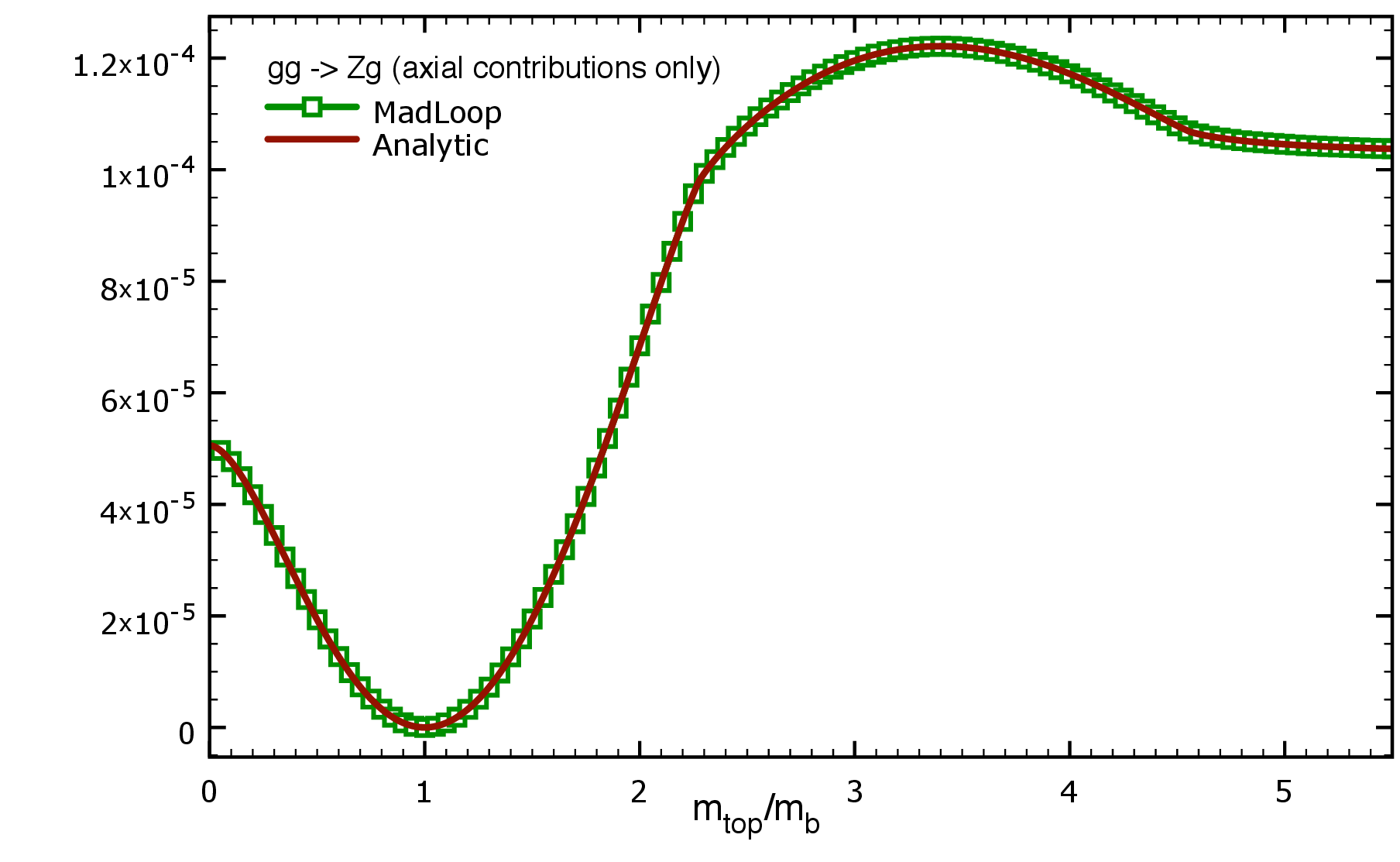, width=0.72\textwidth}
  \end{center}
  \vspace{-20pt}
  \caption{Square of the one-loop amplitude for $gg\to Zg$ (axial
  coupling only), as a function of $\mtop/m_b$, with the top and bottom 
  quarks circulating in the loop.}
      \label{gg_Zg_axial}
\end{figure}

When one considers the axial couplings of the $Z$ boson, there is an
anomalous contribution independent of the mass of the fermion that
circulates in the loop. The most straightforward way to circumvent
the problem of the anomaly is that of considering both the top and
the bottom quarks as loop particles. The bottom mass is kept fixed 
and equal to 20~GeV. We cannot set $m_b=0$ since the computation of 
ref.~\cite{vanderBij:1988ac} is performed assuming non-zero quark
masses. On the \MadLoop\ side, we have studied the bottom-mass
dependence, and found no peculiar numerical behaviours. We have
chosen $m_b=20$~GeV when presenting our results in order to be
able to explore both regions $\mtop\gg m_b$ and $\mtop\ll m_b$
(see fig.~\ref{gg_Zg_axial}). By setting $\mtop=80$~GeV we obtain:
\begin{center}
\begin{tabular}{cll}\toprule
$gg \to Zg$ axial & \MadLoop\ & Ref.~\cite{vanderBij:1988ac}
\\\midrule
$c_0$ & \texttt{\phantom{-}1.17192023257760489E-004}&\texttt{\phantom{-}1.1719202325756625E-004}
\\\bottomrule
\end{tabular}
\end{center}
Analogously to what was done in fig.~\ref{gg_Zg_vector}, we present in
fig.~\ref{gg_Zg_axial} the results for the axial part as a function
of $\mtop$. In the decoupling limit $\mtop\to \infty$, we are left
with the non-anomalous part of the contribution due to the bottom quark. 
As is expected, the contributions due to top and bottom quarks exactly
cancel each other when $\mtop=m_b$. We have also checked that this is
the case for various value of $m_b$ (including the large-mass limit).

We conclude this section by mentioning that we have checked all possible
gluon crossings of the process considered in this section. As for all other
cases in which gluons appear, gauge independence (eq.~(\ref{gaugeinv})) 
is automatically checked.

\subsubsection{The process $ug \to t\bb d$ (four-flavour $t$-channel 
single-top production)}
The $t$-channel single-top process in the four-flavour scheme is an
interesting case from the point of view of NLO
computations~\cite{Campbell:2009ss,Campbell:2009gj}. Even though it is
a $2\to3$ process at the Born level, the colour-singlet exchange of 
the $W$-boson in the $t$ channel ``disconnects'' NLO corrections to 
the light and heavy quark lines, essentially removing all pentagon diagrams. 
The complexity here arises then from the fact that both the top 
and the bottom quarks have to be treated as massive.

In order to generate this process with \MadLoop, a minimal amount of
manual work is required. This is due to the fact that in the present
version of the code the possibility is only given to specify the initial 
and final states, but not intermediate particles. Therefore, when
entering \mbox{$ug \to t\bb d$} as an input to \MadLoop\ we also
obtain the $s$-channel contributions, which we eventually set
equal to zero by hand. We use the following input parameters:
\begin{center}
\begin{tabular}{cl|cl}\toprule
Parameter & value & Parameter & value
\\\midrule
$\as$ & \texttt{0.118} & $n_{lf}$ & \texttt{4}
\\
$m_Z$ & \texttt{91.188} & $\mu$ & \texttt{91.188}
\\
$m_W$ & \texttt{80.419} & $\alpha^{-1}$ & \texttt{132.506980}
\\
$\mtop$ & \texttt{174.3} & $m_b$ & \texttt{4.5}
\\
$\sin^2\theta_W$ & $1-m_W^2/m_Z^2$ & $\Gamma_W$ & \texttt{2.0476}
\\\bottomrule
\end{tabular}
\end{center}
and a diagonal CKM matrix. The kinematic configuration is:
\begin{small}
  \begin{align*}
    p_u&=\textrm{(250, 0, 0, 250)}\\
    p_g&=\textrm{(250, 0, 0, -250)}\\
    p_t&=\textrm{(255.3729192644455, 29.17335589243201, 159.7715722928748, -91.96084974966891)}\\
    p_{\bar{b}}&=\textrm{(177.5248259329844, -66.11648748945143, -111.8173550700313, 120.9144450003231)}\\
    p_d&=\textrm{(67.10225480257022, 36.94313159701945, -47.95421722284348, -28.95359525065417)}
  \end{align*}
\end{small}
and we present the result in the 't~Hooft-Veltman scheme. 
We compare the results obtained by \MadLoop\ with the virtual 
corrections computed in ref.~\cite{Campbell:2009ss} and implemented 
in MCFM. We find excellent agreement:
\begin{center}
\begin{tabular}{cll}\toprule
$ug \to t\bar{b}d$& \multirow{2}*{\MadLoop}& \multirow{2}*{Ref.~\cite{Campbell:2009ss}}
\\
($t$-channel) &&
\\\midrule
$a_0$ & \texttt{\phantom{-}7.79629086614075984E-007}&\texttt{\phantom{-}7.79629086614075031E-007}
\\
$c_{-2}$ & \texttt{-8.29693789210587181E-008}&\texttt{-8.29693789210586651E-008}
\\
$c_{-1}$ & \texttt{\phantom{-}2.15034348206562335E-007}&\texttt{\phantom{-}2.15034348206885610E-007}
\\
$c_0$ & \texttt{\phantom{-}2.31517097632403642E-007}&\texttt{\phantom{-}2.31517097628348630E-007}
\\\bottomrule
\end{tabular}
\end{center}

\subsubsection{The process $u\bd(\to W^+)\to\nu_e e^+ b\bb$ with massive $b$}
This process has been computed very recently\footnote{The older computation
by Cordero, Reina and Wackeroth in ref.~\cite{Cordero:2006sj} does not
include spin correlations, i.e.~the $W$ is produced unpolarized
and on-shell.} by Badger, Campbell and Ellis 
and implemented in MCFM~\cite{Badger:2010mg}. Even 
though this part of the MCFM code is still private (but will be made 
public soon), John Campbell kindly provided us with a 
phase-space point to check against \MadLoop.
We have used the following input parameters:
\begin{center}
\begin{tabular}{cl|cl}\toprule
Parameter & value & Parameter & value
\\\midrule
$\as$ & \texttt{0.118298} & $n_{lf}$ & \texttt{4}
\\
$\sin^2\theta_W$ & \texttt{0.223} & $\mu$ & $m_W+2m_b$
\\
$m_W$ & \texttt{80.44} & 
 $G_F$ & \texttt{1.16639}$\cdot$\texttt{10}$^\texttt{-5}$ $\textrm{ GeV}^{-2}$
\\
$\mtop$ & \texttt{172.6} & $m_b$ & \texttt{4.62}
\\
$V_{ud}=V_{cs}$ & \texttt{0.974} & $V_{us}=V_{cd}$ & \texttt{0.227}
\\\bottomrule
\end{tabular}
\end{center}
and the kinematic configuration:
\begin{small}
  \begin{align*}
    p_u&=\textrm{(105.1910162427912, 0, 0, 105.1910162427912)}\\
    p_{\bar{d}}&=\textrm{(1315.106152715695, 0, 0, -1315.106152715695)}\\
    p_{\nu_e}&=\textrm{(552.4486825382789, 159.0272780479767, 133.6457759793429, -511.9069038778545)}\\
    p_{e^+}&=\textrm{(148.3926012456811, 43.98296570595704, -5.568920346765623, -141.6151473478021)}\\
    p_b&=\textrm{(348.0072091647683, 68.78221202106434, 5.036723885587666, -341.0737632135319)}\\
    p_{\bar{b}}&=\textrm{(371.4486760097574, -271.7924557749981, -133.1135795181650, -215.3193220337149)}
  \end{align*}
\end{small}
The finite part is given in the DR scheme. The $b$ quark mass is
set equal to zero in closed fermion loops as well as for 
$\as$ renormalization (while the exact top-quark-mass dependence 
is kept everywhere). We obtain:
\begin{center}
\begin{tabular}{cll}\toprule
$u\bd(\to W^+)\to\nu_e e^+ b\bb$ & \MadLoop\ & MCFM~\cite{Badger:2010mg}
\\\midrule
$a_0$ & \texttt{\phantom{-}1.79478780194792655E-007}&\texttt{\phantom{-}1.794787732024681E-007}
\\
$c_{-2}$ & \texttt{-9.01109441027896061E-009}&\texttt{\phantom{-}--}
\\
$c_{-1}$ & \texttt{\phantom{-}8.46671286067843610E-008}&\texttt{\phantom{-}--}
\\
$c_0$ & \texttt{\phantom{-}1.54008876289213431E-007}&\texttt{\phantom{-}1.540098268053898E-007}
\\\bottomrule
\end{tabular}
\end{center}
The residues of the double and single poles have been checked
against those returned by \MadFKS, and perfect agreement has been found.

\subsection{Processes with two vector bosons}
\subsubsection{The process $d\bar{d} \to W^+W^- \to \nu_e e^+ e^- \bar{\nu}_e$}

This process, that contributes to the NLO corrections to 
fully-decayed $W^+W^-$ production (i.e, all spin correlations are
included), has been implemented in MCFM~\cite{Campbell:1999ah}, 
using the virtual amplitudes calculated in ref.~\cite{Dixon:1998py}. 
The computation of ref.~\cite{Dixon:1998py} includes the singly-resonant 
contributions where an off-shell photon or $Z$ boson ``decays'' into a 
pair of $W$ bosons. However, it does not include diagrams in which the 
off-shell photon or $Z$ boson ``decays'' to leptons and one of the leptons 
radiates a $W$ boson. This latter contribution is however kinematically 
highly suppressed w.r.t.~the others, and its neglect is a very good
approximation for most physics applications. Such a contribution
is included in the \MadLoop\ result, but in
order not to consider it and be consistent with ref.~\cite{Dixon:1998py}
we simply set the couplings of photons and $Z$ bosons to leptons 
equal to zero.

We have used the following input parameters:
\begin{center}
\begin{tabular}{cl|cl}\toprule
Parameter & value & Parameter & value
\\\midrule
$\as$ & \texttt{0.118} & $n_{lf}$ & \texttt{5}
\\
$m_Z$ & \texttt{91.1876} & $\mu$ & \texttt{91.1876}
\\
$m_W$ & \texttt{80.44} & $\alpha^{-1}$ & \texttt{132.6844139}
\\
$\sin^2\theta_W$ & $1-m_W^2/m_Z^2$ & $\Gamma_W$ & \texttt{2.1054}
\\\bottomrule
\end{tabular}
\end{center}
and kinematic configuration:
\begin{small}
  \begin{align*}
    p_d&=\textrm{(39.534683750772302, 0, 0, 39.534683750772302)}\\
    p_{\bar{d}}&=\textrm{(546.24075297747743, 0, 0, -546.24075297747743)}\\
    p_{\nu_e}&=\textrm{(188.27600670927578, 3.8276243346653374, -38.361733789650529, -184.28668257634874)}\\
    p_{e^+}&=\textrm{(295.10612392593191, 49.617890129404948, 30.642119343108476, -289.28662236587513)}\\
    p_{e^-}&=\textrm{(41.828055877825669, -7.1022701637404531, -30.841911801229820, -27.348135100677510)}\\
    p_{\bar{\nu}_e}&=\textrm{(60.565250215216373, -46.343244300329829, 38.561526247771873, -5.7846291838037445)}
  \end{align*}
\end{small}
The finite part is given in the DR scheme. We obtain:
\begin{center}
\begin{tabular}{cll}\toprule
  $d\bar{d}\to \nu_e e^+ e^- \bar{\nu}_e$& \MadLoop\ &MCFM~\cite{Dixon:1998py,Campbell:1999ah}
\\\midrule
$a_0$ & \texttt{\phantom{-}1.11000204402873114E-004}&\texttt{\phantom{-}1.11000204410578607E-004}
\\
$c_{-2}$ & \texttt{-5.55897408896383675E-006}&\texttt{\phantom{-}--}
\\
$c_{-1}$ & \texttt{-4.67335122692354957E-006}&\texttt{\phantom{-}--}
\\
$c_0$ & \texttt{\phantom{-}2.24254527912372296E-005}&\texttt{\phantom{-}2.24254527928672022E-005}
\\\bottomrule
\end{tabular}
\end{center}
The residues of the double and single poles have been checked
against those returned by \MadFKS, and perfect agreement has been found.

\subsubsection{The processes $u\bu\to W^+W^-b\bb$ and
$gg\to W^+W^-b\bb$ with massless $b$\label{qqwwbb}}
$W$-boson pair plus $b$-quark pair production is among the different result
presented in ref.~\cite{vanHameren:2009dr} by A.~van Hameren \etal.
The comparisons performed in this section constitute therefore
a check of the implementation of the \CutTools\ software in a 
framework different from that of HELAC-1Loop~\cite{Bevilacqua:2010ve}
(see sect.~\ref{uubbbb} and~\ref{uuttbb} for a similar comparison).
The following set of parameters is used:
\begin{center}
\begin{tabular}{cl|cl}\toprule
Parameter & value & Parameter & value
\\\midrule
$g_{\sss S}$ & \texttt{1} & $n_{lf}$ & \texttt{5}
\\
$m_{top}$ & \texttt{174.0} & $m_{W}$ & \texttt{80.419}
\\
$\Gamma_{top}$ & \texttt{0} & $m_b$ & \texttt{0}
\\
$m_Z$ & \texttt{91.188} & $\Gamma_{Z}$ & \texttt{2.44140351}
\\
$G_F$ & \texttt{1.1663910e-05} & 
 $\sin^2{\theta_W} $ & $1-\frac{m^2_W}{m^2_Z}$
\\
$\alpha$ & $\sqrt{2}G_Fm^2_W\sin^2{\theta_W}/\pi$ & 
 $\mu$ & \texttt{500.0}
\\\bottomrule
\end{tabular}
\end{center}
The CKM matrix is diagonal and the Higgs channel (i.e., diagrams
that contain the ``decay'' $H\to W^+W^-$) is not
included\footnote{\MadLoop\ can easily compute this contribution, 
but it is very small and anyhow not included in the computation 
performed by Van Hameren \etal.}.  
We have chosen the following kinematic configuration:
\begin{small}
  \begin{align*}
    p_1=&\textrm{( 250}&&\textrm{, 0}&&\textrm{, 0}&&\textrm{, \phantom{-}250}&&\textrm{)}\\
    p_2=&\textrm{( 250}&&\textrm{, 0}&&\textrm{, 0}&&\textrm{, -250}&&\textrm{)}\\
    p_{W^+}=&\textrm{( 154.8819879118765}&&\textrm{, \phantom{-}22.40377113462118}&&\textrm{, -16.53704884550758}&&\textrm{, \phantom{-}129.4056091248114}&&\textrm{)}\\
    p_{W^-}=&\textrm{( 126.4095336206695}&&\textrm{, \phantom{-}92.64238702192333}&&\textrm{, -0.4920930146078141}&&\textrm{, \phantom{-}30.48443210132545}&&\textrm{)}\\
    p_b=&\textrm{( 174.1159068988160}&&\textrm{, -71.68369328357026}&&\textrm{, \phantom{-}6.716416578342183}&&\textrm{, -158.5329205583824}&&\textrm{)}\\
   p_{\bar{b}}=&\textrm{( 44.59257156863792}&&\textrm{, -43.36246487297426}&&\textrm{, \phantom{-}10.31272528177322}&&\textrm{, -1.357120667754454}&&\textrm{)}
  \end{align*}
\end{small}
with $p_1$ and $p_2$ the momenta of the initial-state partons coming
from the left ($u$ or $g$) and from the right ($\bu$ or $g$) respectively.
The finite part is given in the 't~Hooft-Veltman scheme. 
For consistency with ref.~\cite{vanHameren:2009dr}, no UV counterterms 
are included except the one relevant to top-mass renormalization. We obtain:
\begin{center}
\begin{tabular}{cll}\toprule
  $u\bar{u}\to W^+W^-b\bar{b} $& \MadLoop\ & Ref.~\cite{vanHameren:2009dr}
  \\\midrule
$a_0$ & \texttt{\phantom{-}2.338047209268890E-008}&\texttt{\phantom{-}2.338047130649064E-008}
\\
$c_{-2}$ & \texttt{-2.493920703542680E-007}&\texttt{-2.493916939359002E-007}
\\
$c_{-1}$ & \texttt{-4.885901939046758E-007}&\texttt{-4.885901774740355E-007}
\\
$c_0$ & \texttt{-2.775800623041098E-007}&\texttt{-2.775787767591390E-007}
\\\midrule
$gg\to W^+W^-b\bar{b} $& &
\\\midrule   
$a_0$ & \texttt{\phantom{-}1.549795815702494E-008}&\texttt{\phantom{-}1.549794572435312E-008}
\\
$c_{-2}$ & \texttt{-2.686312747217639E-007}&\texttt{-2.686310592221201E-007}
\\
$c_{-1}$ & \texttt{-6.078687041491385E-007}&\texttt{-6.078682316434646E-007}
\\
$c_0$ & \texttt{-5.519004042667462E-007}&\texttt{-5.519004727276688E-007}
\\\bottomrule
\end{tabular}
\end{center}
The $Z$-boson decay width is not specified in ref.~\cite{vanHameren:2009dr},
and this is most likely the reason for which we find a relative difference
of ${\cal O}(10^{-6})$ between \MadLoop\ and 
ref.~\cite{vanHameren:2009dr}\footnote{We point out, in fact, that the 
two results for the residue of the double pole can be made to coincide 
by multiplying the \MadLoop\ results by 
$\sigma_{\text{BornHELAC}}/\sigma_{\text{BornMadLoop}}$.}.
Since the agreement is however already quite satisfactory, we have
refrained from investigating this point further.

As a final remark it should be noted that this process is not yet suitable 
for {\em full} phase-space integration with \MadLoop. 
Due to possible intermediate top quarks, which can go on-shell, 
the top width should be taken into account. To do this in a gauge-independent 
and consistent way, a scheme  such as e.g.~the complex-mass one needs to 
be implemented\footnote{Using the complex-mass scheme, the phase-space 
integration for the process $pp \to W^+W^-b\bb$ at the NLO has recently 
been performed by two groups~\cite{Denner:2010jp,Bevilacqua:2010qb}.}. 
For a discussion on this point, see sects.~\ref{sec:limits} 
and~\ref{sec:outlook}.

\subsection{Processes with a single Higgs boson}

\subsubsection{The process $bg \to Hb$}
The process of SM Higgs boson production in association with a bottom
quark was computed by Campbell \etal\ in ref.~\cite{Campbell:2002zm}. 
The corresponding computer code is publicly available in the MCFM
package. In our comparison, we use the following input parameters:
\begin{center}
\begin{tabular}{cl|cl}\toprule
Parameter & value & Parameter & value
\\\midrule
$\as$ & \texttt{0.118} & $m_H$ & \texttt{120} 
\\
 $v$ & \texttt{246.2185}~\textrm{GeV} & $\mu$ & \texttt{91.188}
\\
$m_b$ & \texttt{0} & $\overline{m}_b(\mu)$ & \texttt{2.937956}
\\\bottomrule
\end{tabular}
\end{center}
with $\overline{m}_b(\mu)$  being used for the calculation
of the Yukawa coupling. We adopt the following kinematic configuration:
\begin{small}
  \begin{align*}
    p_b&=\textrm{(250, 0, 0, 250)}\\
    p_g&=\textrm{(250, 0, 0, -250)}\\
    p_H&=\textrm{(264.4, -83.84841332241601, -86.85350630148753, -202.3197272300720)}\\
    p_b&=\textrm{(235.6, 83.84841332241599, 86.85350630148751, 202.3197272300720)}
  \end{align*}
\end{small}
The finite part is given in the 't~Hooft-Veltman scheme. 
We obtain:
\begin{center}
\begin{tabular}{cll}\toprule
$bg \to Hb$& \MadLoop\ & Ref.~\cite{Campbell:2002zm}
\\\midrule
$a_0$ & \texttt{\phantom{-}3.11285493284766746E-007}&\texttt{\phantom{-}3.11285493372811162E-007}
\\
$c_{-2}$ & \texttt{-3.31275018959845830E-008}&\texttt{-3.31275018959846227E-008}
\\
$c_{-1}$ & \texttt{\phantom{-}6.99063829676915201E-008}&\texttt{\phantom{-}6.99063829676930686E-008}
\\
$c_0$ & \texttt{-1.41076086675311370E-007}&\texttt{-1.4107608671538634E-007}
\\\bottomrule
\end{tabular}
\end{center}

\subsubsection{The process $gb \to H^-t$}
Charged-Higgs boson production in association with a top quark was
first calculated in ref.~\cite{Zhu:2001nt}. Recently, the computation
has been redone in two independent ways in the context of the implementation
of this process in the MC@NLO framework~\cite{Frixione:2002ik,Weydert:2009vr} 
for a generic 2HDM model. We use the latter result in our comparisons.
We adopt the following input parameters:
\begin{center}
\begin{tabular}{cl|cl}\toprule
Parameter & value & Parameter & value
\\\midrule
$\as$ & \texttt{0.10751760258646566} & $m_b$ & \texttt{0}
\\
$\mtop$ & \texttt{174.3} & $\mu$ & \texttt{174.3}
\\
$m_H$ & \texttt{120} & $a$ & \texttt{-0.0170580225049951247} 
\\
$b$ & \texttt{0} &  & 
\\\bottomrule
\end{tabular}
\end{center}
with $a$ and $b$ the coefficients entering the $Htb$ vertex according
to the conventions of ref.~\cite{Weydert:2009vr}:
\beqn
G_{H^-tb}=iV_{tb}\left(a-b\gamma_5\right)\,,
\eeqn
with a diagonal CKM matrix. We point out that the $R_2$ SM vertices 
are sufficient for the computation of this process.
We have chosen the following kinematic configuration:
\begin{small}
  \begin{align*}
    p_g&=\textrm{(200, 0, 0, 200)}\\
    p_b&=\textrm{(200, 0, 0, -200)}\\
    p_{H^-}&=\textrm{(180.0243875, -120.4281794945461, -1.755237425897029, 59.18405883308687)}\\
    p_t&=\textrm{(219.9756125, -120.4281794945461, -1.755237425897029, 59.18405883308687)}
  \end{align*}
\end{small}
The finite part is given in the 't~Hooft-Veltman scheme. We obtain:
\begin{center}
\begin{tabular}{cll}\toprule
$gb \to H^-t$& \MadLoop\ & Ref.~\cite{Weydert:2009vr}
\\\midrule
$a_0$ & \texttt{\phantom{-}1.10048820395828282E-005}&\texttt{\phantom{-}1.10048820395828078E-005}
\\
$c_{-2}$ & \texttt{-8.16032006344512711E-007}&\texttt{\phantom{-}--}
\\
$c_{-1}$ & \texttt{\phantom{-}2.77298585886145253E-007}&\texttt{\phantom{-}--}
\\
$c_0$ & \texttt{\phantom{-}5.32062254695102591E-007}&\texttt{\phantom{-}5.32062219706480764E-007}
\\\bottomrule
\end{tabular}
\end{center}
The residues of the double and single poles have been checked
against those returned by \MadFKS, and perfect agreement has been found.

\subsubsection{The processes $u\bu\to t\bt H$
and $gg\to t\bt H$}
Two groups have calculated these two subprocesses
that contribute to $t\bar{t}H$ hadroproduction at the NLO in
QCD -- see refs.~\cite{Reina:2001bc,Dawson:2002tg}
and refs.~\cite{Beenakker:2001rj,Beenakker:2002nc}.
However, their codes are not publicly available. We therefore
compare \MadLoop\ results with those obtained with the HELAC-1Loop
code~\cite{vanHameren:2009dr}.
We use the following input parameters:
\begin{center}
\begin{tabular}{cl|cl}\toprule
Parameter & value & Parameter & value
\\\midrule
$\as$ & \texttt{0.1076395107858145} & $m_H$ & \texttt{130} 
\\
$\mtop$ & \texttt{172.6} & $\mu$ & \texttt{172.6}
\\
$v$ & \texttt{246.21835258713082} &  & 
\\\bottomrule
\end{tabular}
\end{center}
with the running mass of the top entering the Yukawa coupling set equal
to the pole mass. We have chosen the following kinematic
configuration:
\begin{small}
  \begin{align*}
    p_{u,g}&=\textrm{(250, 0, 0, 250)}\\
    p_{\bu,g}&=\textrm{(250, 0, 0, -250)}\\
    p_t&=\textrm{(181.47665951104506, 20.889486679044587, -50.105625289561424, 14.002628607367491)}\\
    p_{\bar{t}}&=\textrm{(182.16751255202476, -36.023358488530903, 22.118891298530357, -40.091332234320859)}\\
    p_H&=\textrm{(136.35582793693018, 15.133871809486299, 27.986733991031045, 26.088703626953386)}
  \end{align*}
\end{small}
The finite part is given in the 't~Hooft-Veltman scheme.
We obtain:
\begin{center}
\begin{tabular}{cll}\toprule
  $u\bu\to t\bar{t}H$ & \MadLoop\ & Ref.~\cite{vanHameren:2009dr}
  \\\midrule
$a_0$ & \texttt{\phantom{-}4.07927424576157583E-005}&\texttt{\phantom{-}4.07927080724888850E-005}
\\
$c_{-2}$ & \texttt{-1.86356048126662262E-006}&\texttt{-1.86355892362235005E-006}
\\
$c_{-1}$ & \texttt{-1.14634232495081623E-006}&\texttt{-1.14634136678800731E-006}
\\
$c_0$ & \texttt{-1.06894889909139909E-005}&\texttt{-1.06894800561762434E-005}
\\\midrule
$gg\to t\bar{t}H$& & 
\\\midrule   
$a_0$ & \texttt{\phantom{-}1.13589882608476193E-005}&\texttt{\phantom{-}1.13589786860990613E-005}
\\
$c_{-2}$ & \texttt{-1.16756954296913433E-006}&\texttt{-1.16756856706412638E-006}
\\
$c_{-1}$ & \texttt{\phantom{-}5.81128096324302437E-007}&\texttt{\phantom{-}5.81127610592268124E-007}
\\
$c_0$ & \texttt{\phantom{-}1.75265433284702459E-006}&\texttt{\phantom{-}1.75265286791477312E-006}
\\\bottomrule
\end{tabular}
\end{center}
We point out that it is likely that the agreement between the two codes
could be further improved, since we are aware of differences beyond 
single-precision accuracy between the input parameters used in the two codes.

\section{MadLoop technical details\label{sec:MLtechn}}

\subsection{Example of filtering of loop diagrams\label{sec:filter}}
In this appendix, we illustrate a simple example of the procedure
implemented in \MadLoop\ for the generation of one-loop diagrams
through \Lcut\ diagrams. We work in QCD with one light flavour, which
we identify with the $u$ quark, and consider the process
\beqn
e^+e^-\;\longrightarrow\;u\bu\,.
\label{eeuu}
\eeqn
As discussed in sect.~\ref{sec:genfil}, the associated \Lcut\ processes
are:
\beqn
e^+e^-&\longrightarrow& g^\star g^\star u\bu\,,
\label{gguu}
\\
e^+e^-&\longrightarrow& u^\star \bu^\star u\bu\,,
\label{uuuu}
\eeqn
since ghosts do not contribute.
\begin{figure}[thb]
 \begin{center}
~\hskip -2.5truecm\vskip 2.0truecm\hskip -2.5truecm
  \epsfig{figure=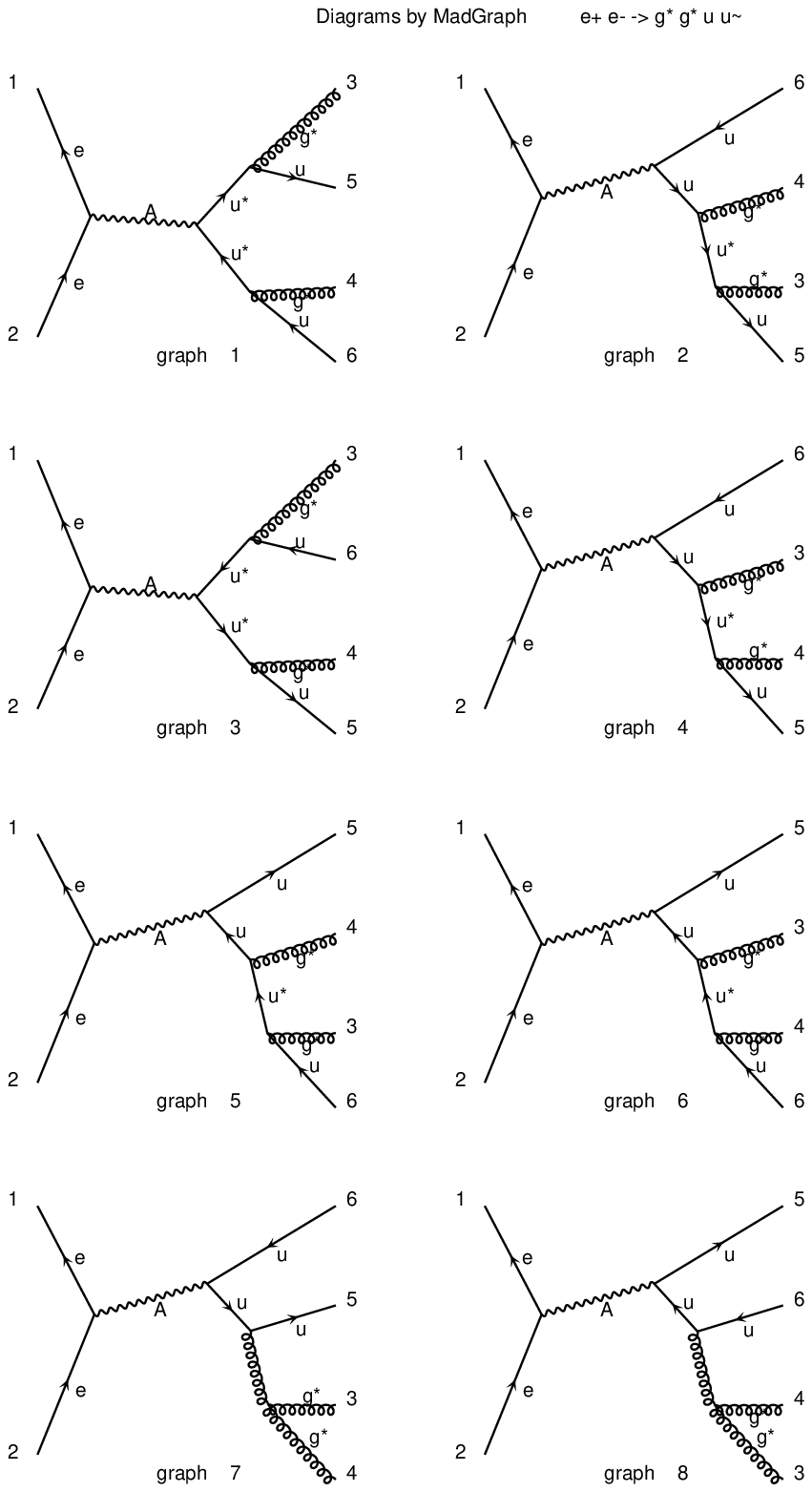,width=0.48\textwidth}
  \epsfig{figure=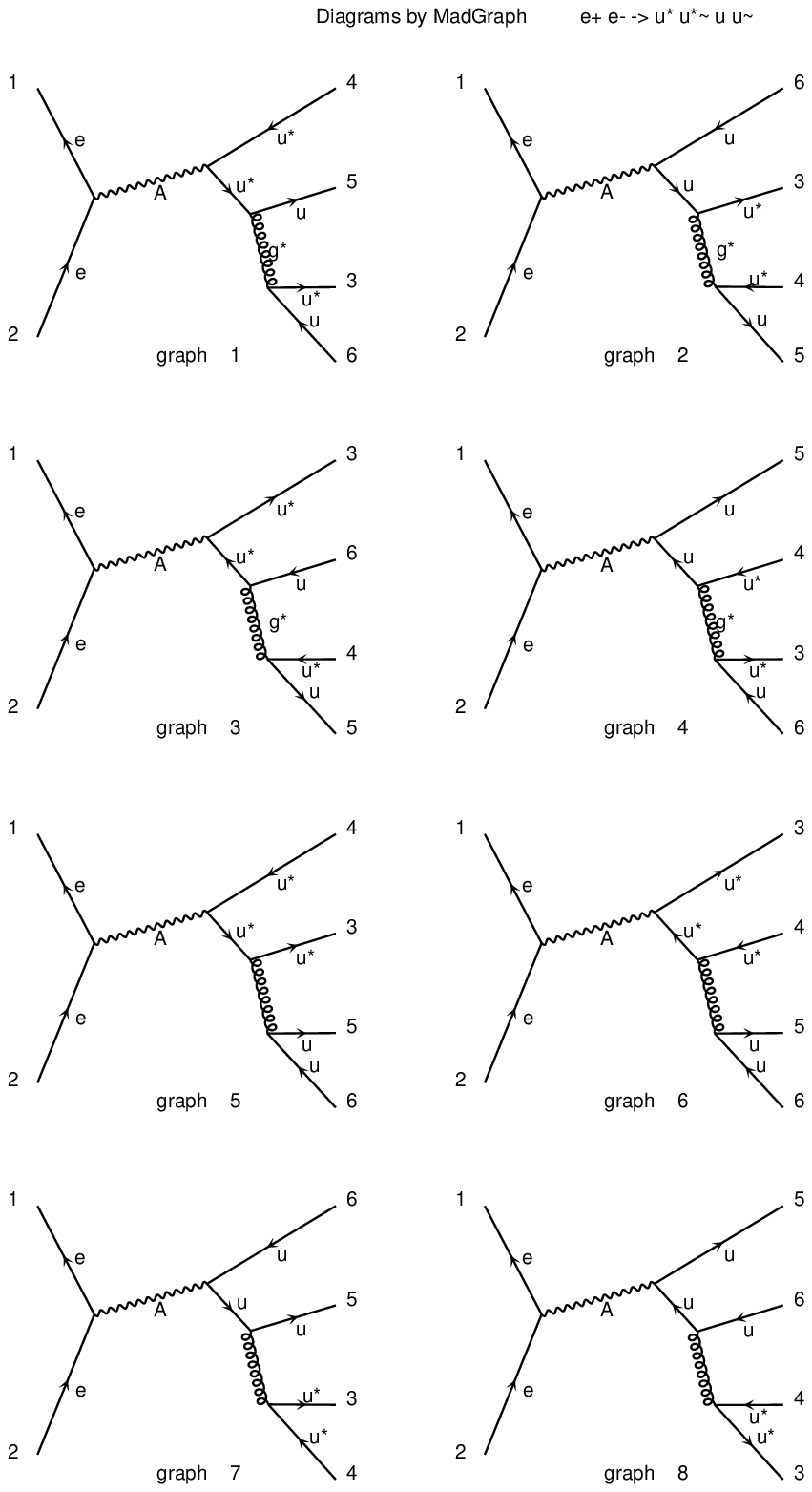,width=0.48\textwidth}
\caption{\label{fig:Lcutdiag} 
\Lcut\ diagrams of the processes in eqs.~(\ref{gguu}) and~(\ref{uuuu}).
}
 \end{center}
\end{figure}
The \Lcut\ diagrams are shown in fig.~\ref{fig:Lcutdiag}, and the
corresponding diagram identities are reported in tables~\ref{tab:Lcutglu}
and~\ref{tab:Lcutqrk}. These diagram identities are constructed in
the following way. Firstly, the \Lcut\ particles $\qpart$ and $\qbpart$ 
(with $(\qpart,\qbpart)=(g^\star,g^\star)$ and 
$(\qpart,\qbpart)=(u^\star,\bu^\star)$ for eqs.~(\ref{gguu}) 
and~(\ref{uuuu}) respectively) are assigned momenta equal to $\qloop$ 
and $-\qloop$ respectively. This implies that the loop momentum flows
from particles $\qbpart$ to particle $\qpart$. Secondly, for any given
\Lcut\ diagram, one starts from particle $\qpart$ and, by following 
the loop flow backwards, writes down either the identity of a loop
particle, or a symbol $T$ associated unambiguously with a tree structure
attached to the loop. The diagram identity is completed when particle
$\qbpart$ is encountered. As already anticipated in sect.~\ref{sec:genfil},
it is not necessary to keep track of whether a loop particle is a fermion
or an antifermion when defining diagram identities. In fact, this distinction
is meaningless if a fermion is not attached to an external fermion line
(i.e., in the case of a closed fermion loop), since in such a case it 
depends solely on the orientation of the loop momentum, which is easier
to keep track of than the fermion/antifermion identity. This observation
obviously applies to the case of ghost loops as well.
\begin{table}
\begin{center}
\begin{tabular}{cccl}\toprule
Diagram \# & ID & Topology & Action
\\\midrule
g.1 & $g^\star T_1 u^\star T_2 u^\star T_3 g^\star$ & triangle & keep
\\
g.2 & $g^\star T_1 u^\star T_4 g^\star$ & bubble & discard
\\
g.3 & $g^\star T_3 u^\star T_2 u^\star T_1 g^\star$ & triangle & 
 discard $(\equiv g.1)$
\\
g.4 & $g^\star T_4 u^\star T_1 g^\star$ & bubble & discard
\\
g.5 & $g^\star T_3 u^\star T_5 g^\star$ & bubble & discard
\\
g.6 & $g^\star T_5 u^\star T_3 g^\star$ & bubble & discard
\\
g.7 & $g^\star T_6 g^\star$ & tadpole & discard
\\
g.8 & $g^\star T_7 g^\star$ & tadpole & discard
\\\bottomrule
\end{tabular}
\end{center}
\caption{\label{tab:Lcutglu}
Identities of \Lcut\ diagrams for the process in eq.~(\ref{gguu}).
}
\end{table}
\begin{table}
\begin{center}
\begin{tabular}{cccl}\toprule
Diagram \# & ID & Topology & Action
\\\midrule
q.1 & $u^\star T_3 g^\star T_1 u^\star T_2 u^\star$ & triangle & 
 discard $(\equiv g.1)$
\\
q.2 & $u^\star T_4 g^\star T_1 u^\star$ & bubble & discard
\\
q.3 & $u^\star T_2 u^\star T_3 g^\star T_1 u^\star$ & triangle & 
 discard $(\equiv g.1)$
\\
q.4 & $u^\star T_3 g^\star T_5 u^\star$ & bubble & discard
\\
q.5 & $u^\star T_8 u^\star T_2 u^\star$ & bubble & keep
\\
q.6 & $u^\star T_2 u^\star T_8 u^\star$ & bubble & 
 discard $(\equiv q.5)$
\\
q.7 & $u^\star T_6 u^\star$ & tadpole & discard
\\
q.8 & $u^\star T_7 u^\star$ & tadpole & discard
\\\bottomrule
\end{tabular}
\end{center}
\caption{\label{tab:Lcutqrk}
Identities of \Lcut\ diagrams for the process in eq.~(\ref{uuuu}).
}
\end{table}
\begin{table}
\begin{center}
\begin{tabular}{cl}\toprule
Tree & Particle content
\\\midrule
$T_1$ & $u$ 
\\
$T_2$ & $\gamma e^+e^-$ 
\\
$T_3$ & $\bu$ 
\\
$T_4$ & $u(\gamma e^+e^-)\bu$ 
\\
$T_5$ & $\bu(\gamma e^+e^-)u$ 
\\
$T_6$ & $gu\big[u(\gamma e^+e^-)\bu\big]$ 
\\
$T_7$ & $g\bu\big[\bu(\gamma e^+e^-)u\big]$ 
\\
$T_8$ & $gu\bu$ 
\\\bottomrule
\end{tabular}
\end{center}
\caption{\label{tab:trees}
Tree structures used in tables~\ref{tab:Lcutglu} and~\ref{tab:Lcutqrk}.
The brackets indicate sub-trees, and are inserted here for the sole
purpose of simplifying the reading of the diagrams.}
\end{table}
On the other hand, when a fermion in the loop is attached to an external 
fermion line, the information on its particle/antiparticle identity is 
implicitly included in that relevant to the tree structure $T$ to which 
the external fermion belongs\footnote{It may not be trivial to exploit 
efficiently this information in the presence of four-fermion effective 
vertices. For theories with such vertices, 
it may in fact be convenient to keep track of the
fermion/antifermion identities in the loop. We defer the discussion of this
point to a future work.}. The difference between these two situations
is the reason why in the case of a purely-fermion (or ghost) 
loop one must {\em not}
consider as being equivalent two diagrams whose identities are equal up 
to mirror symmetry -- such two diagrams differ by the orientation of
the loop momentum, and for fermion or ghost loops both orientations 
must be considered\footnote{This seems to imply that one would keep both
orientations of a closed-fermion or ghost bubble. This is not the case, 
since the two corresponding \Lcut\ diagrams are identical up to a
cyclic permutation.}. Finally, as described in sect.~\ref{sec:genfil},
diagrams are filtered out according to the properties of their identities
under cyclic permutation and (possibly) mirror symmetry. It is not
difficult to see that this procedure allows one to associate symmetry
factors with loop diagrams as customary in QCD -- they are all equal to one,
except in the case of a gluon bubble, where such factor is equal to one-half.
This renders it trivial to take them into account in \MadLoop.

By following the general rules outlined above, the reader can work
out the diagram identities reported in table~\ref{tab:Lcutglu} 
(table~\ref{tab:Lcutqrk}) using the diagrams depicted in the 
left (right) panel of fig.~\ref{fig:Lcutdiag}. By doing so, one is
naturally led to introduce the tree structures reported in
table~\ref{tab:trees}. When filtering,
we (arbitrarily) begin from diagram \#1 of the process in
eq.~(\ref{gguu}), moving eventually on to diagrams associated with
the process in eq.~(\ref{uuuu}). A diagram is kept or filtered out
as indicated in the last columns of tables~\ref{tab:Lcutglu} 
and~\ref{tab:Lcutqrk}. All bubbles on external lines and tadpoles
are discarded by definition. Diagram g.3 is identical to diagram
g.1 up to mirror symmetry, while diagrams q.1 and q.3 are identical
to diagram g.1 up to cyclic permutations. Diagram q.6 is identical to
diagram q.5 up to a cyclic permutation. Diagram q.5 represents a
closed fermion bubble on an internal line; it must be taken into account,
but its contribution is equal to zero being proportional to the trace of
a single Gell-Mann matrix. We are thus left out with
one triangle loop diagram (arising from sewing diagram g.1) that
contributes to the one-loop corrections to eq.~(\ref{eeuu}), which
is of course the well-known result.

\subsection{UV renormalization counterterms\label{sec:UVcnts}}
As anticipated in sect.~\ref{sec:R2}, for UV renormalization we
use a scheme which subtracts the massless modes according to 
$\overline{\text{MS}}$, and the massive ones at zero momentum
(see e.g. ref.~\cite{Beenakker:2002nc}). In sect.~\ref{sec:R2}, we also
pointed out that for the cases we consider all UV counterterms except that 
relevant to mass renormalization give a contribution to eq.~(\ref{VR2UV}) 
which is proportional to the Born amplitude squared. We denote by $\CA$, 
$\CF$ and $\TF$ the usual colour factors. $\NC$ is the number of colours,
and $n_{lf}$ and $n_{hf}$ are the numbers of light and heavy flavours
respectively that circulate in the loops (a quark is by definition 
heavy if it has a non-zero mass). We make use of the prefactor 
$N_{\epsilon}$ defined as follows:
\beqn
N_{\epsilon}=\frac{1}{16\pi^2}(4\pi)^{\epsilon}\Gamma(1+\epsilon)\,.
\eeqn
If the Born cross section is of order $\as^b$, the contribution 
to eq.~(\ref{VR2UV}) due to strong-coupling renormalization reads:
\beqn
V^{\rm UV}_{\as}=b \abs{\ampnt}^2 \gs^2 N_{\epsilon} 
\left[ \frac{4}{3\epsilon}\TF n_{lf} - \frac{11}{3\epsilon}\CA
+ \frac{4}{3\epsilon}\TF
\sum_{\{n_{hf}\}} \left( \frac{\muR^2}{m_{hf}^2} \right)^{\epsilon} 
\right]\,,
\label{UVas}
\eeqn
where the sum in the third term on the r.h.s.~runs over all heavy 
flavours that circulate in the loops. The contribution due to the 
renormalization of  the Yukawa couplings reads:
\beqn
V^{\rm UV}_{\rm yuk}=-\abs{\ampnt}^2 \gs^2 N_{\epsilon} 2 \CF 
\left[ \frac{3}{\epsilon}n_{\text{yuk},l}+  
\left( 4 +\frac{3}{\epsilon} \right)\sum_{\{n_{\text{yuk},h}\}} 
\left( \frac{\muR^2}{m_{\text{yuk},h}^2} \right)^{\epsilon}  \right],
\label{UVyuk}
\eeqn
with $n_{\text{yuk},l}$ and $n_{\text{yuk},h}$ the number of
Yukawa vertices with massless and massive particles respectively.
Colour singlets and massless colour triplets do not require any 
wave-function renormalization. The gluon wave function is renormalized
only if there are massive colour triplets fermions running in the loop. 
Denoting by $n_g$ the number of external gluons at Born level, the
contribution to eq.~(\ref{VR2UV}) due to gluon wave-function
renormalization reads:
\beqn
V^{\rm UV}_{g_{\text{wf}}}=-n_g \abs{\ampnt}^2 \gs^2 N_{\epsilon} \TF 
\;\frac{4}{3\epsilon} 
\sum_{\{n_{hf}\}} \left( \frac{\muR^2}{m_{hf}^2} \right)^{\epsilon}\,.
\label{UVgwf}
\eeqn
The wave-function renormalization of the external massive quarks
(denoted $\text{ext}_{hf}$) gives the following contribution:
\beqn
V^{\rm UV}_{\text{ext}_{hf}}=- \abs{\ampnt}^2 \gs^2 N_{\epsilon} \CF 
\left( 4+ \frac{3}{\epsilon} \right) 
\sum_{\{ \text{ext}_{hf} \}}  
\left( \frac{\muR^2}{m_{\text{ext}_{hf}}^2} \right)^{\epsilon}\,.
\label{UVextm}
\eeqn
Finally, for carrying out mass renormalization for a quark line
with momentum $k$, mass $m$, and colour indices $i$ and $j$, one 
uses the mass insertion given by the following equation:
\beqn
\label{UVEnd}
\mathcal{G}_{ij}^{{\rm UV}_{\delta m}}(k)&=&
\frac{i\delta_{ik}}{\slashed{k}-m}(-i\delta m)
\frac{i\delta_{kj}}{\slashed{k}-m}\,,
\label{UVmassins}
\eeqn
with
\beqn
\delta m&=&\gs^2 \CF N_\epsilon \left( \frac{\muR^2}{m^2}\right)^{\epsilon}  
\left( 4 + \frac{3}{\epsilon}\right) m\,.
\eeqn

\end{document}